\documentclass[twocolumn,linenumbers]{aastex631}
\usepackage{CLASSY_Preamble}
\graphicspath{{./}{figures/}}
\nolinenumbers
\begin{document}

\submitjournal{AASJournal ApJ}
\shorttitle{CLASSY~II}
\shortauthors{James et al.}

\title{CLASSY II: A technical Overview of the COS Legacy Archive Spectroscopic SurveY\footnote{
Based on observations made with the NASA/ESA Hubble Space Telescope,
obtained from the Data Archive at the Space Telescope Science Institute, which
is operated by the Association of Universities for Research in Astronomy, Inc.,
under NASA contract NAS 5-26555.}}

\author[0000-0002-0786-7307]{Bethan L. James}
\affiliation{ESA for AURA \\
Space Telescope Science Institute \\
3700 San Martin Drive, Baltimore, MD 21218, USA}

\author[0000-0002-4153-053X]{Danielle A. Berg}
\affiliation{Department of Astronomy \\ 
The University of Texas at Austin \\
2515 Speedway, Stop C1400, Austin, TX 78712, USA}

\author[0000-0003-0834-4150]{Teagan King}
\affiliation{Space Telescope Science Institute \\ 
3700 San Martin Drive, Baltimore, MD 21218, USA}

\author[0000-0001-9594-161X]{David J. Sahnow}
\affiliation{Space Telescope Science Institute \\ 
3700 San Martin Drive, Baltimore, MD 21218, USA}

\author[0000-0003-2589-762X]{Matilde Mingozzi}
\affiliation{Space Telescope Science Institute, 3700 San Martin Drive, Baltimore, MD 21218, USA}
\collaboration{45}{and the CLASSY Team:}

\author[0000-0002-0302-2577]{John Chisholm}
\affiliation{Department of Astronomy, The University of Texas at Austin, 2515 Speedway, Stop C1400, Austin, TX 78712, USA}

\author[0000-0003-1127-7497]{Timothy Heckman}
\affiliation{Center for Astrophysical Sciences, Department of Physics \& Astronomy, Johns Hopkins University, Baltimore, MD 21218, USA}

\author[0000-0001-9189-7818]{Crystal L. Martin}
\affiliation{Department of Physics, University of California, Santa Barbara, Santa Barbara, CA 93106, USA}

\author[0000-0000-0000-0000]{Dan P. Stark}
\affiliation{Steward Observatory, The University of Arizona, 933 N Cherry Ave, Tucson, AZ, 85721, USA}

\author[0000-0003-4137-882X]{Alessandra Aloisi}
\affiliation{Space Telescope Science Institute, 3700 San Martin Drive, Baltimore, MD 21218, USA}

\author[0000-0001-5758-1000]{Ricardo O. Amor\'{i}n}
\affiliation{Instituto de Investigaci\'{o}n Multidisciplinar en Ciencia y Tecnolog\'{i}a, Universidad de La Serena, Raul Bitr\'{a}n 1305, La Serena 2204000, Chile}
\affiliation{Departamento de Astronom\'{i}a, Universidad de La Serena, Av. Juan Cisternas 1200 Norte, La Serena 1720236, Chile}

\author[0000-0002-2644-3518]{Karla Z. Arellano-C\'{o}rdova}
\affiliation{Department of Astronomy, The University of Texas at Austin, 2515 Speedway, Stop C1400, Austin, TX 78712, USA}

\author[0000-0003-1074-4807]{Matthew Bayliss}
\affiliation{Department of Physics, University of Cincinnati, Cincinnati, OH 45221, USA}

\author[0000-0002-3120-7173]{Rongmon Bordoloi}
\affiliation{Department of Physics, North Carolina State University, 421 Riddick Hall, Raleigh, NC 27695-8202, USA}

\author[0000-0003-4359-8797]{Jarle Brinchmann}
\affiliation{Instituto de Astrof\'{i]}sica e Ci\^{e}ncias do Espa\c{c}o, Universidade do Porto, CAUP, Rua das Estrelas, PT4150-762 Porto, Portugal}

\author[0000-0003-3458-2275]{St\'{e}phane Charlot}
\affiliation{Sorbonne Universit\'{e}, UPMC-CNRS, UMR7095, Institut d'Astrophysique de Paris, F-75014, Paris, France}

\author[0000-0002-2178-5471]{Zuyi Chen}
\affiliation{Steward Observatory, The University of Arizona, 933 N Cherry Ave, Tucson, AZ, 85721, USA}

\author[0000-0002-7636-0534]{Jacopo Chevallard}
\affiliation{Sorbonne Universit\'{e}, UPMC-CNRS, UMR7095, Institut d'Astrophysique de Paris, F-75014, Paris, France}

\author[0000-0003-3334-4267]{Ilyse Clark}
\affiliation{Department of Astronomy, The University of Texas at Austin, 2515 Speedway, Stop C1400, Austin, TX 78712, USA}

\author[0000-0001-9714-2758]{Dawn K. Erb}
\affiliation{Center for Gravitation, Cosmology and Astrophysics, Department of Physics, University of Wisconsin Milwaukee, 3135 N Maryland Ave., Milwaukee, WI 53211, USA}

\author[0000-0001-6865-2871]{Anna Feltre}
\affiliation{INAF - Osservatorio di Astrofisica e Scienza dello Spazio di Bologna, Via P. Gobetti 93/3, 40129 Bologna, Italy}

\author[0000-0001-8587-218X]{Matthew Hayes}
\affiliation{Stockholm University, Department of Astronomy and Oskar Klein Centre for Cosmoparticle Physics, AlbaNova University Centre, SE-10691, Stockholm, Sweden}

\author[0000-0002-6586-4446]{Alaina Henry}
\affiliation{Space Telescope Science Institute, 3700 San Martin Drive, Baltimore, MD 21218, USA}

\author[0000-0003-4857-8699]{Svea Hernandez}
\affiliation{Space Telescope Science Institute, 3700 San Martin Drive, Baltimore, MD 21218, USA}

\author[0000-0002-6790-5125]{Anne Jaskot}
\affiliation{Department of Astronomy, Williams College, USA}

\author[0000-0001-8152-3943]{Lisa J. Kewley}
\affiliation{Research School of Astronomy and Astrophysics, Australian National University, Cotter Road, Weston Creek, ACT 2611, Australia; ARC Centre of Excellence for All Sky Astrophysics in 3 Dimensions (ASTRO 3D), Canberra, ACT 2611, Australia}

\author[0000-0002-5320-2568]{Nimisha Kumari}
\affiliation{Space Telescope Science Institute, 3700 San Martin Drive, Baltimore, MD 21218, USA}

\author[0000-0003-2685-4488]{Claus Leitherer}
\affiliation{Space Telescope Science Institute, 3700 San Martin Drive, Baltimore, MD 21218, USA}

\author[0000-0003-1354-4296]{Mario Llerena}
\affiliation{Instituto de Investigaci\'{o}n Multidisciplinar en Ciencia y Tecnolog\'{i}a, Universidad de La Serena, Raul Bitr\'{a}n 1305, La Serena 2204000, Chile}

\author[0000-0003-0695-4414]{Michael Maseda}
\affiliation{Leiden Observatory, Leiden University, PO Box 9513, 2300 RA, Leiden, The Netherlands}

\author[0000-0003-2804-0648]{Themiya Nanayakkara}
\affiliation{Swinburne University of Technology, Melbourne, Victoria, AU}

\author[0000-0002-1049-6658]{Masami Ouchi}
\affiliation{Institute for Cosmic Ray Research, The University of Tokyo, Kashiwa-no-ha, Kashiwa 277-8582, Japan}

\author[0000-0003-0390-0656]{Adele Plat}
\affiliation{Steward Observatory, The University of Arizona, 933 N Cherry Ave, Tucson, AZ, 85721, USA}

\author[0000-0003-1435-3053]{Richard W. Pogge}
\affiliation{Department of Astronomy, The Ohio State University, 140 W 18th Avenue, Columbus, OH 43210, USA}
\affiliation{Center for Cosmology \& AstroParticle Physics, The Ohio State University, 191 W Woodruff Avenue, Columbus, OH 43210}

\author[0000-0002-5269-6527]{Swara Ravindranath}
\affiliation{Space Telescope Science Institute, 3700 San Martin Drive, Baltimore, MD 21218, USA}

\author[0000-0002-7627-6551]{Jane R. Rigby}
\affiliation{Observational Cosmology Lab, Code 665, NASA Goddard Space Flight Center, 8800 Greenbelt Rd, Greenbelt, MD 20771, USA}

\author[0000-0002-9136-8876]{Claudia Scarlata}
\affiliation{Minnesota Institute for Astrophysics, University of Minnesota, 116 Church Street SE, Minneapolis, MN 55455, USA}

\author[0000-0002-9132-6561]{Peter Senchyna}
\affiliation{Carnegie Observatories, 813 Santa Barbara Street, Pasadena, CA 91101, USA}

\author[0000-0003-0605-8732]{Evan D. Skillman}
\affiliation{Minnesota Institute for Astrophysics, University of Minnesota, 116 Church Street SE, Minneapolis, MN 55455, USA}

\author[0000-0002-4834-7260]{Charles C. Steidel}
\affiliation{Cahill Center for Astronomy and Astrophysics, California Institute of Technology, MC249-17, Pasadena, CA 91125, USA}

\author[0000-0002-4834-7260]{Allison L. Strom}
\affiliation{The Observatories of the Carnegie Institution for Science, 813 Santa Barbara Street, Pasadena, CA 91101, USA}

\author[0000-0001-6958-7856]{Yuma Sugahara}
\affiliation{National Astronomical Observatory of Japan, 2-21-1 Osawa, Mitaka, Tokyo 181-8588, Japan}
\affiliation{Waseda Research Institute for Science and Engineering, Faculty of Science and Engineering, Waseda University, 3-4-1, Okubo, Shinjuku, Tokyo 169-8555, Japan}

\author[0000-0003-3903-6935]{Stephen M. Wilkins}
\affiliation{Astronomy Centre, University of Sussex, Falmer, Brighton BN1 9QH, UK}

\author[0000-0001-8289-3428]{Aida Wofford}
\affiliation{Instituto de Astronom\'{i}a, Universidad Nacional Aut\'{o}noma de M\'{e}xico, Unidad Acad\'{e}mica en Ensenada, Km 103 Carr. Tijuana-Ensenada, Ensenada 22860, M\'{e}xico}

\author[0000-0002-9217-7051]{Xinfeng Xu}
\affiliation{Center for Astrophysical Sciences, Department of Physics \& Astronomy, Johns Hopkins University, Baltimore, MD 21218, USA}

\suppressAffiliations
\correspondingauthor{Bethan L. James} 
\email{bjames@stsci.edu}



\begin{abstract}\nolinenumbers
The COS Legacy Archive Spectroscopic SurveY (CLASSY) is designed to provide the community with a spectral atlas of 45 nearby star-forming galaxies which were chosen to cover similar properties as those seen at high-$z$ ($z>6$). The prime high level science product of CLASSY is accurately coadded UV spectra, ranging from $\sim$1000--2000~\AA, derived from a combination of archival and new data obtained with HST's Cosmic Origins Spectrograph (COS). This paper details the multi-stage technical processes of creating this prime data product, and the methodologies involved in extracting, reducing, aligning, and coadding far-ultraviolet (FUV) and near-ultraviolet (NUV) spectra. We provide guidelines on how to successfully utilize COS observations of extended sources, despite COS being optimized for point sources, and best-practice recommendations for the coaddition of UV spectra in general. Moreover, we discuss the effects of our reduction and coaddition techniques in the scientific application of the CLASSY data. In particular, we find that accurately accounting for flux calibration offsets can affect the derived properties of the stellar populations, while customized extractions of NUV spectra for extended sources are essential for correctly diagnosing the metallicity of galaxies via \sfCTh\ nebular emission. Despite changes in spectral resolution of up to $\sim$25\%\ between individual datasets (due to changes in the COS line spread function), no adverse affects were observed on the difference in velocity width and outflow velocities of isolated absorption lines when measured in the final combined data products, owing in-part to our signal-to-noise regime of S/N$<20$. 

\end{abstract}


\section{Introduction} \label{sec:intro}

The COS Legacy Archive Spectroscopic SurveY (CLASSY) is a Hubble Space Telescope (HST) treasury program with the Cosmic Origins Spectrograph (COS) dedicated to creating the first high-resolution FUV spectral catalog of star-forming galaxies at $z\sim0$, with a sample of 45 targets selected to span properties seen at high-$z$ ($z>6$) i.e., low-metallicity, low-mass, high star formation rate (SFR), and high ionization parameter ($U$-parameter) systems (Berg et al. 2022, submitted, \citetalias{berg22} hereafter). The spectra detect a suite of emission and absorption lines over 1200--2000~\AA\ that fully characterise the stellar populations, outflows (in absorption), and nebular conditions present in the bright star-forming knots that are within the 2\farcs5 diameter COS aperture. Using a total of 312 orbits of COS FUV and NUV spectroscopy (135 new, 177 archival), CLASSY will be responsible for delivering an atlas of high level science products to the community. In addition to an atlas of coadded FUV-to-NUV high signal-to-noise (an average S/N$\gtrsim5$ at 1500\AA\, with S/N$_{1500}\gtrsim5$ for 40\%\ of the overall sample) spectra for each target, covering the G130M+G160M+G185M/G225M gratings of HST/COS, CLASSY is designed to provide the properties of the stellar continuum, absorption line profiles, and emission line profiles, for each spectrum. Moreover, the survey will release all scientific measurements and the properties derived from them to the community, including stellar ages, outflow velocities, chemical and physical properties of the neutral and ionized gas, along with UV-based emission line diagnostics. Each of these products will be detailed in forthcoming publications from the CLASSY collaboration. 

The main scientific aim of CLASSY is to provide the community with the necessary toolkit to navigate the upcoming JWST era, when spectral properties of galaxies in the distant Universe finally come to light. As such, each of the targets covered by CLASSY are star-forming galaxies akin to those at high-z \citep[$z<7$, e.g., ][]{steidel14, lefevre15, shapley15, mclure18}, including stellar mass (log(M$_{\star}$/\Mo)$\sim$6 to 10), gas phase metallicity (Z/\Zsol~$\sim$0.03 to 1.2), star formation rate (log(SFR/\Mo yr$^{-1}$)$\sim-3$ to 2), ionization (O$_{32}\footnote{log(\fOTh/\fOTw)}\sim-0.3$ to 1.4), and electron density ($10<$\Ne(cm$^{-3}$)$<1120$). The targets have redshifts between $z$=0.002--~0.182, whose UV sources have Galaxy Evolution Explorer (GALEX) FUV magnitudes between $m_{FUV}=15.3-19.2$, ranging from 0\farcs11--1\farcs6 in diameter (as measured from the COS NUV acquisition images). 

The Cosmic Origins Spectrograph \citep[COS;][]{green12} onboard the Hubble Space Telescope (HST) is the ideal (and only) instrument capable of providing the high-S/N UV spectra required for CLASSY. Installed during Servicing Mission 4 in May 2009, COS delivers high-sensitivity, medium- and low-resolution spectroscopy in the $\sim815-3200$\,\AA\ wavelength range. COS is comprised of two separate detectors: a cross delay line (XDL) detector sensitive to FUV wavelengths (815--2050\AA) and a multi-anode microchannel array (MAMA) optimized for the NUV wavelengths (1700--3200\AA). Both detectors offer the use of low-dispersion gratings (e.g., G140L for the FUV or G230L for the NUV) or medium-dispersion gratings (e.g., G130M and G160M in the FUV or G185M and G225M in the NUV).  The instrument was designed to study the origins of large scale structure in the Universe, the formation and evolution of galaxies, the origin of stellar and planetary systems, and the cold interstellar medium (ISM).\footnote{\url{https://www.stsci.edu/hst/instrumentation/cos}}. One drawback of COS, in relation to the needs of CLASSY, is that it is an instrument optimized for point-source spectroscopy. 
For example, (1) the standard reduction pipeline adopts an optimized extraction routine that assumes a point-source profile, and (2) the light entering the COS aperture is vignetted beyond the central 0\farcs4 radius - both of which affect spectroscopy of extended sources (defined in the COS Instrument Handbook, IHB\footnote{\url{https://hst-docs.stsci.edu/cosihb}, \citet{COSIHB}}, as targets with light profiles whose GFWHM\footnote{Gaussian full width at half maximum}$>$0\farcs6). Moreover, COS is typically only used for single-grating science and while its data reduction pipeline produces accurate data summed over a single `visit', it does not coadd data between different visits, central wavelength settings of the grating (\texttt{CENWAVE}s), or gratings. 
Due to the non-point-source nature of $\sim$60\%\ of the CLASSY targets, and in order to produce the highest quality high level science products (HLSPs) spanning the full FUV-NUV wavelength range, the CLASSY data went through significant levels of ad-hoc data reduction and analysis. 

As such, this paper is dedicated to describing the processes involved in producing the CLASSY 
coadded spectra and is formatted as follows: 
Section~\ref{sec:reduction} describes the initial reduction techniques including spectral 
extraction in the FUV (Section~\ref{sec:fuv_ext}) and NUV (Section~\ref{sec:nuv_ext}). 
The details of the spectral coaddition are given in Section~\ref{sec:coadd}, 
including wavelength calibration, spectral resolution, flux calibration, coaddition techniques, and the effects of vignetting and aberrations. 
Finally, in Section~\ref{sec:science}, we demonstrate and discuss the effects of the reduction and 
coaddition processes on the scientific results derived from our spectra. 
A detailed description of the CLASSY survey, including coverage of the sample selection, 
observations, and sample properties, is provided in \citet[][hereafter, \PI]{berg22}. 
Any use of the CLASSY spectral atlas should cite this work (\citetalias{james21}) 
and \citetalias{berg22}. 
Here we provide an in-depth overview of the data reduction strategies utilized in CLASSY and consequently, a technical guide to UV-spectroscopic science in general.

\begin{figure*}
\centering
\begin{tabular}{c|c}
\includegraphics[width = 3.0in]{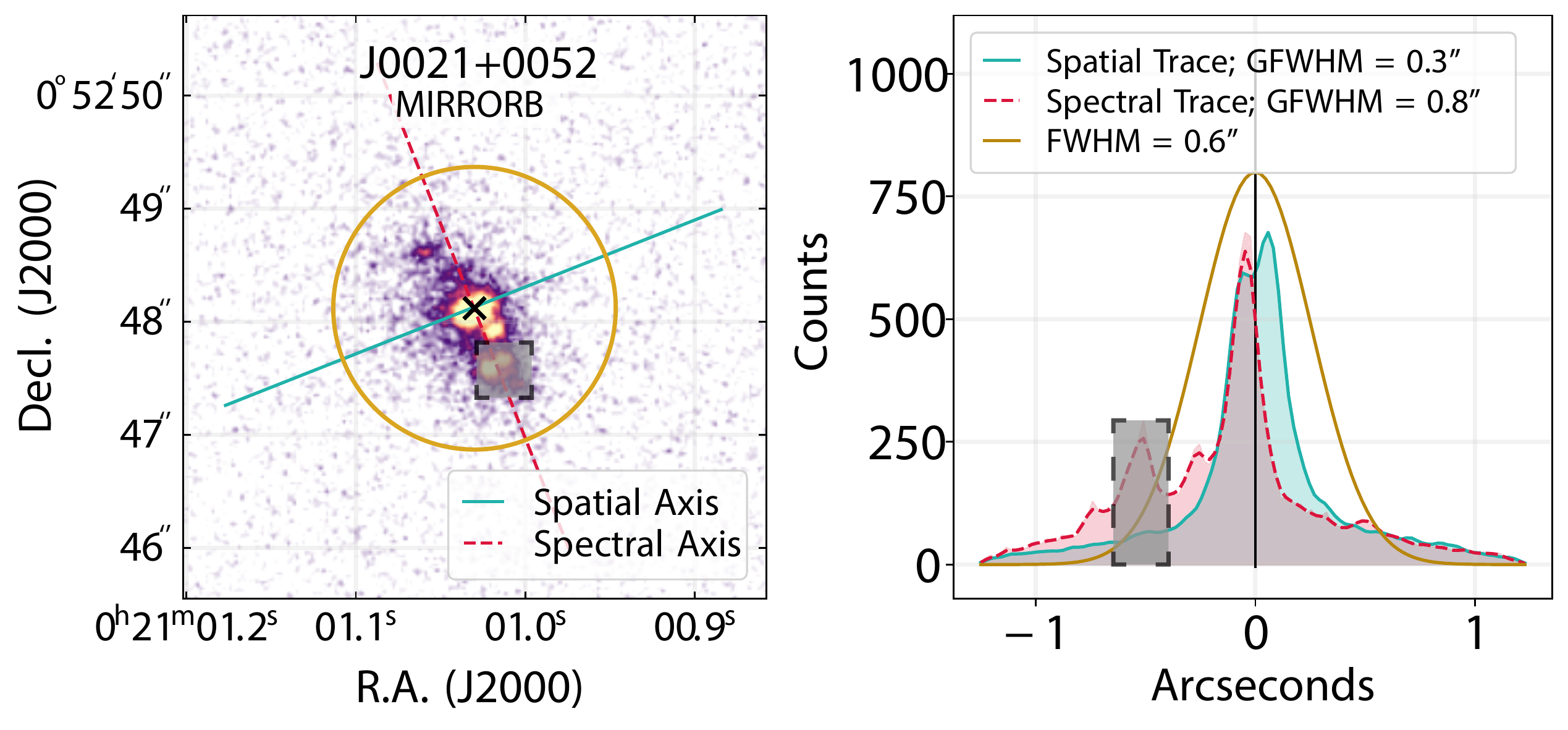} &
\includegraphics[width = 3.0in]{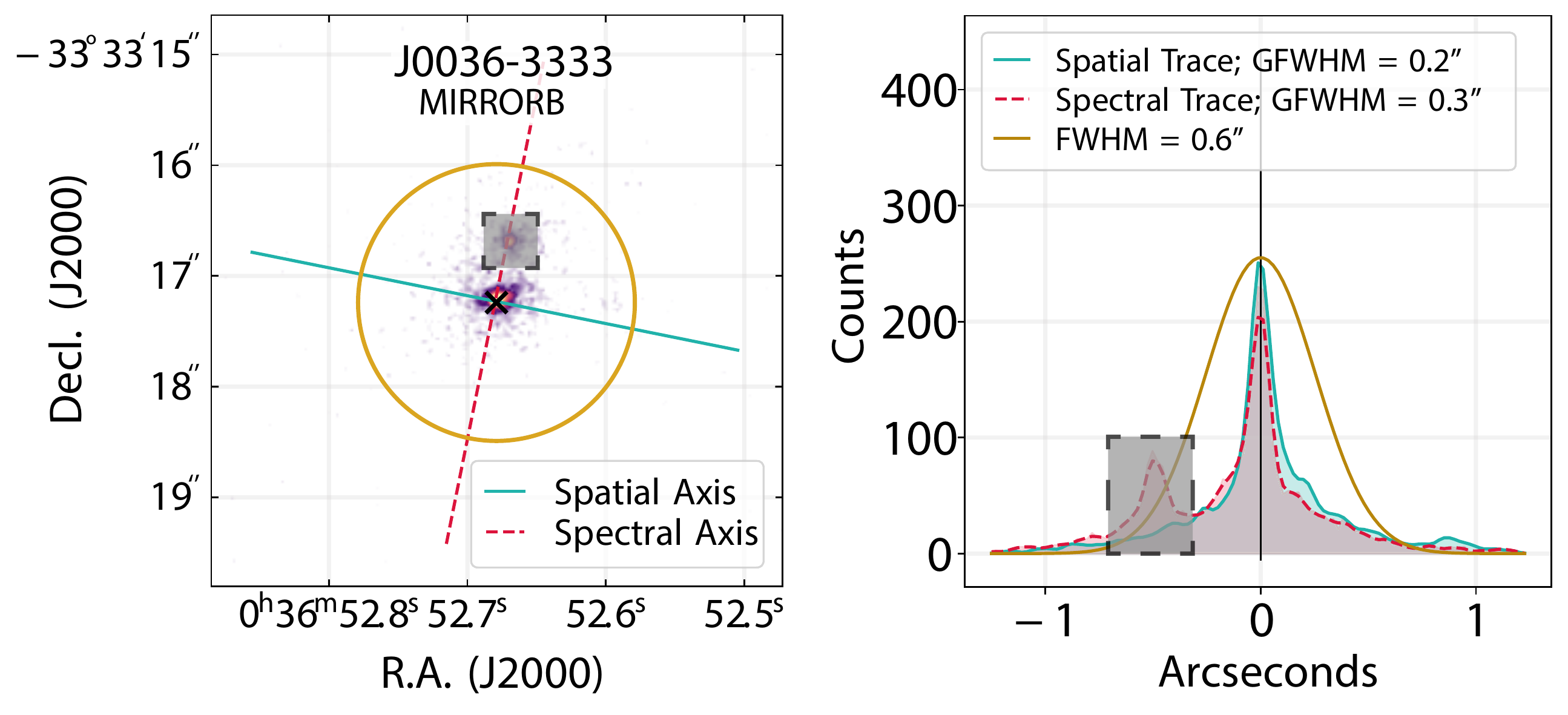}\\ \vspace{-1ex}
\includegraphics[width = 3.0in]{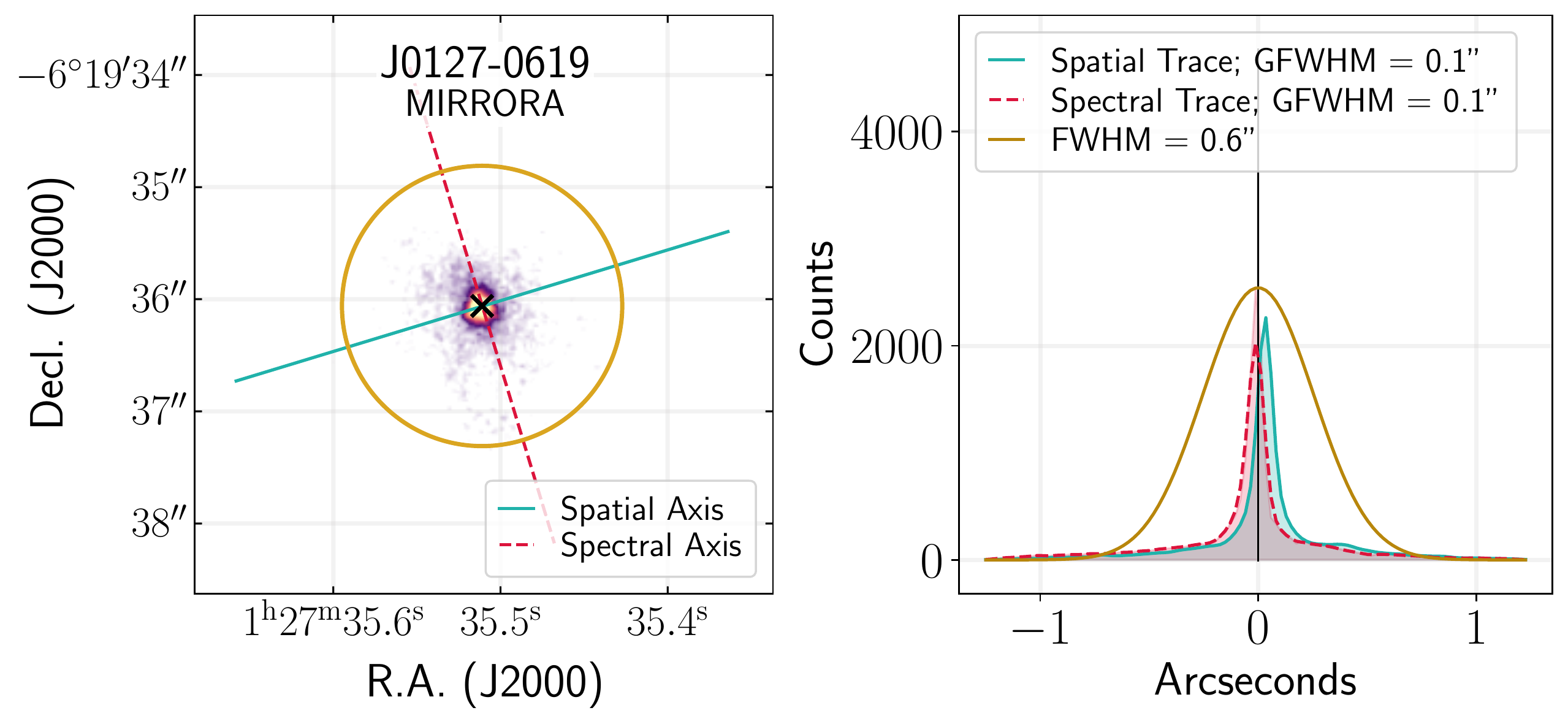} &
\includegraphics[width = 3.0in]{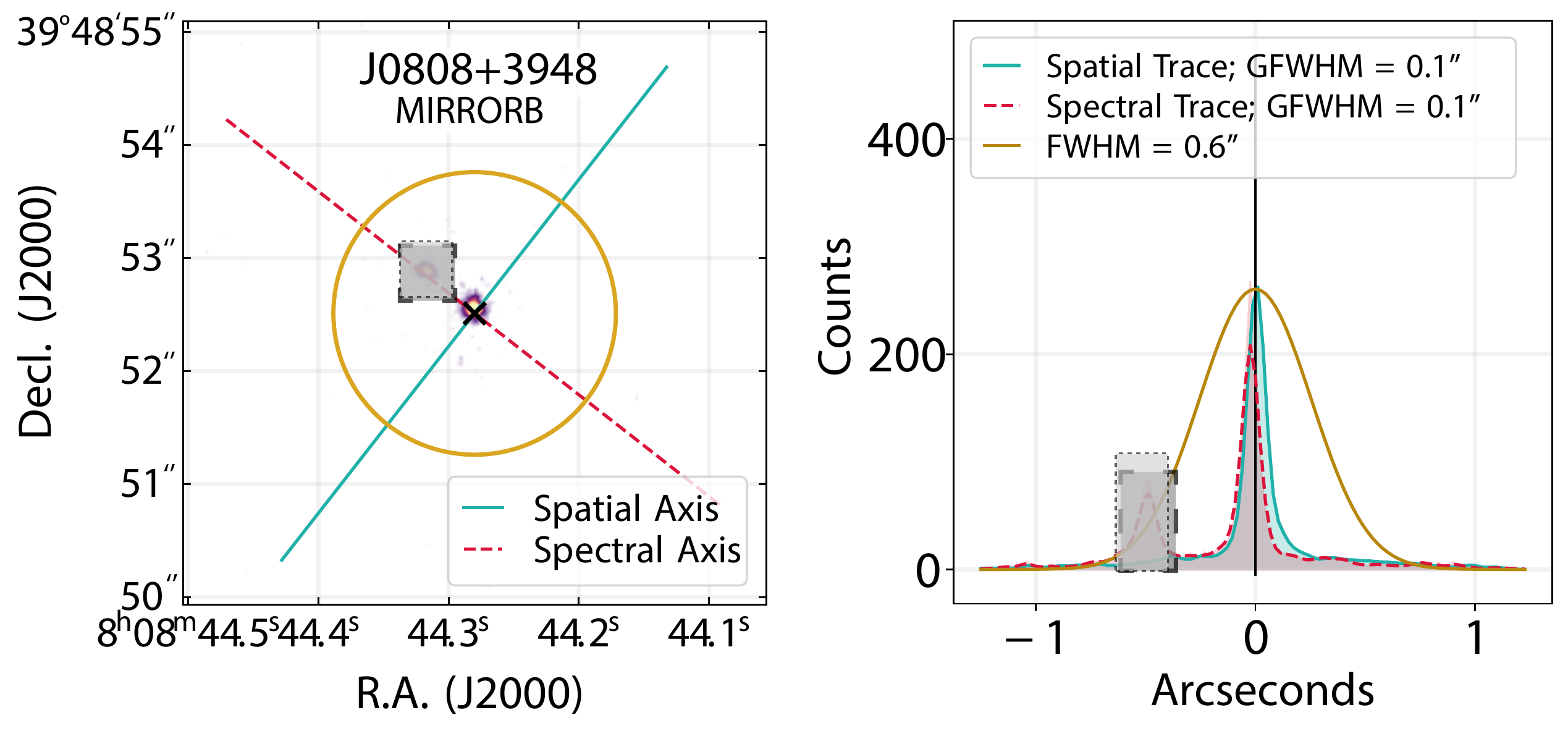}\\ \vspace{-1ex}
\includegraphics[width = 3.0in]{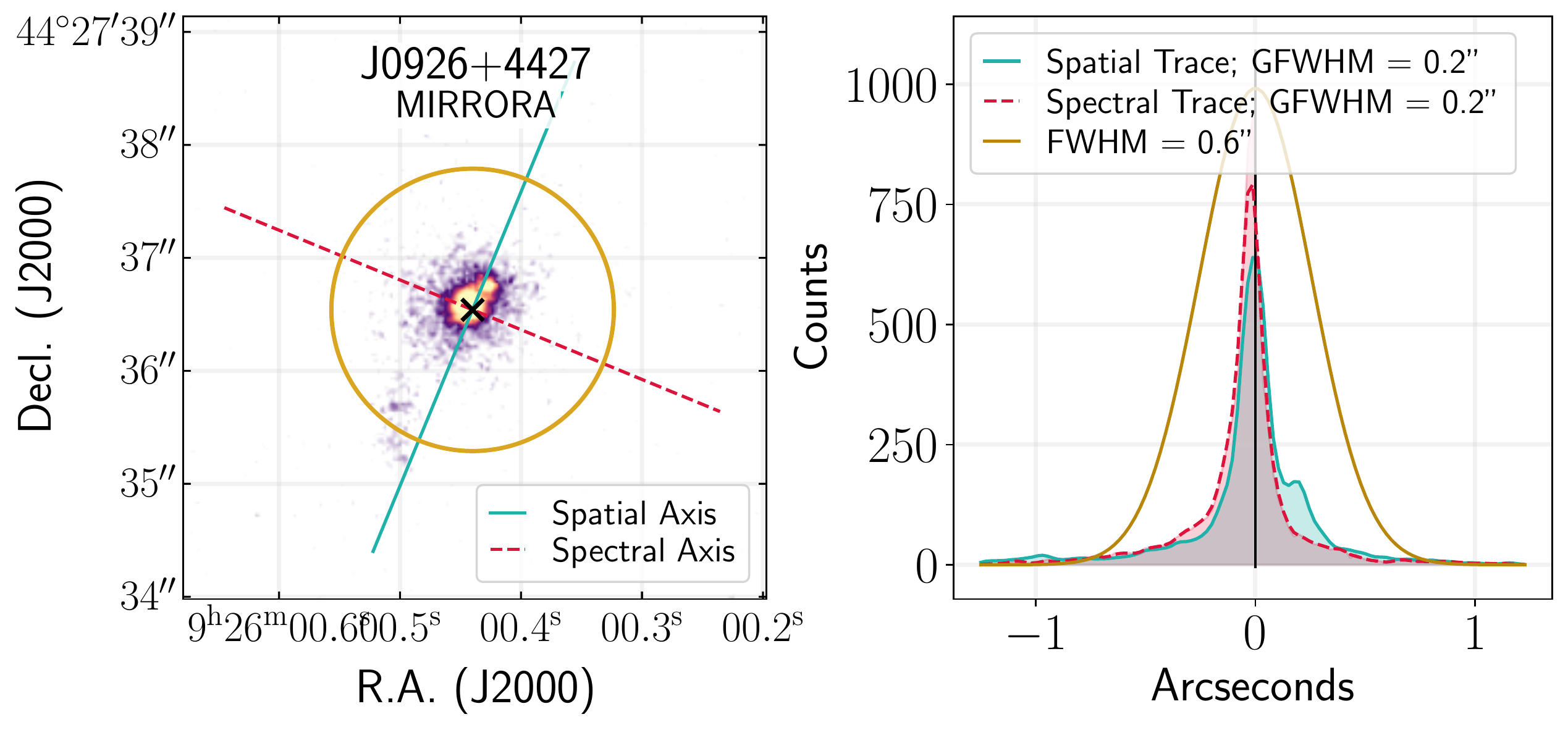} &
\includegraphics[width = 3.0in]{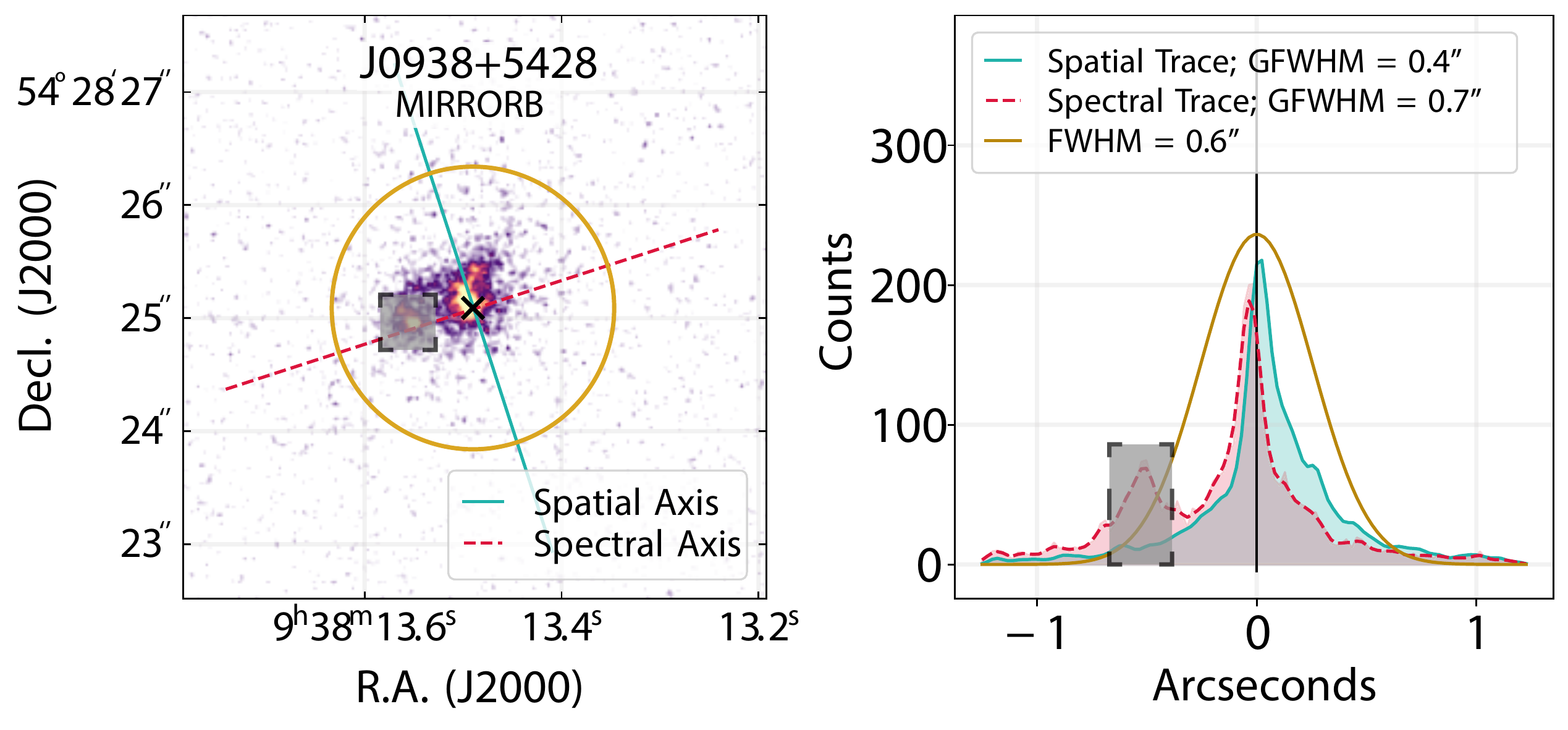}\\ \vspace{-1ex}
\includegraphics[width = 3.0in]{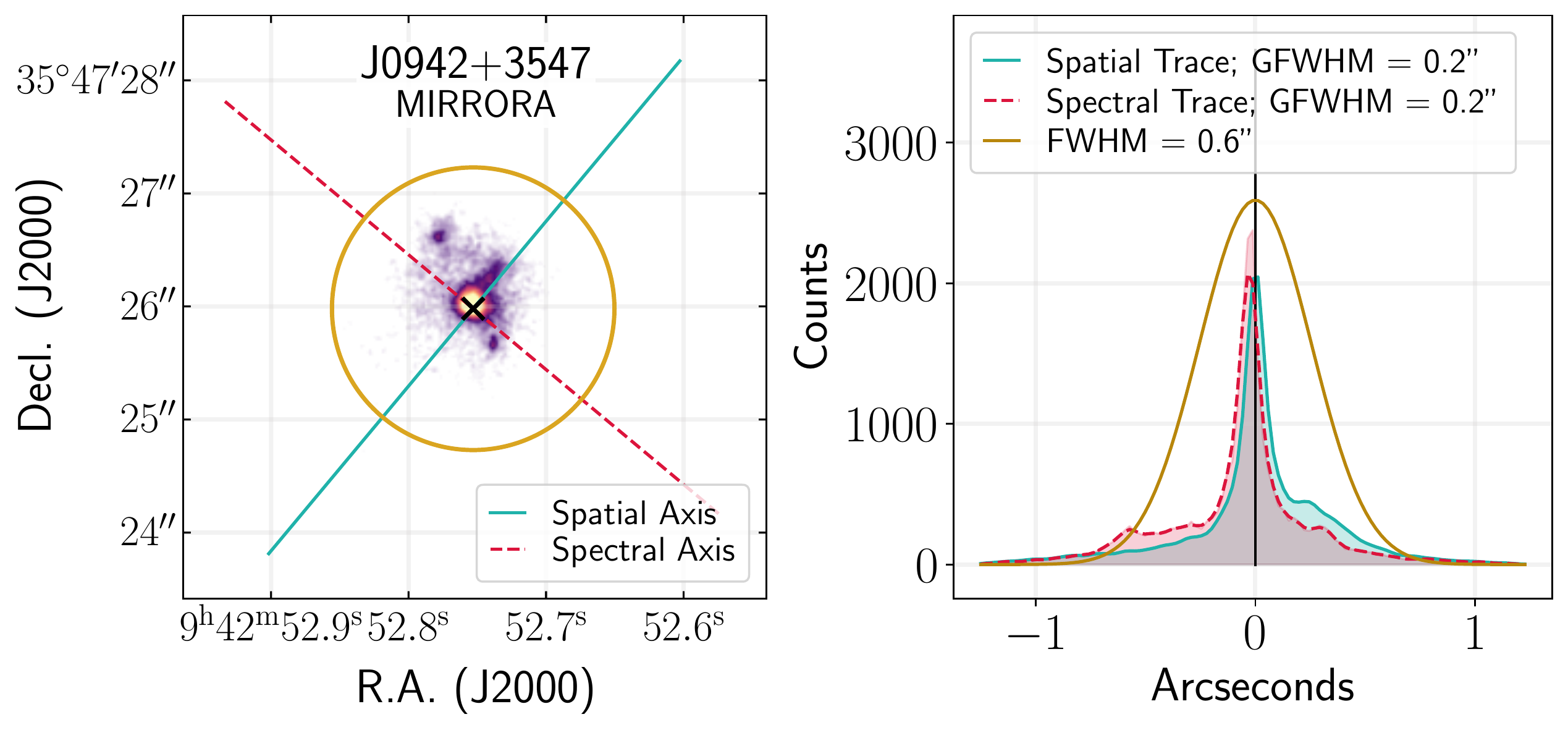} &
\includegraphics[width = 3.0in]{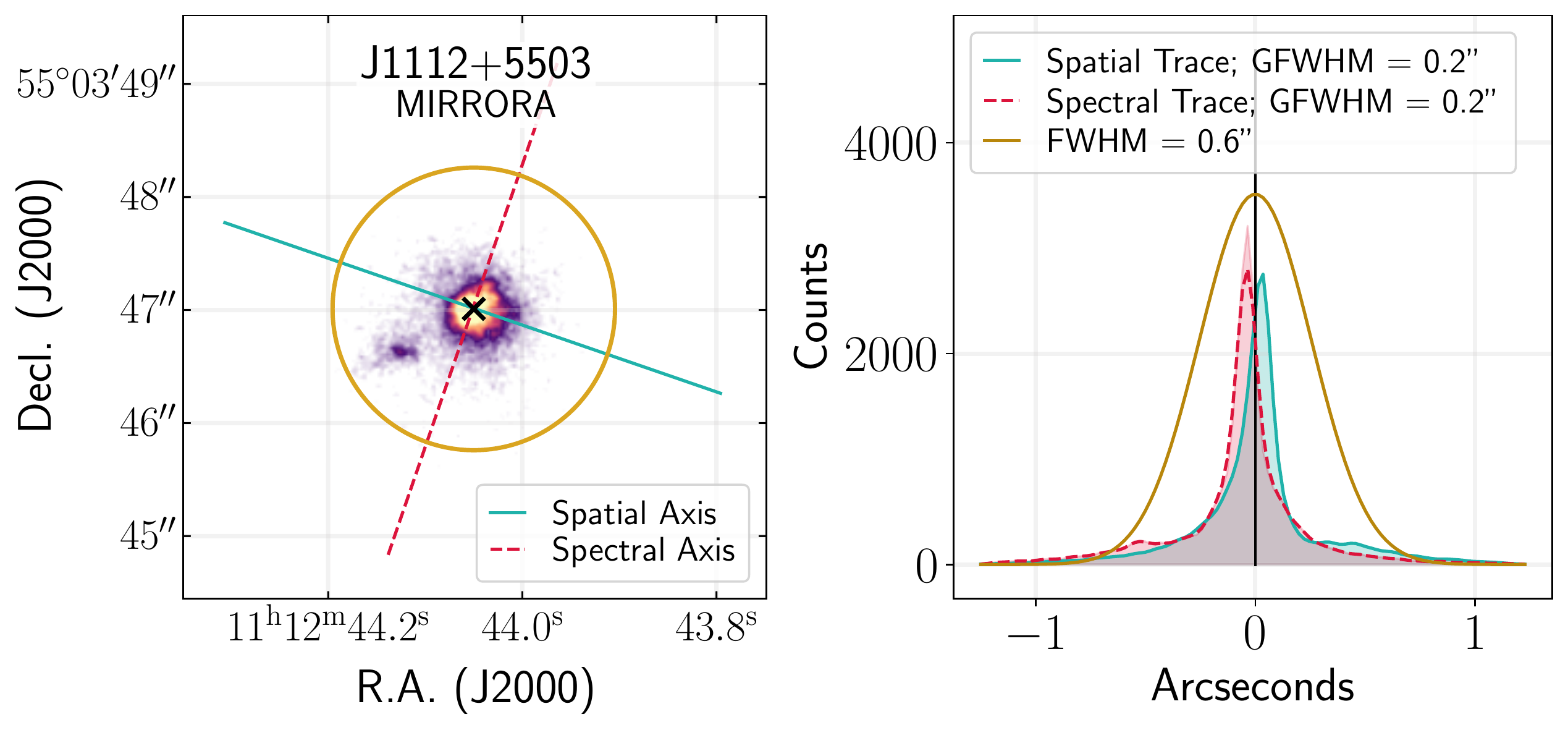}\\ \vspace{-1ex}
\includegraphics[width = 3.0in]{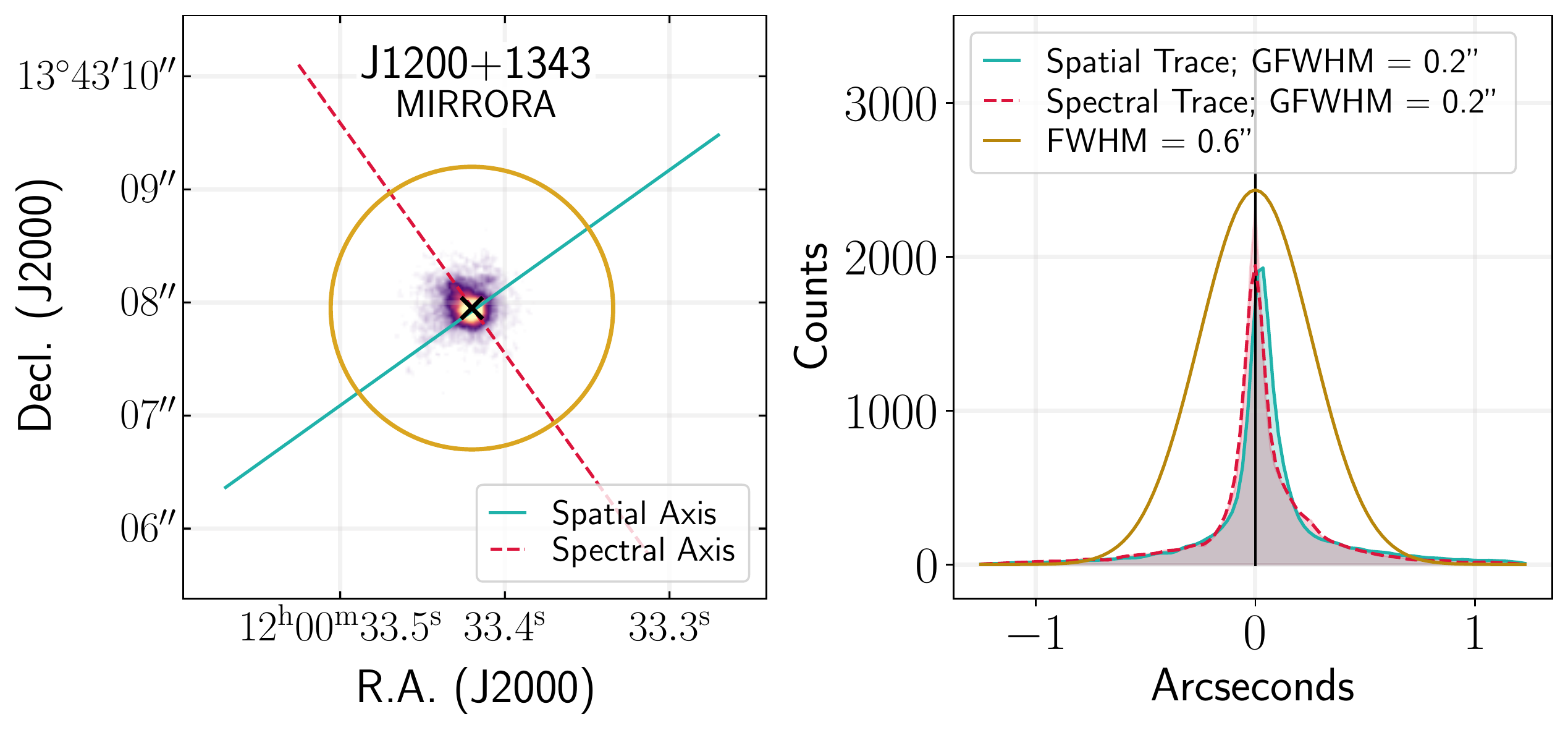} &
\includegraphics[width = 3.0in]{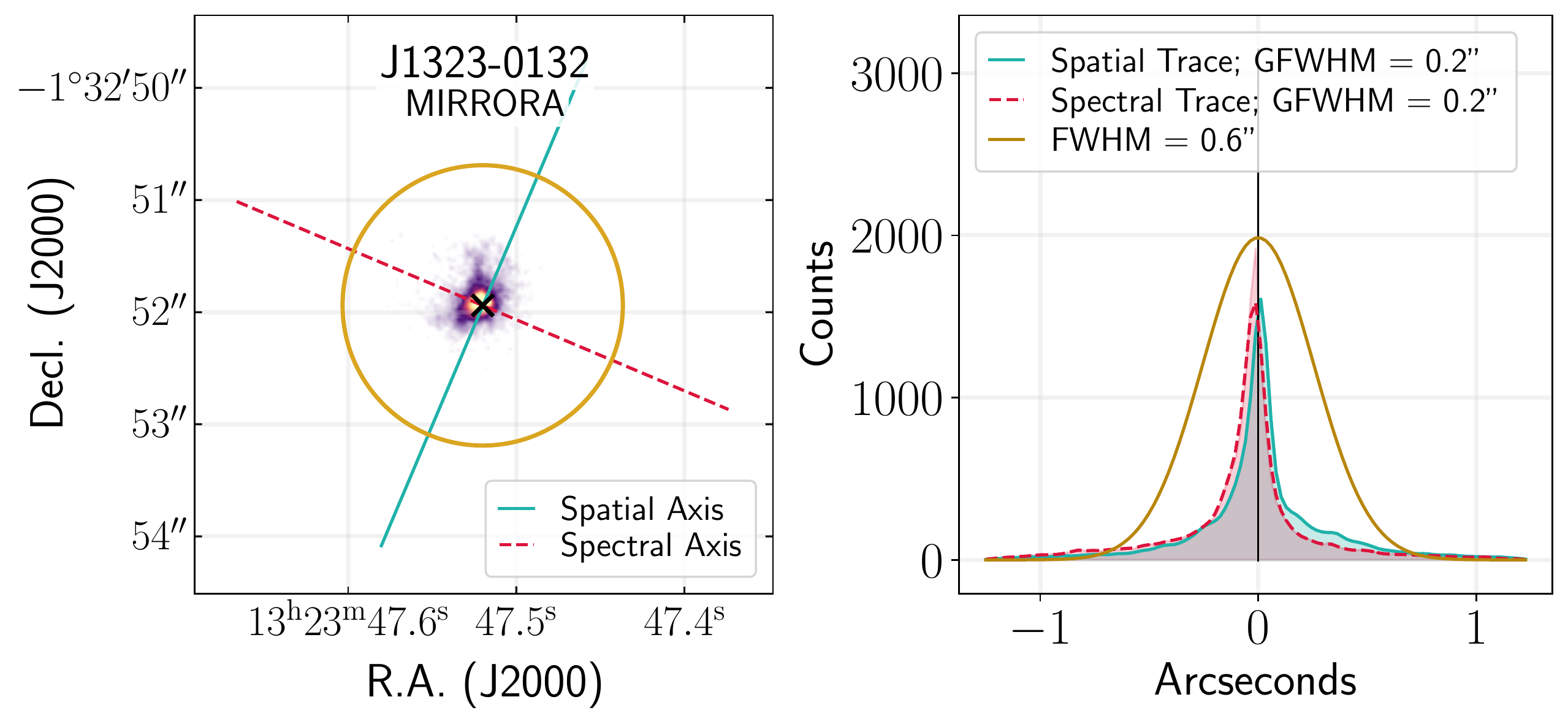}\\ \vspace{-1ex}
\end{tabular}
\caption{CLASSY acquisition images with compact-source light profiles. 
For each galaxy, the NUV target acquisition image from the G130M dataset in shown (left panels), overlaid with the COS 2\farcs5 aperture (gold circle) and labelled according to its MIRRORA or MIRRORB target acquisition configuration. 
The solid-blue and dashed-red lines correspond to the cross-dispersion (PA) 
and dispersion axis of the observations, respectively. 
The collapsed light profiles along the dispersion and cross-dispersion axes are 
show as spectral and spatial traces, respectively (right panels).
The Gaussian FWHM fit to the spectral and spatial traces is given in the legend in comparison
to a profile with FWHM$=$0\farcs6 (gold) which we use for classification purposes. For objects that have been acquired using MIRRORB, the secondary image formed along the spectral axis is marked by a grey opaque box in both panels.}
\label{fig:acq_PS}
\end{figure*}

\renewcommand{\thefigure}{1}
\begin{figure*}
\centering
\begin{tabular}{c|c}
\includegraphics[width = 3.0in]{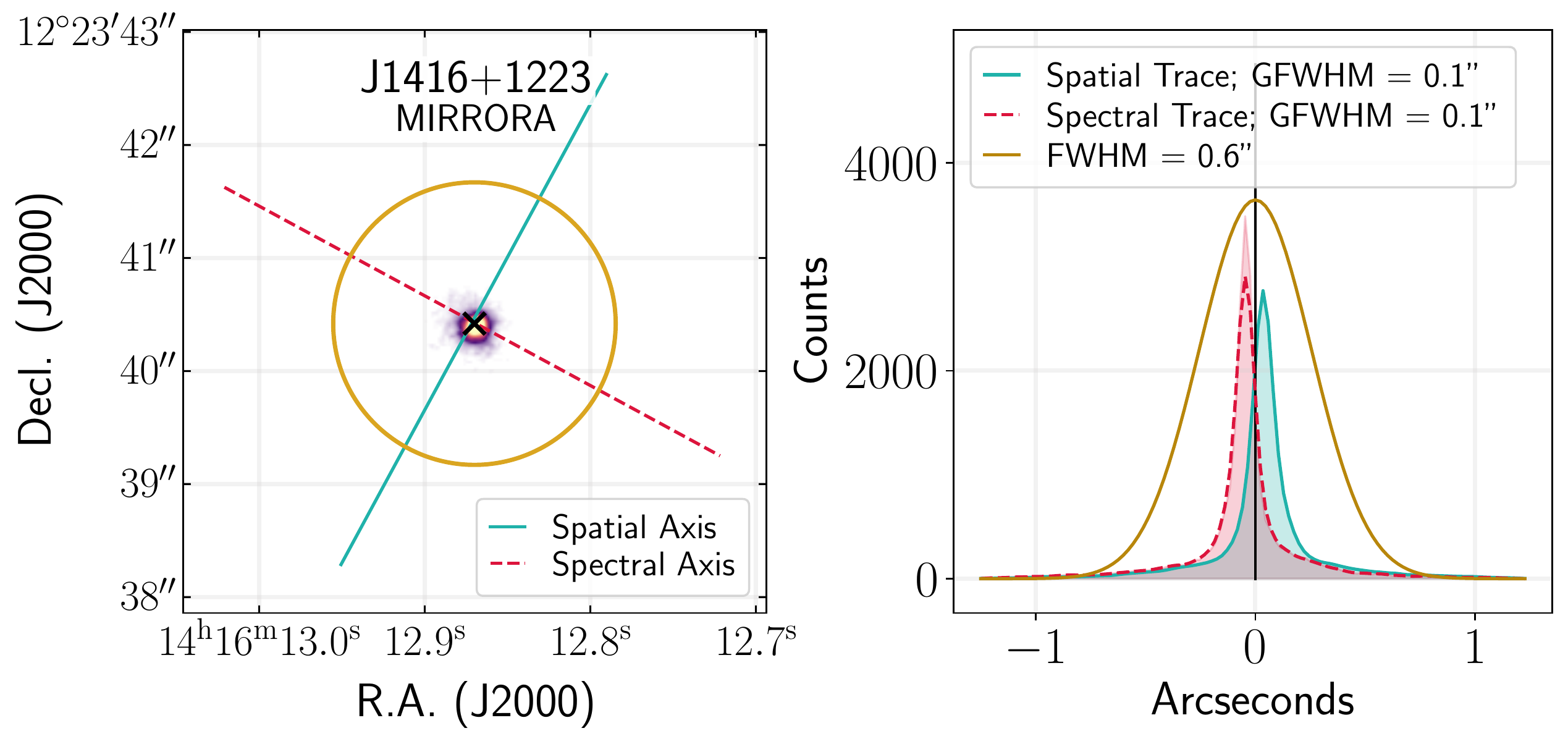} &
\includegraphics[width = 3.0in]{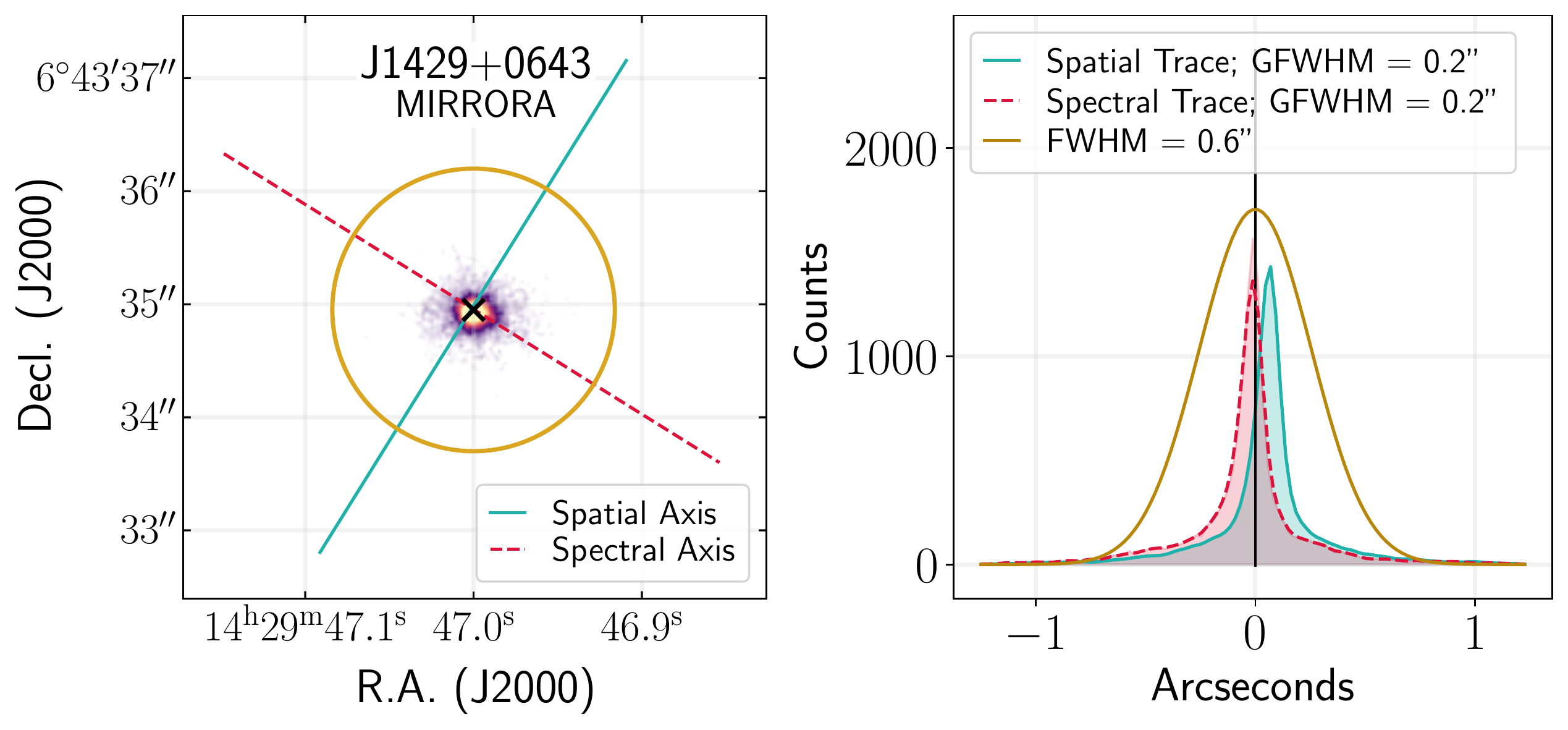}\\ \vspace{-1ex}
\includegraphics[width = 3.0in]{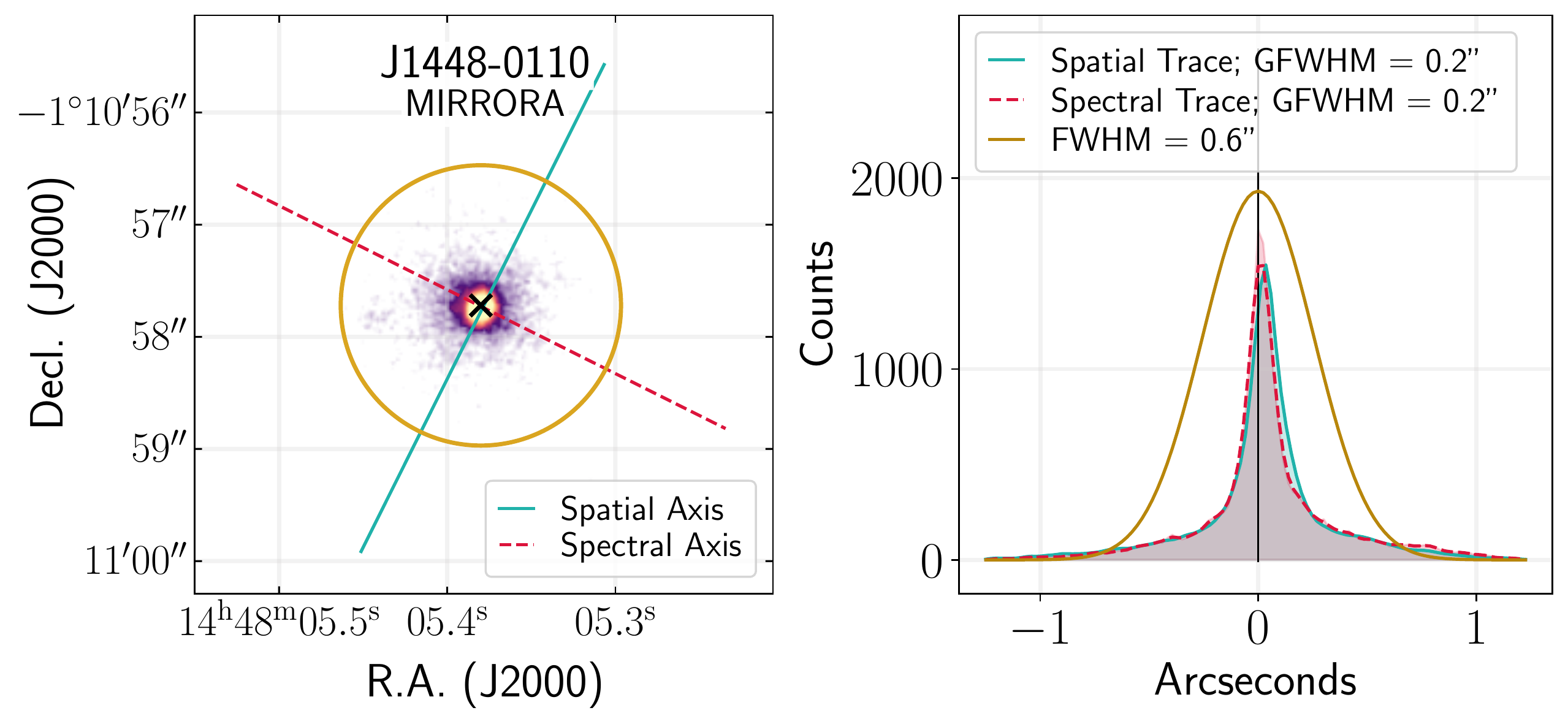} &
\includegraphics[width = 3.0in]{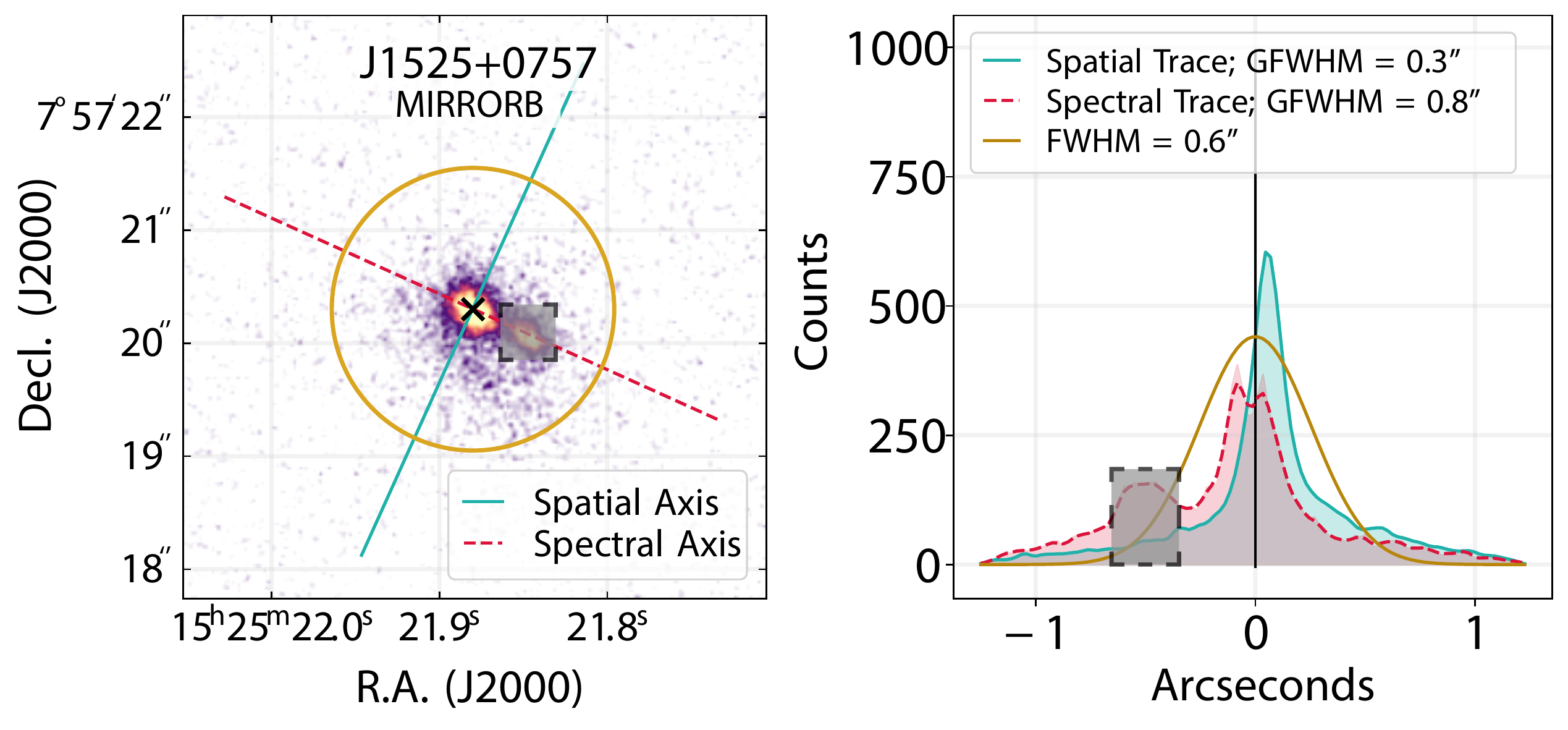}\\ \vspace{-1ex}
\includegraphics[width = 3.0in]{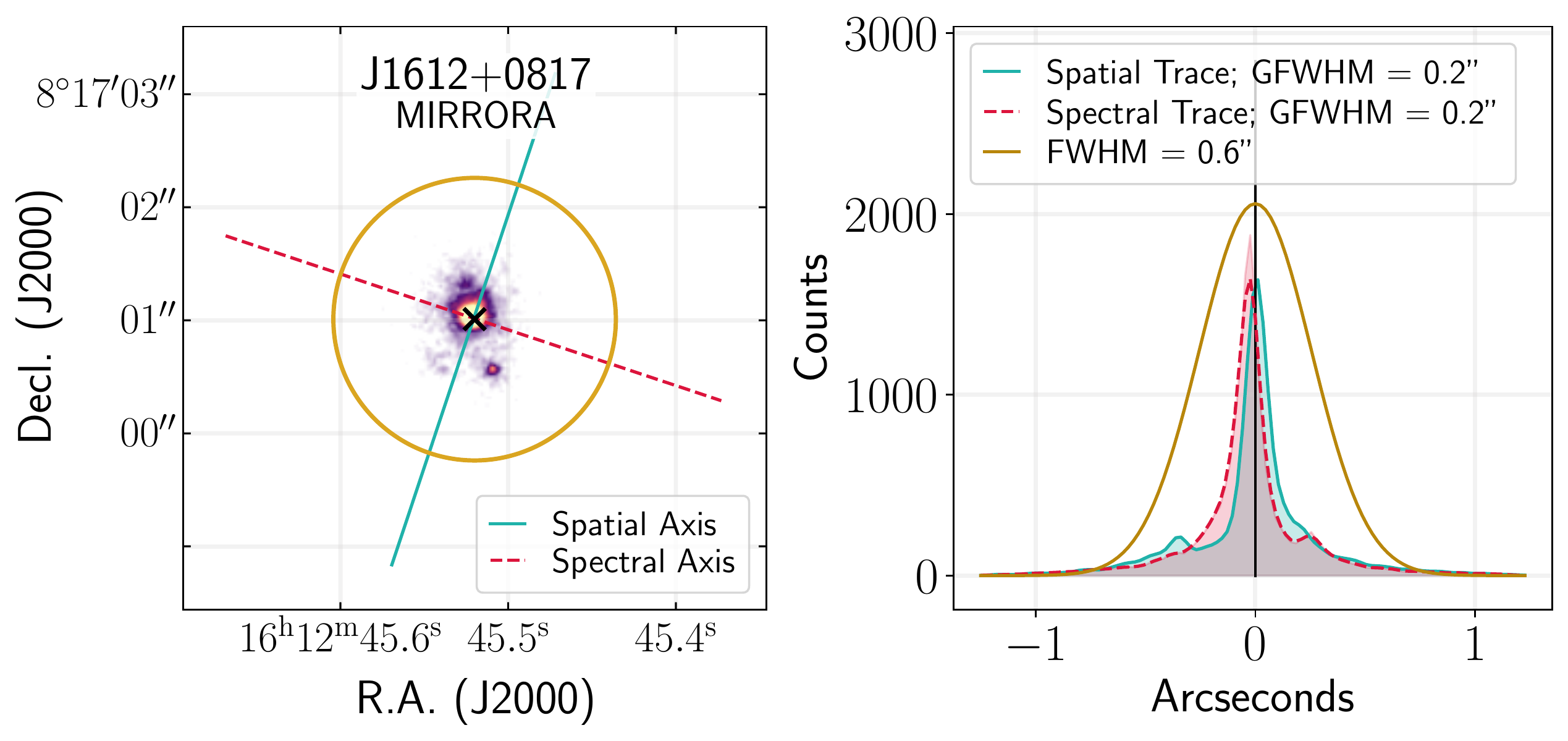} & \\
\end{tabular}
\caption{\it - continued.}
\end{figure*}

\renewcommand{\thefigure}{2}
\begin{figure*}
\begin{tabular}{c|c}
\includegraphics[width = 3.0in]{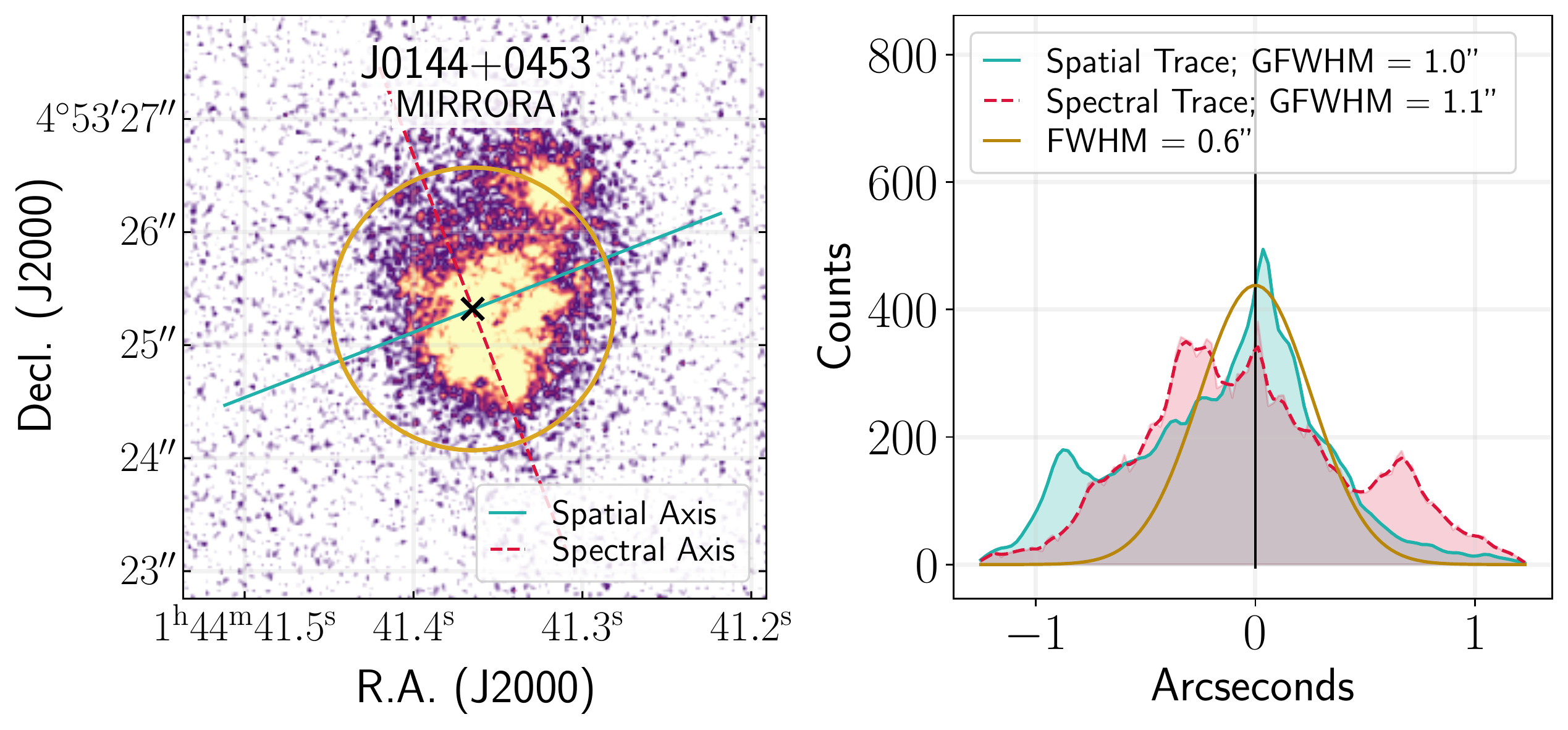} &
\includegraphics[width = 3.0in]{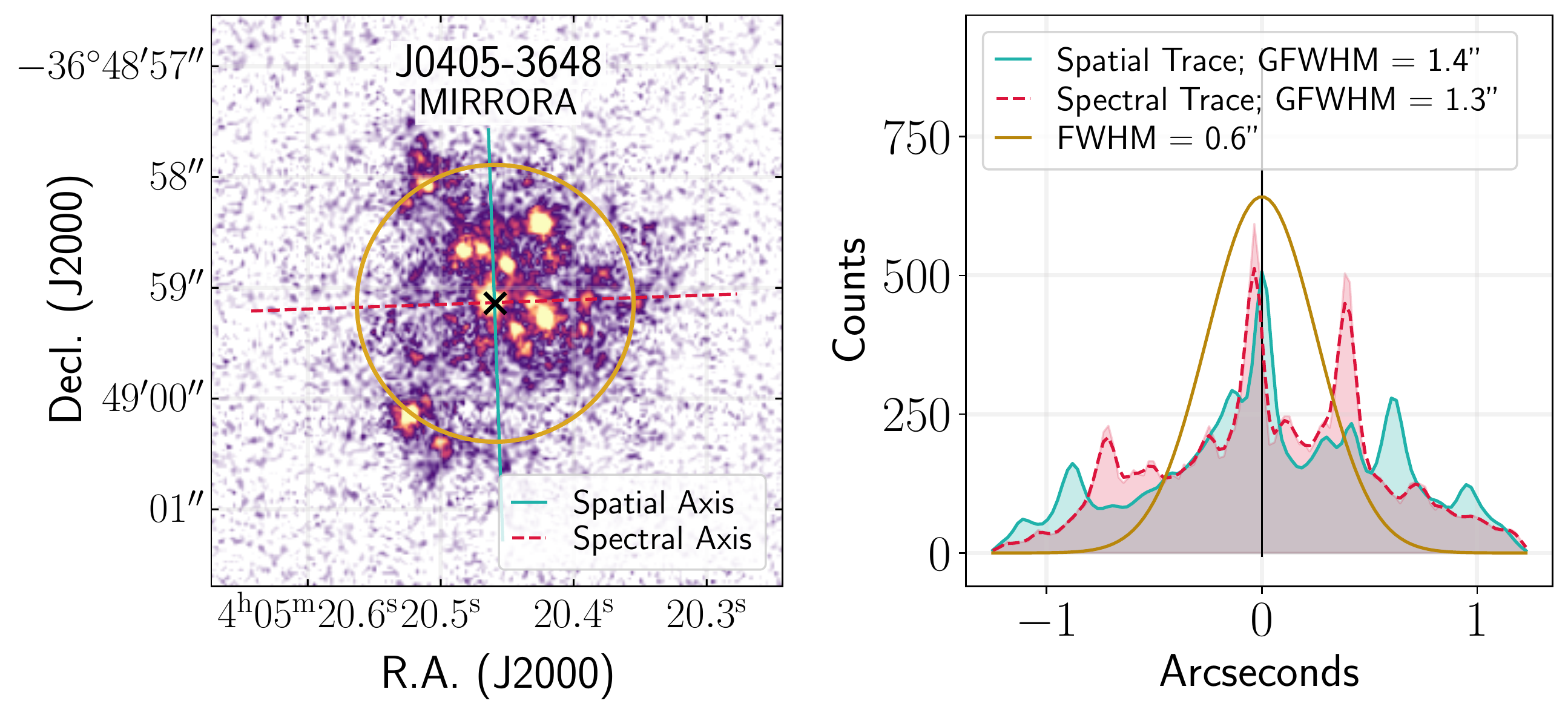}\\ \vspace{-1ex}
\includegraphics[width = 3.0in]{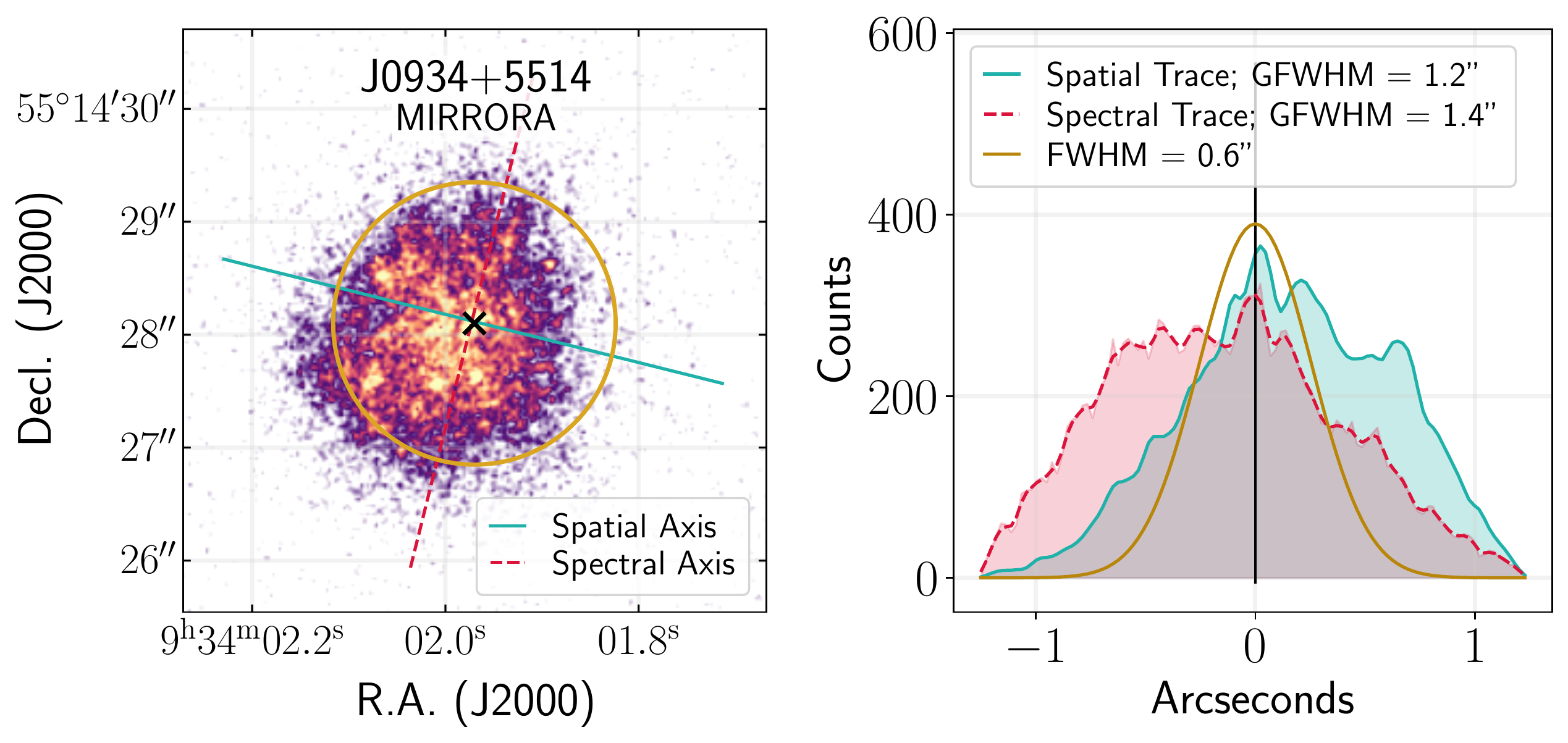} &
\includegraphics[width = 3.0in]{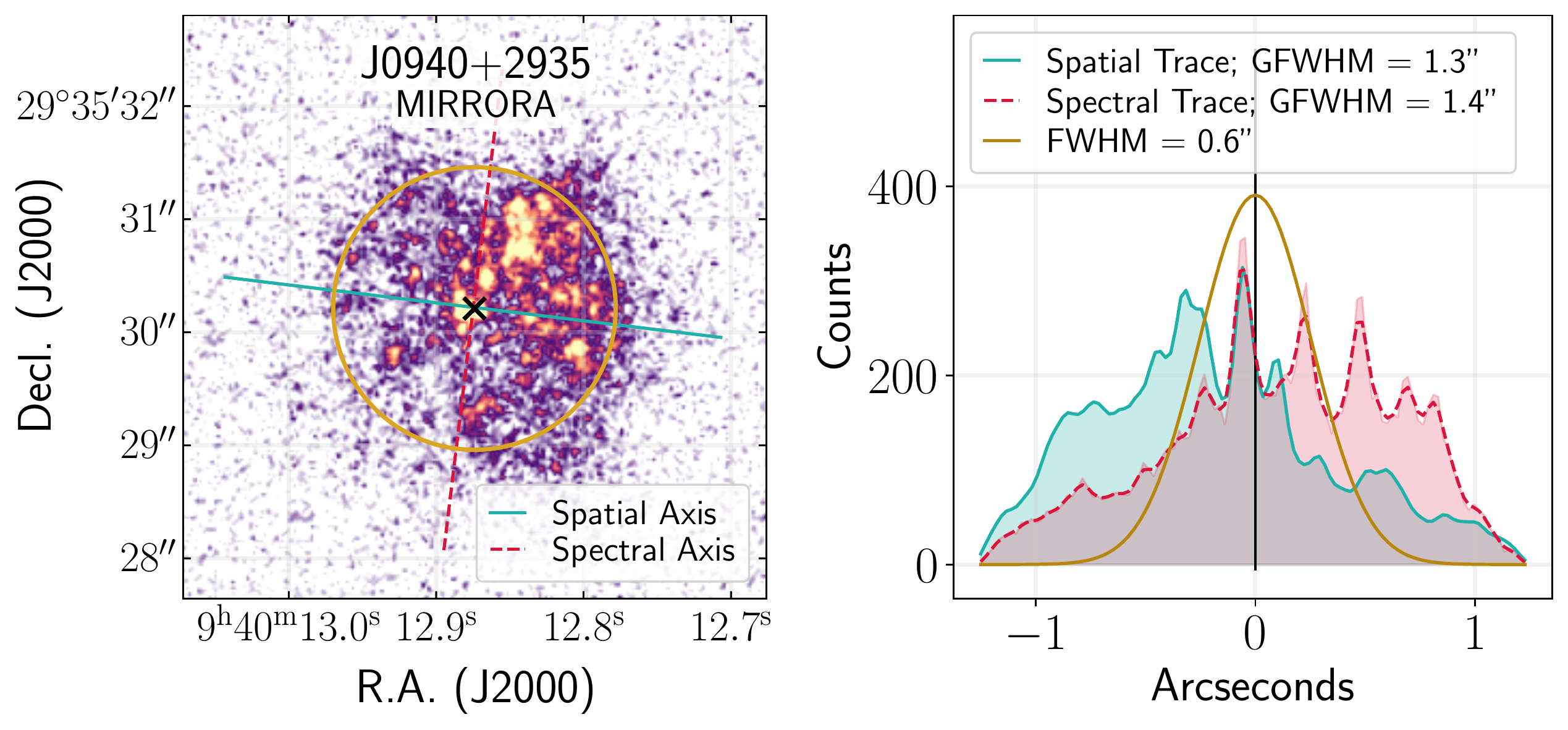}\\ \vspace{-1ex}
\includegraphics[width = 3.0in]{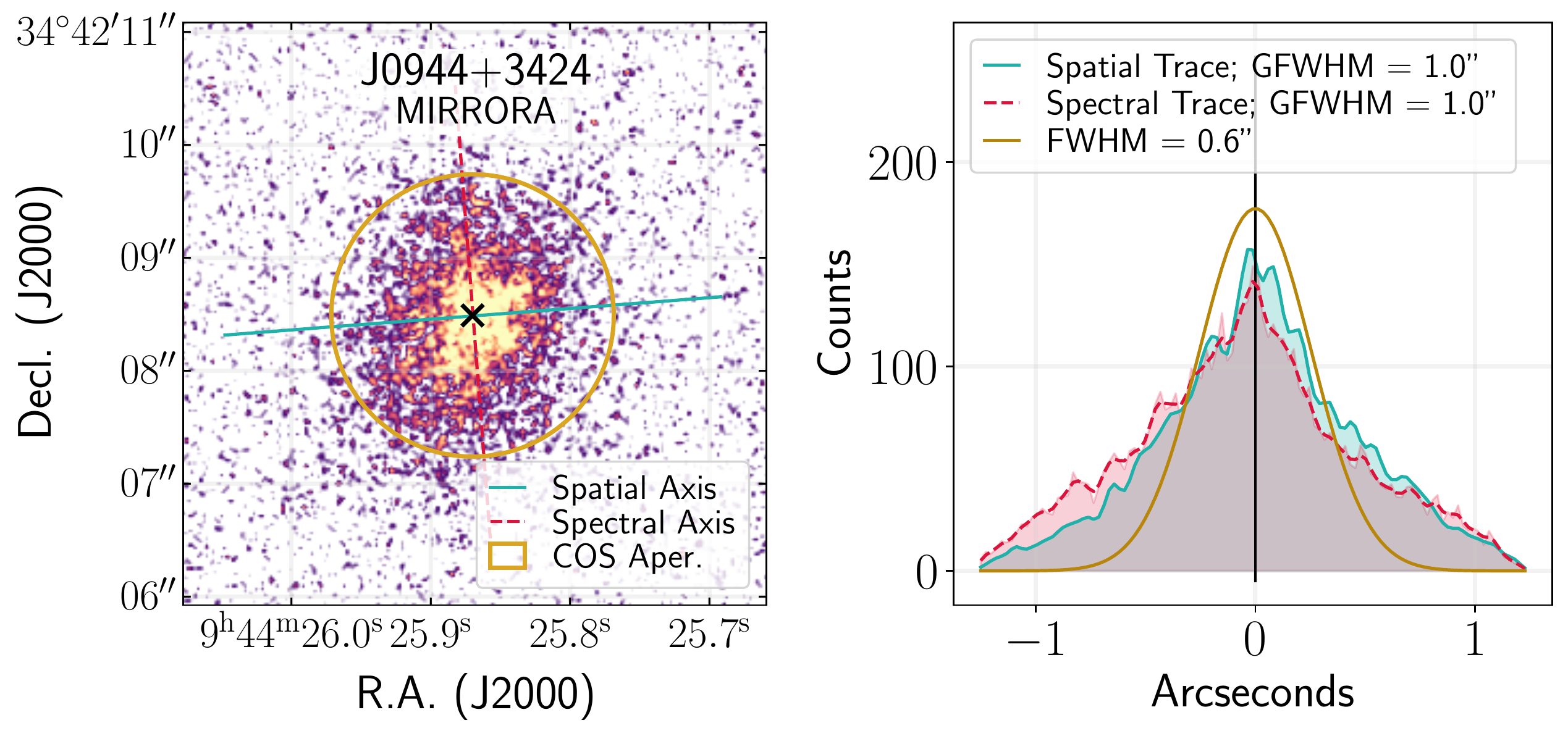} &
\includegraphics[width = 3.0in]{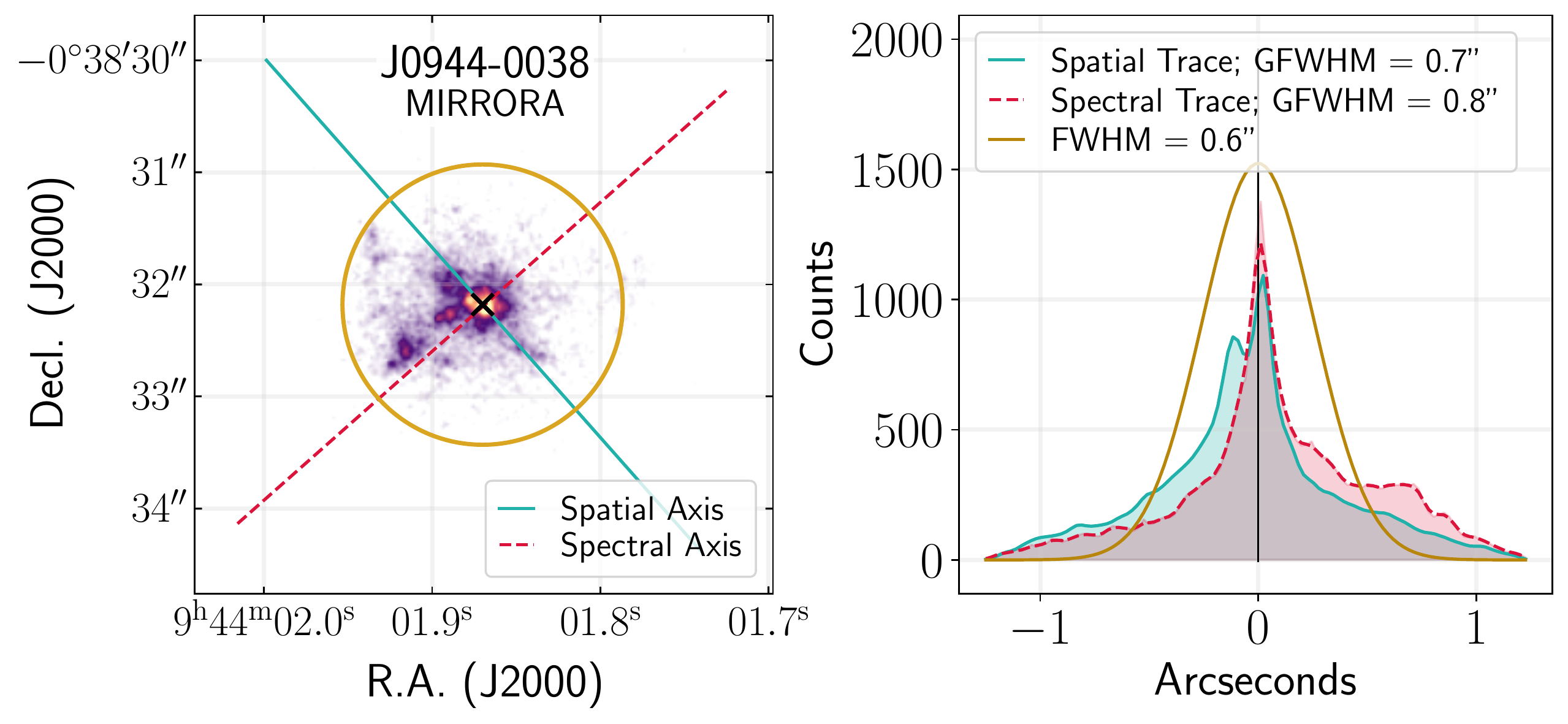}\\ \vspace{-1ex}
\includegraphics[width = 3.0in]{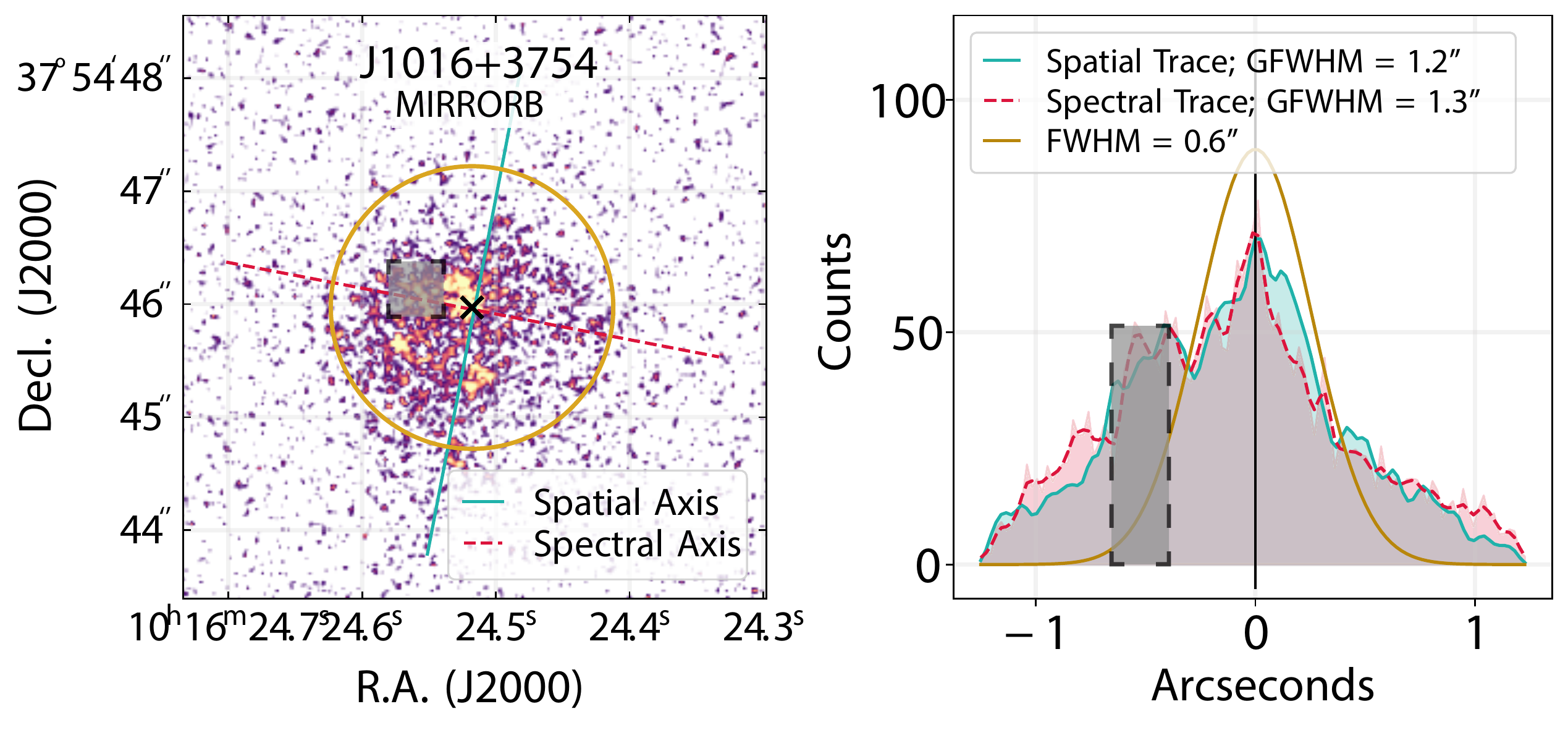} &
\includegraphics[width = 3.0in]{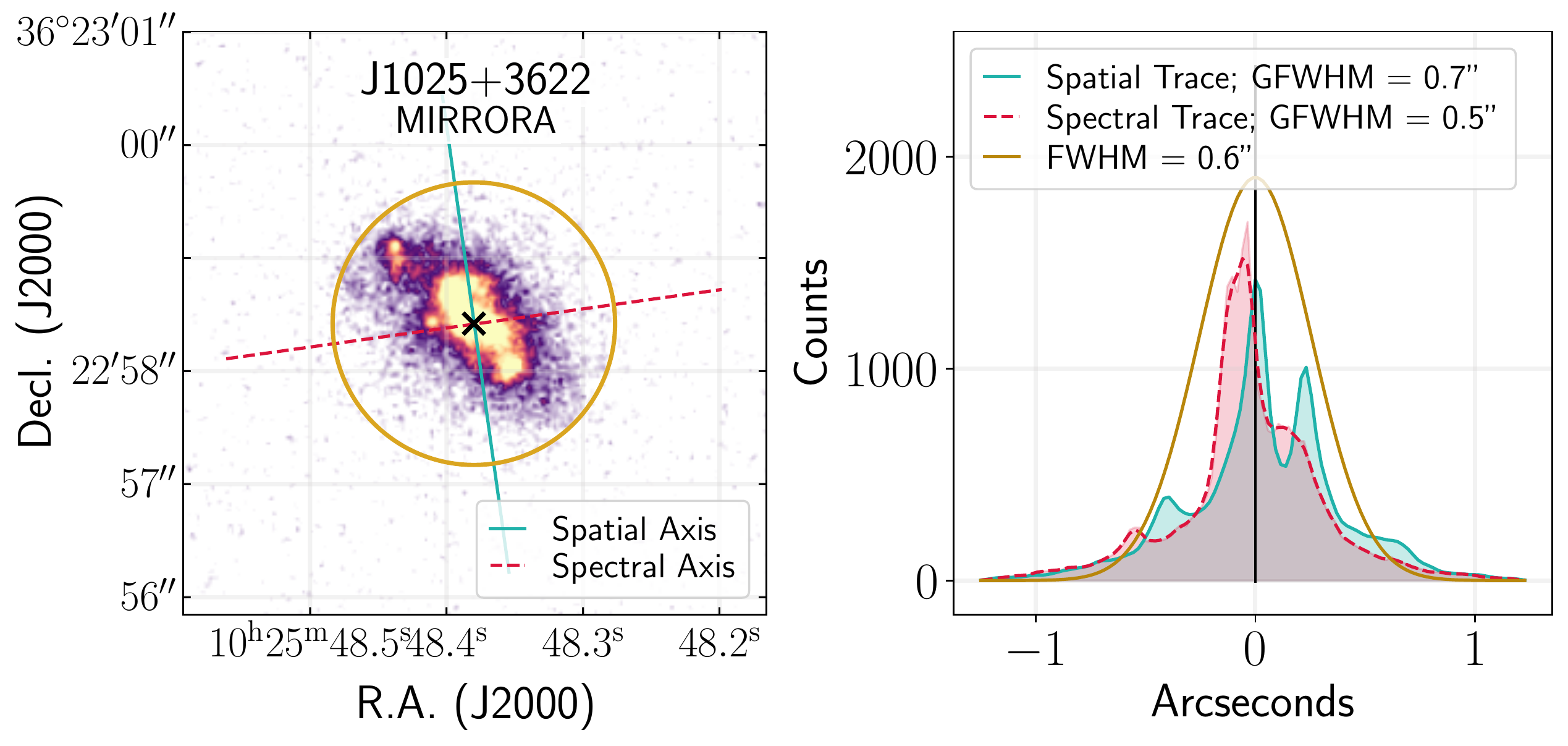}\\ \vspace{-1ex}
\includegraphics[width = 3.0in]{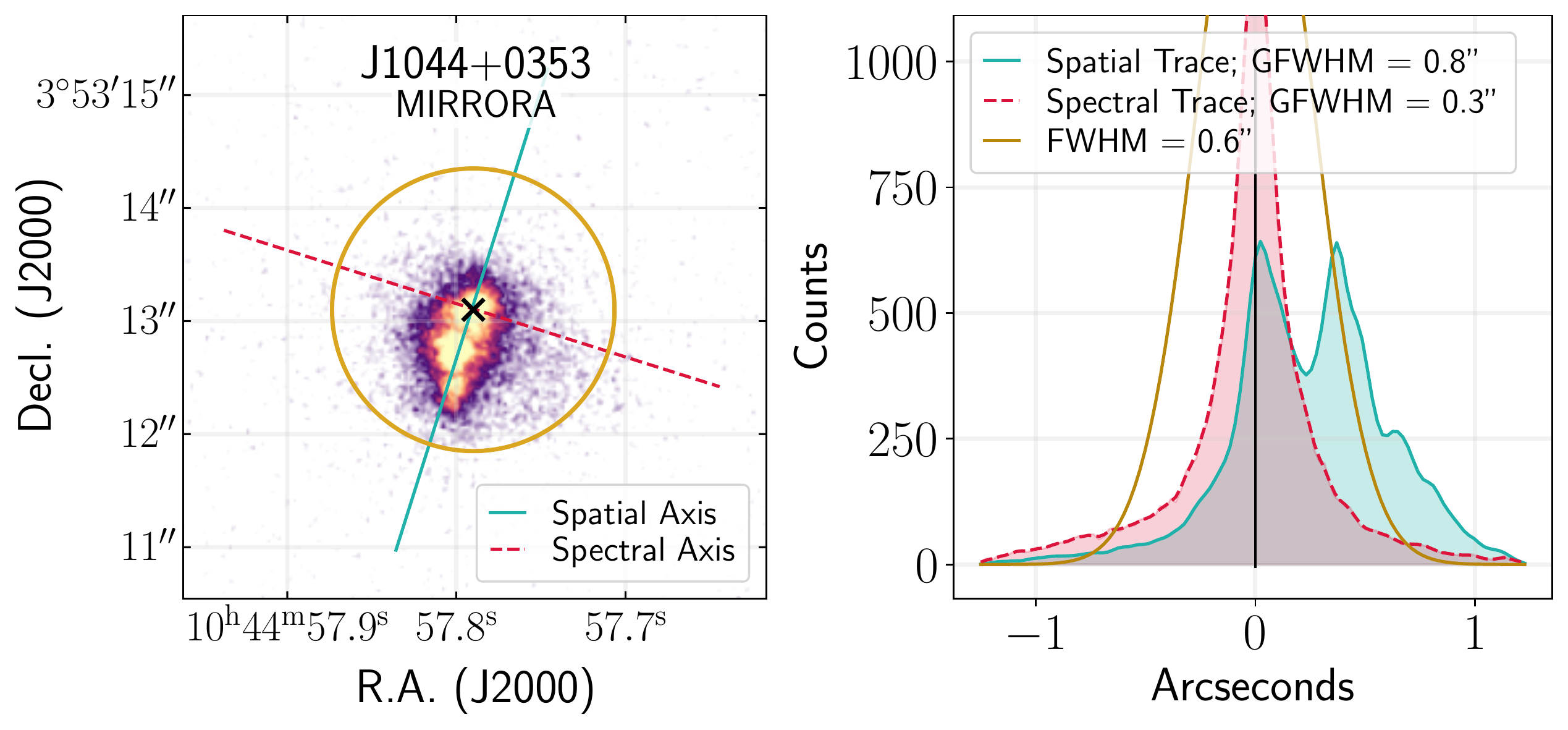} &
\includegraphics[width = 3.0in]{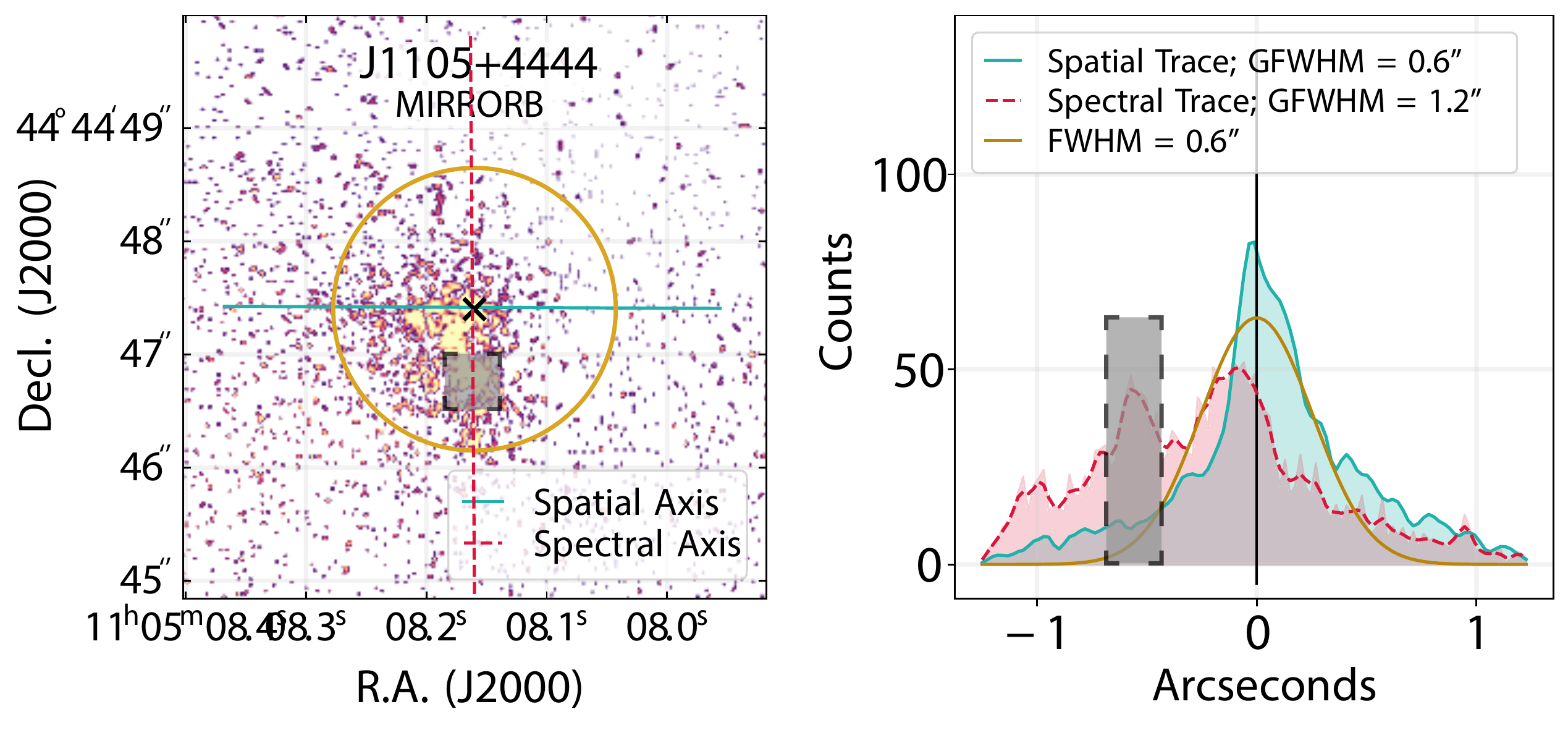}\\ \vspace{-1ex}
\end{tabular}
\caption{Same as Figure~\ref{fig:acq_PS}, here showing targets with extended (GFWHM$>$0\farcs6) light profiles.}
\label{fig:acq_ext}
\end{figure*}

\renewcommand{\thefigure}{2}
\begin{figure*}
\centering
\begin{tabular}{c|c}
\includegraphics[width = 3.0in]{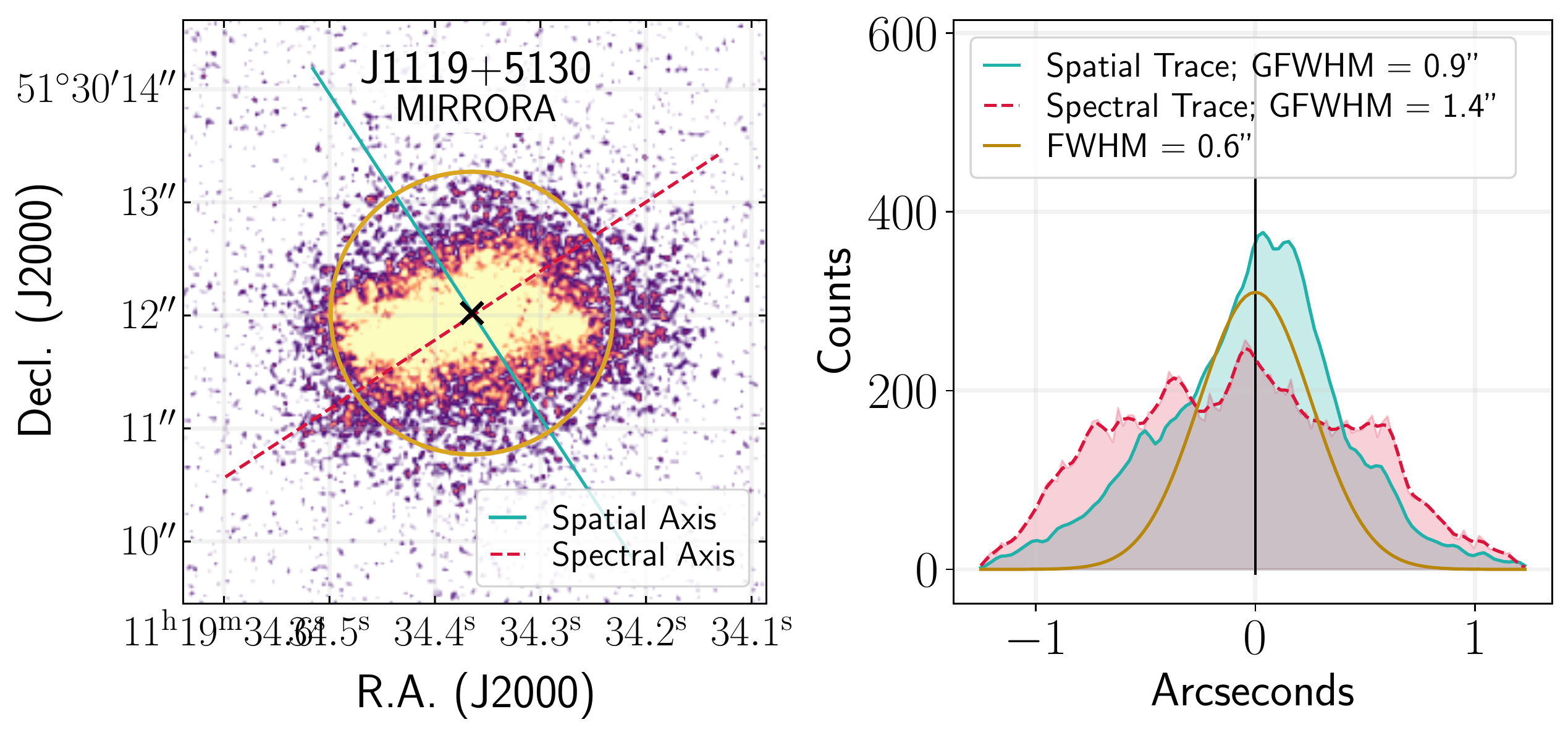} &
\includegraphics[width = 3.0in]{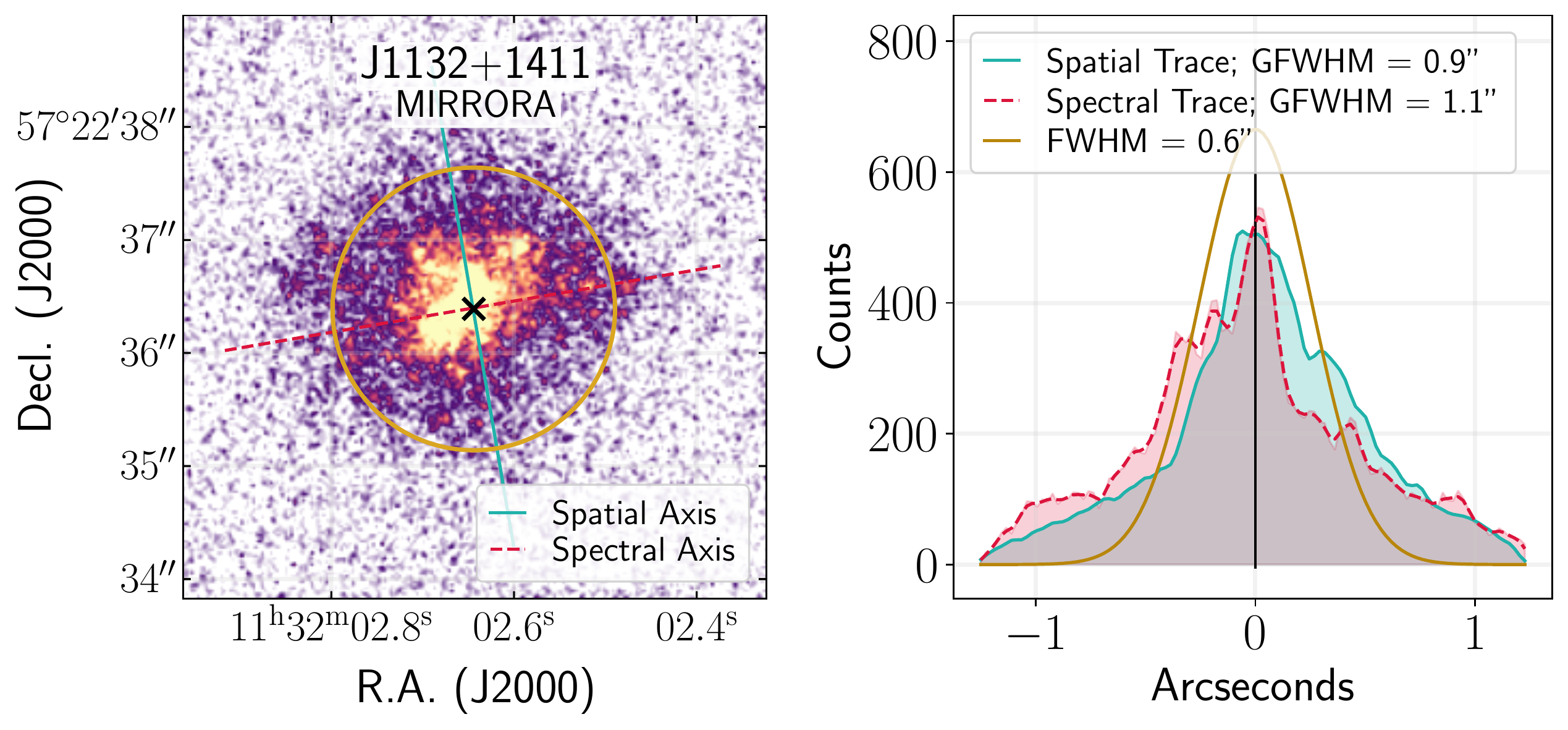}\\ \vspace{-1ex}
\includegraphics[width = 3.0in]{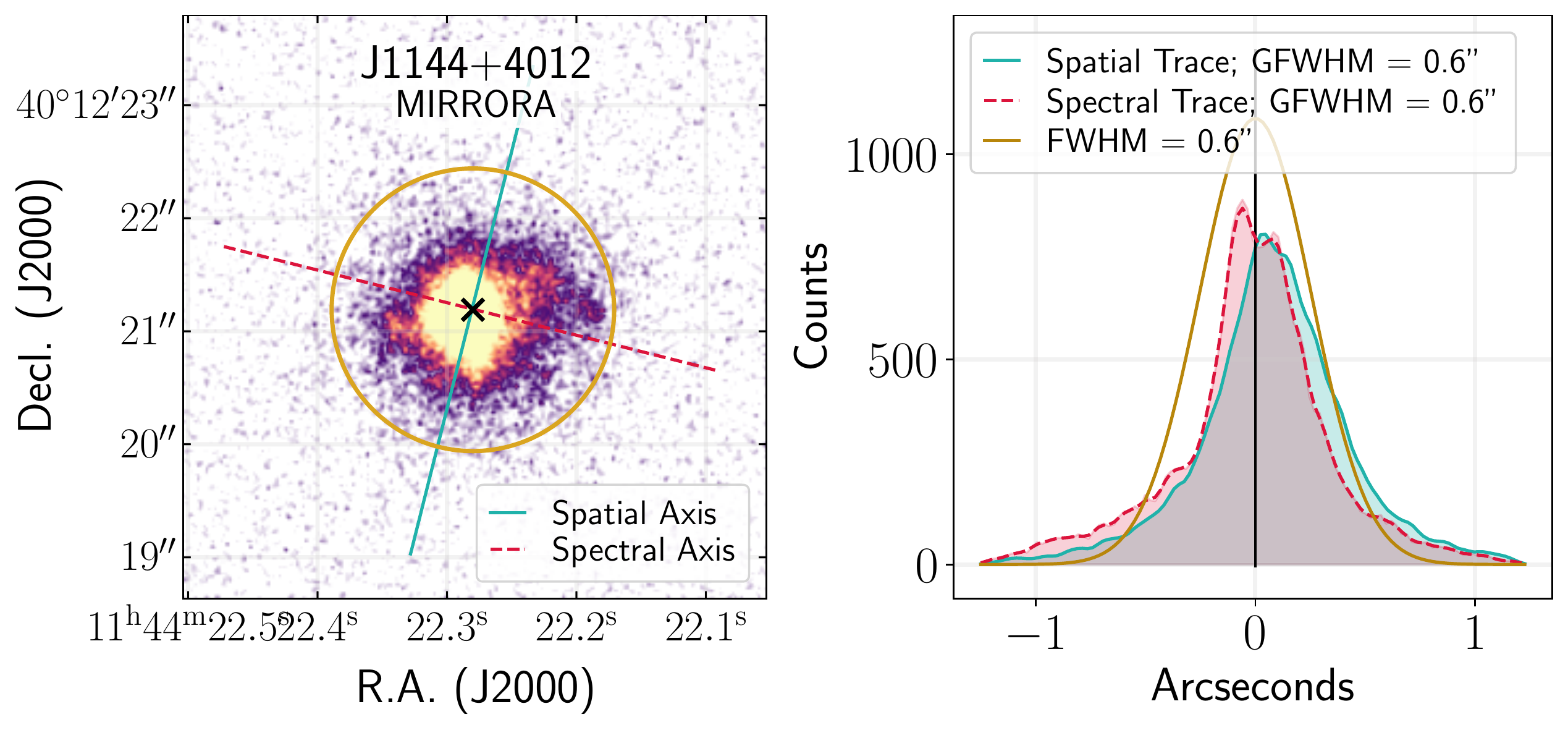} &
\includegraphics[width = 3.0in]{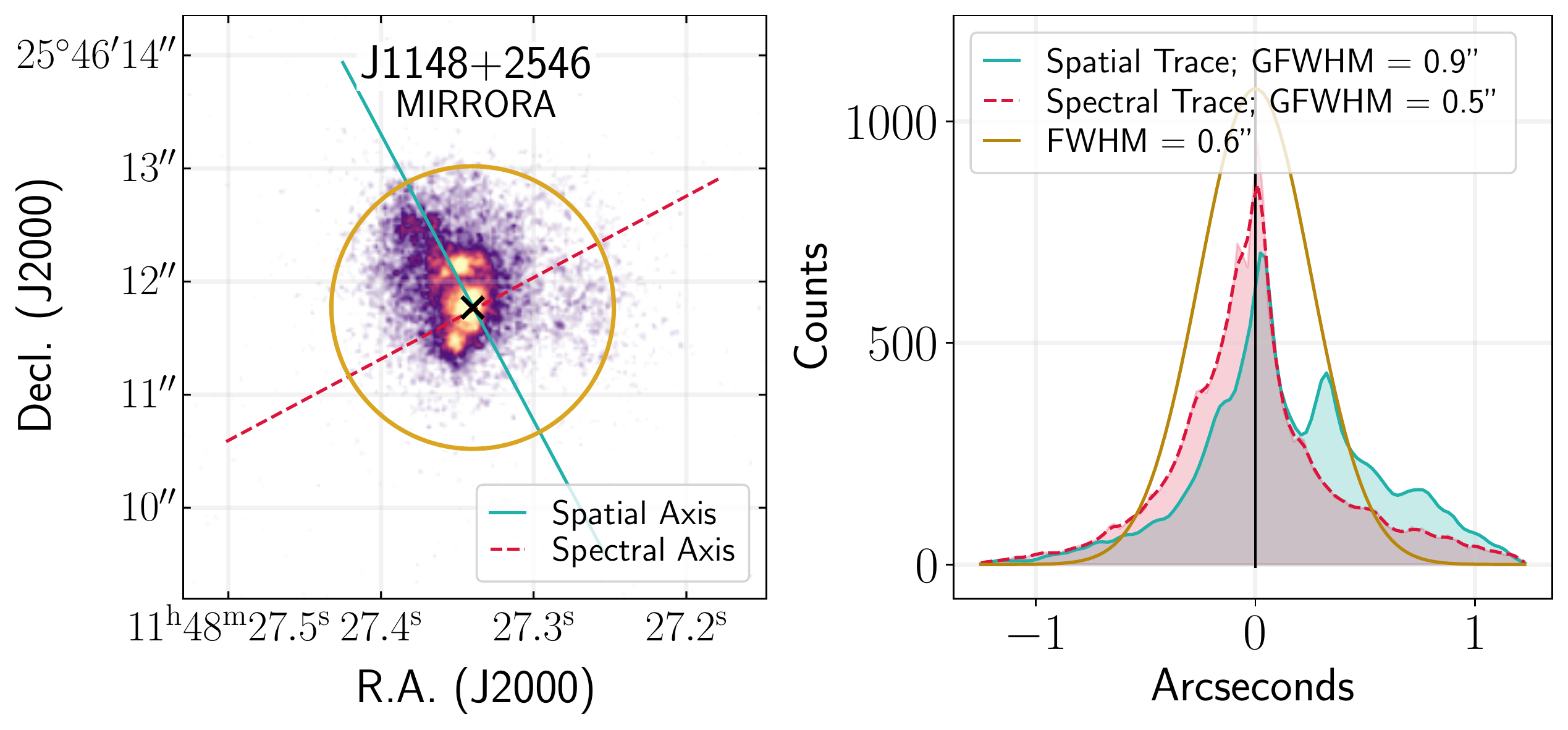}\\ \vspace{-1ex}
\includegraphics[width = 3.0in]{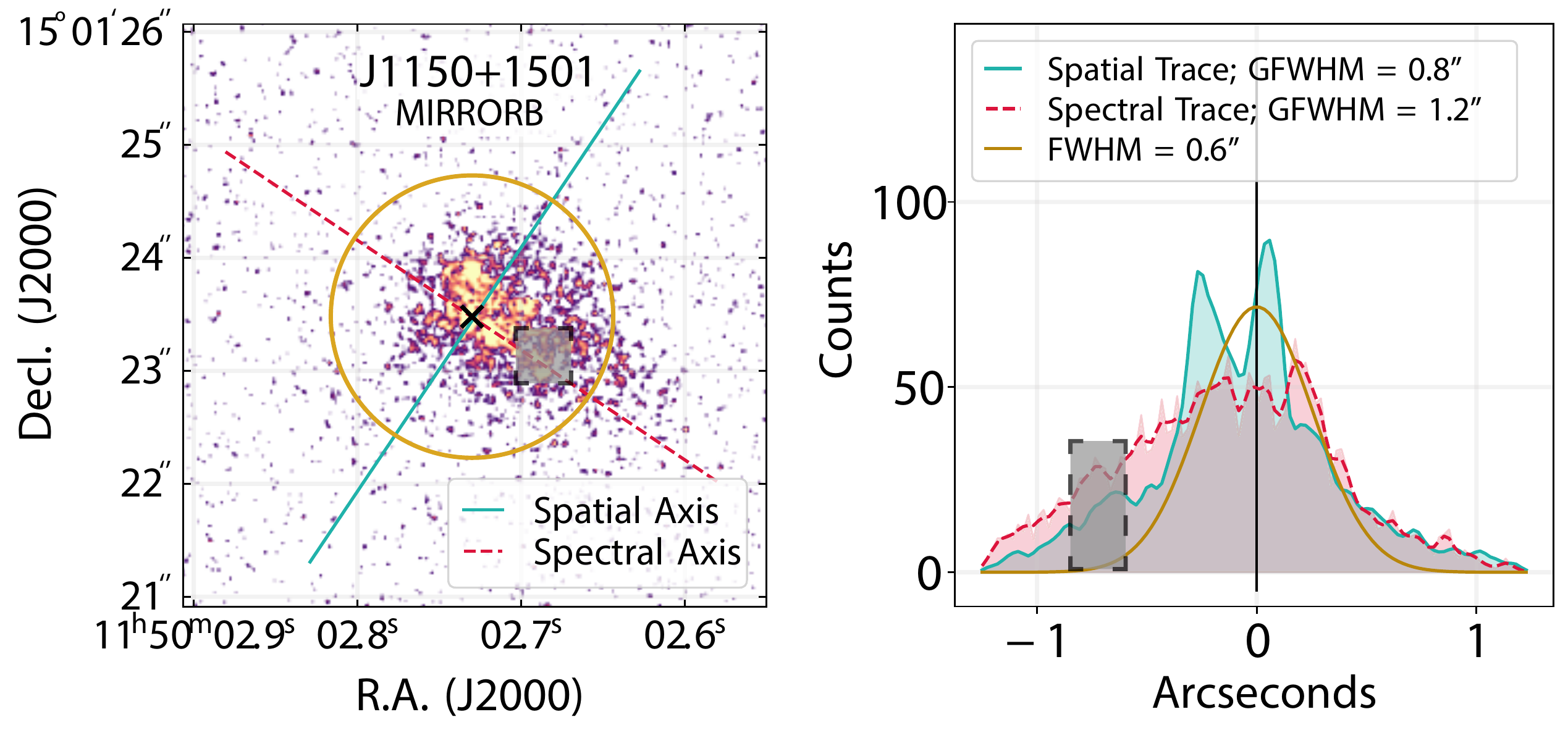} &
\includegraphics[width = 3.0in]{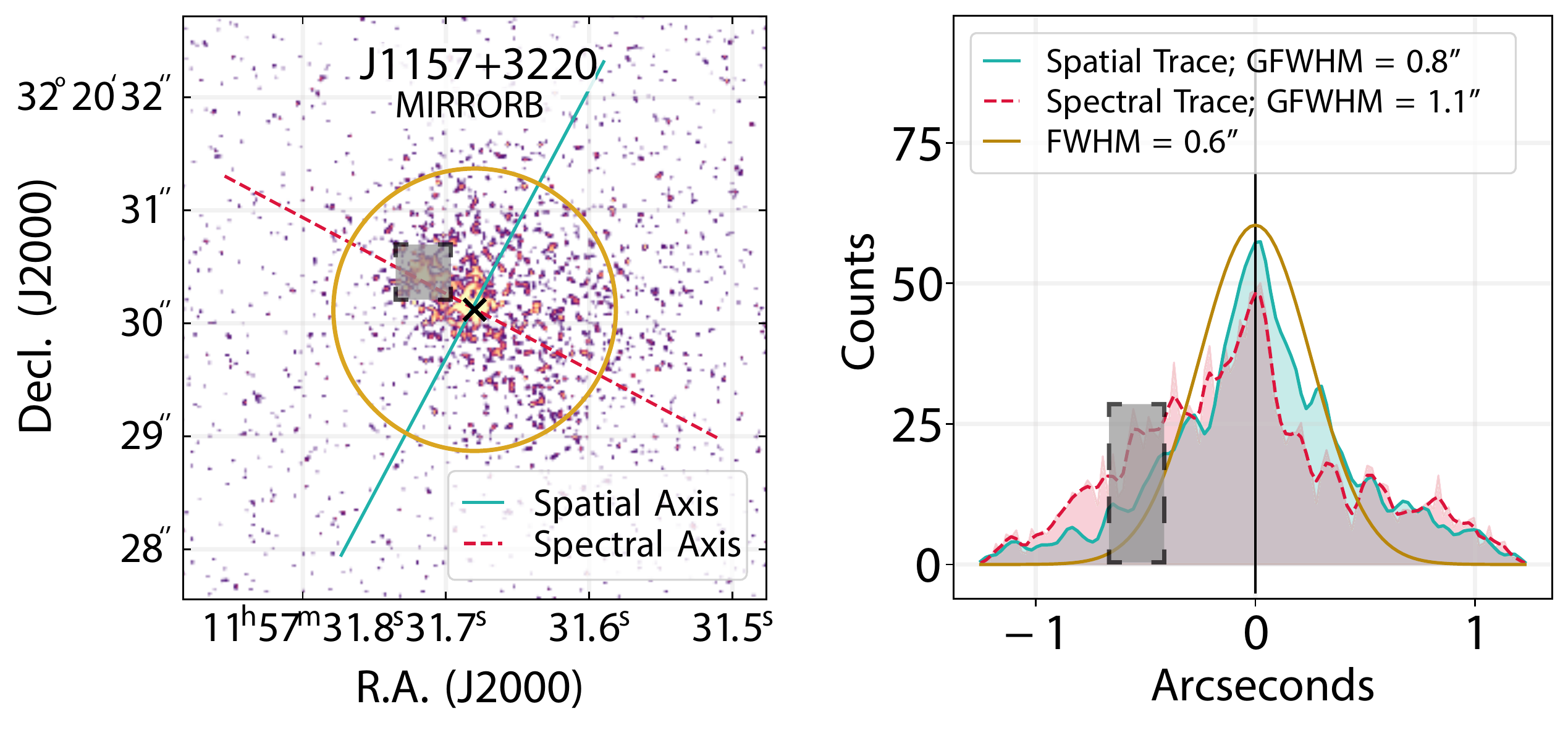}\\ \vspace{-1ex}
\includegraphics[width = 3.0in]{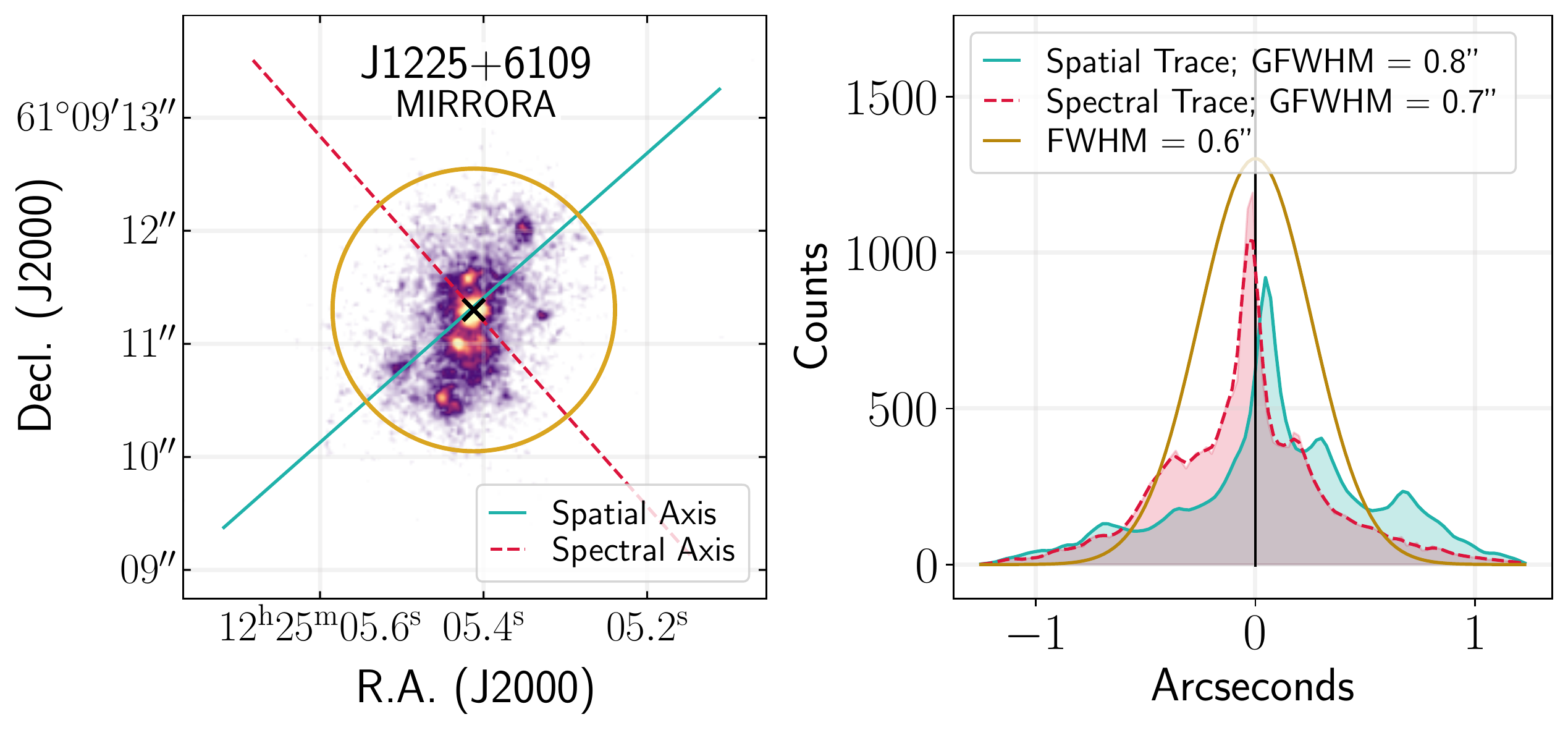} &
\includegraphics[width = 3.0in]{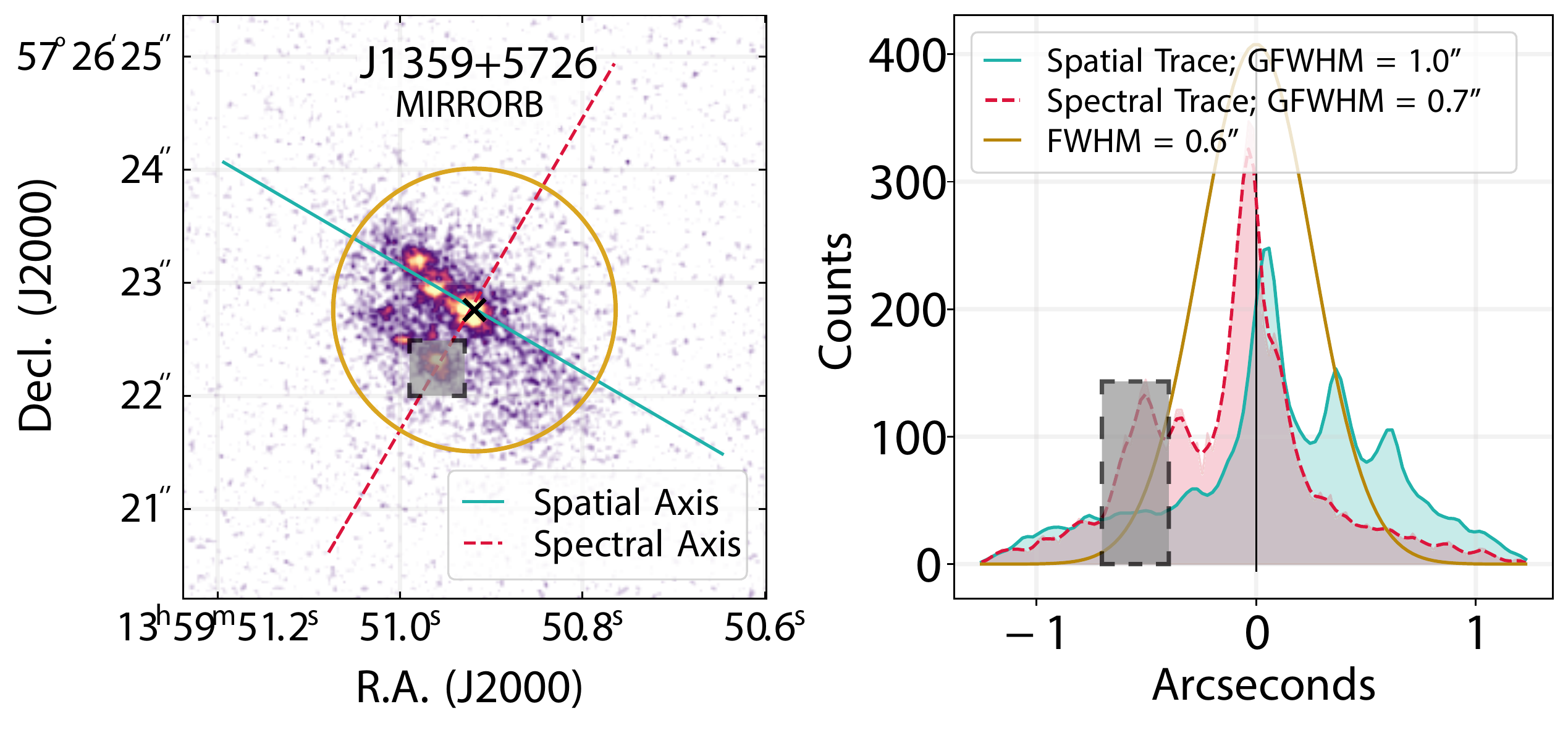}\\ \vspace{-1ex}
\includegraphics[width = 3.0in]{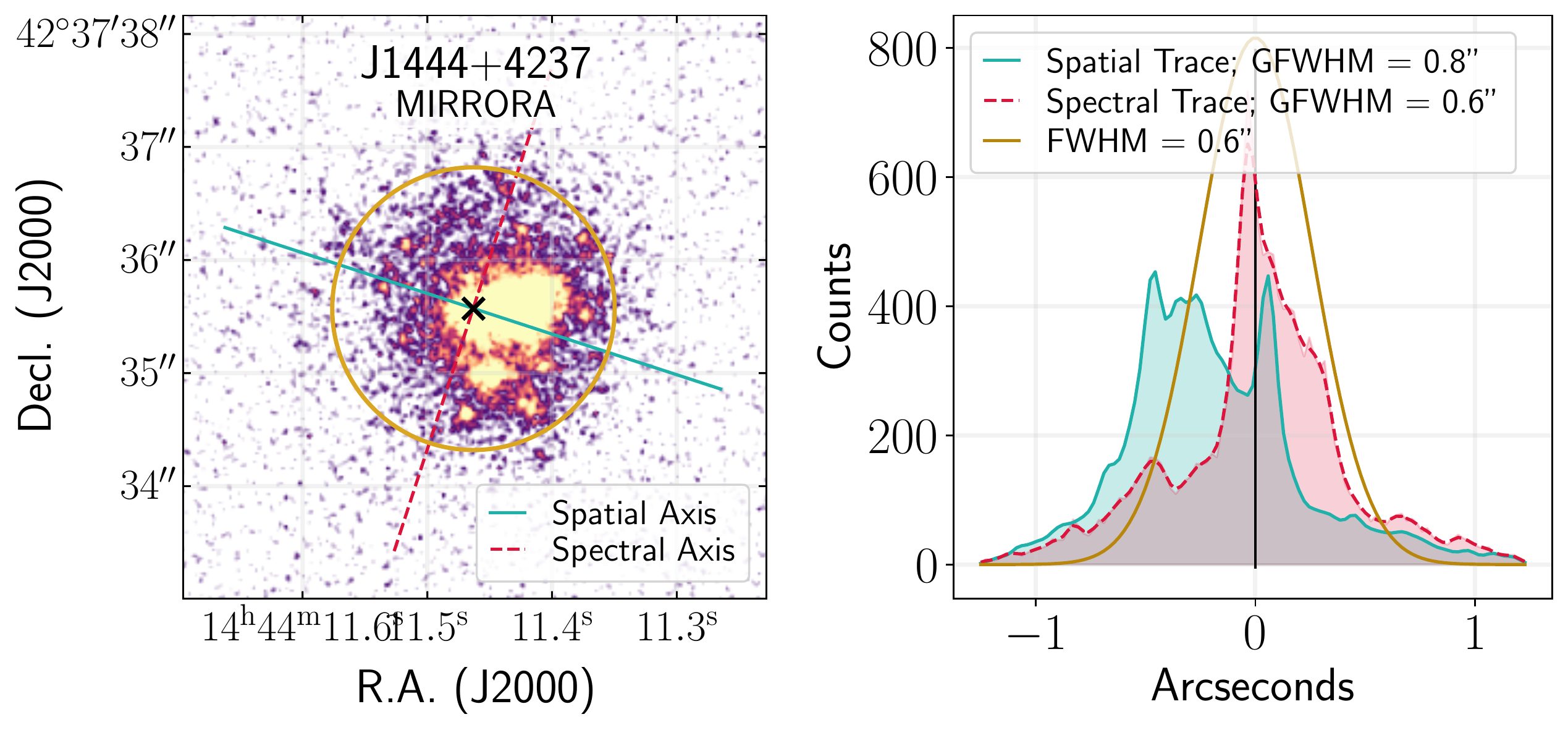} & \\
\end{tabular}
\caption{\textit{ - continued.}}
\end{figure*}

\renewcommand{\thefigure}{3}
\begin{figure*}
\centering
\begin{tabular}{c|c}
\includegraphics[width = 3.0in]{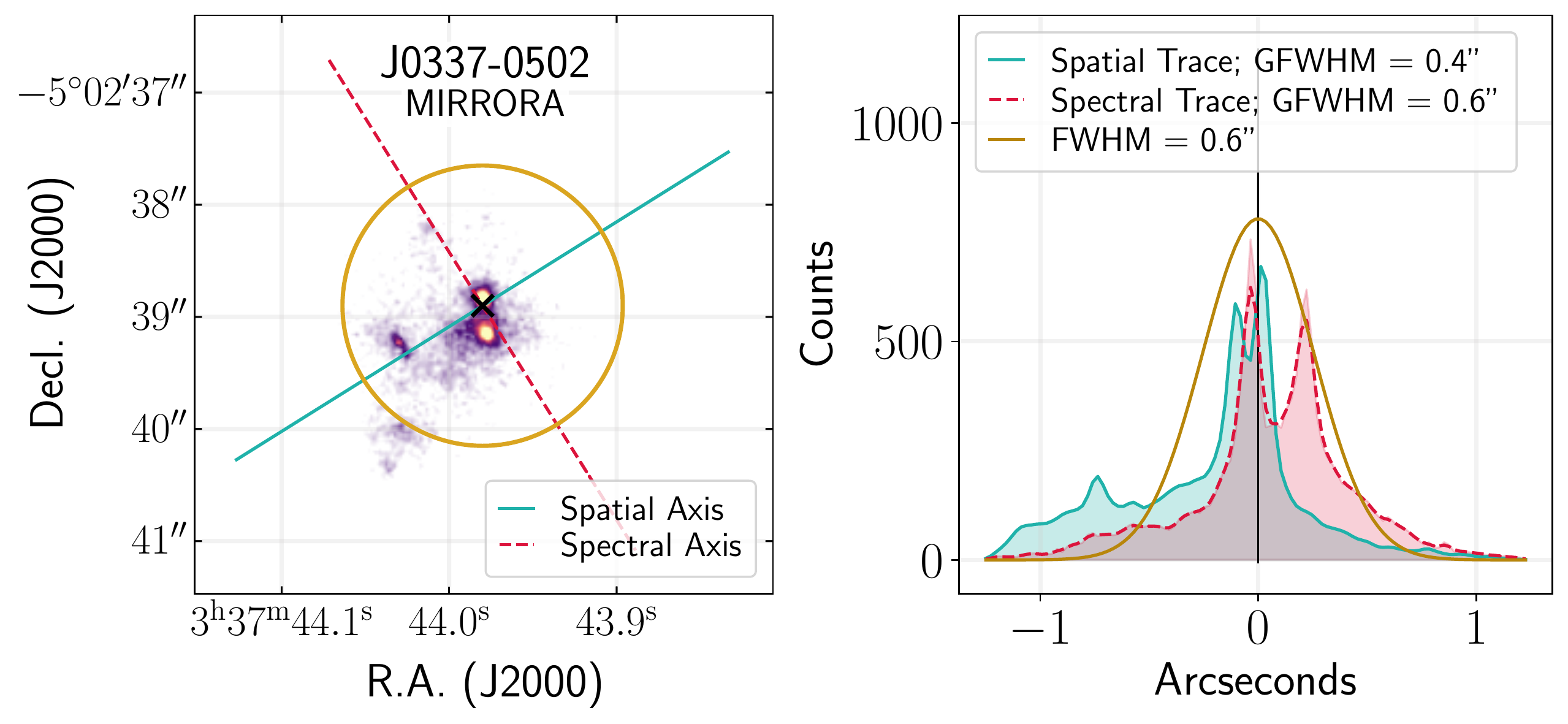} &
\includegraphics[width = 3.0in]{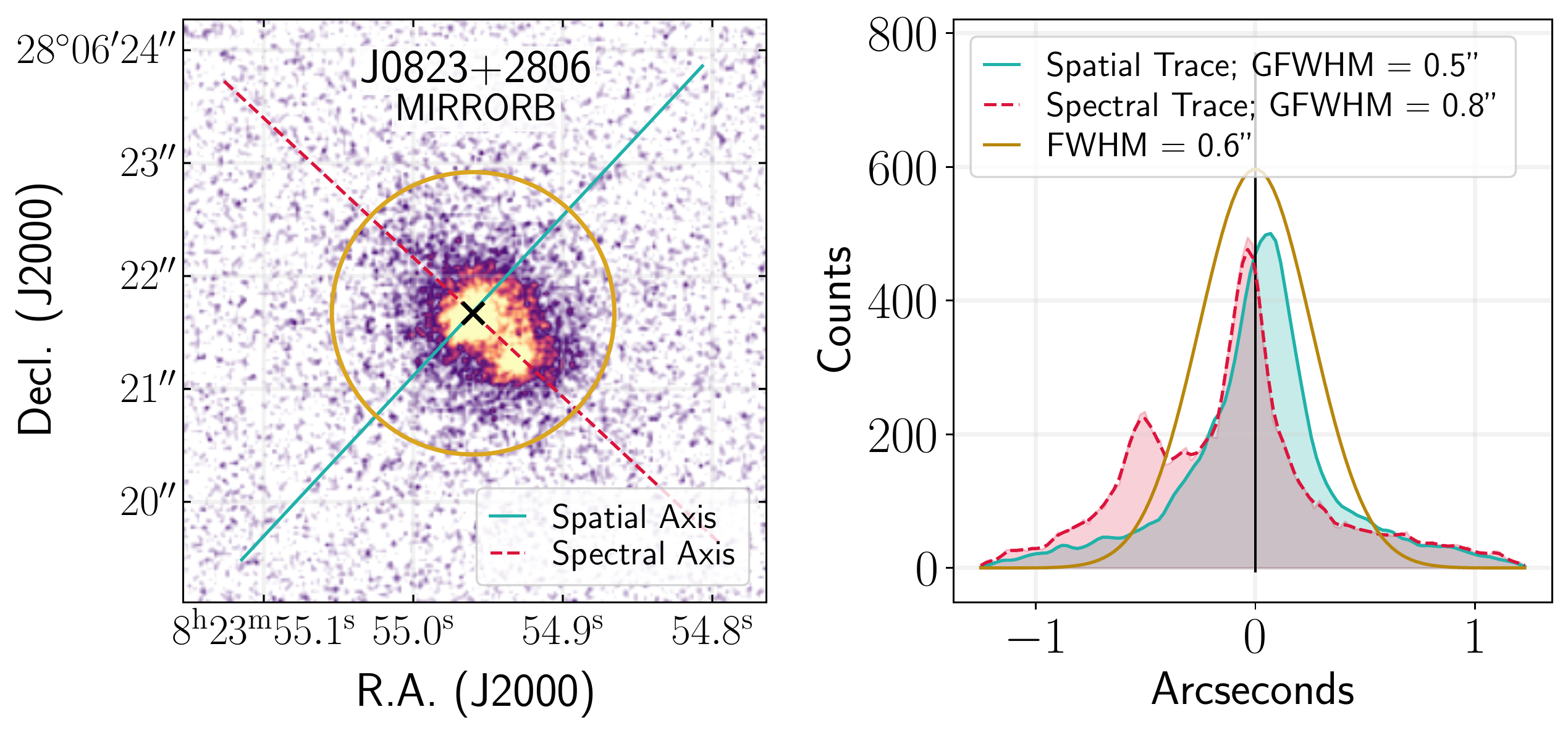}\\ \vspace{-1ex}
\includegraphics[width = 3.0in]{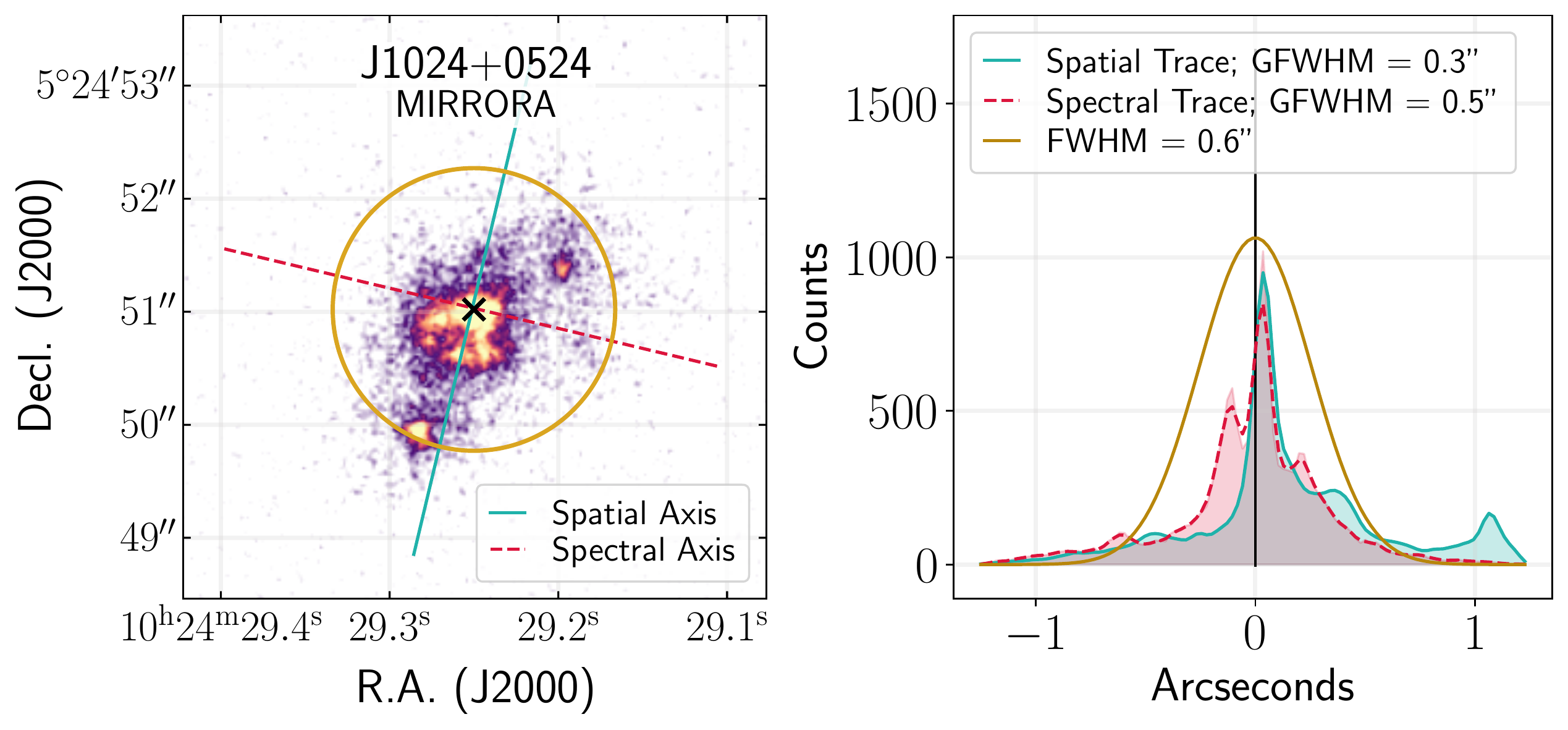} &
\includegraphics[width = 3.0in]{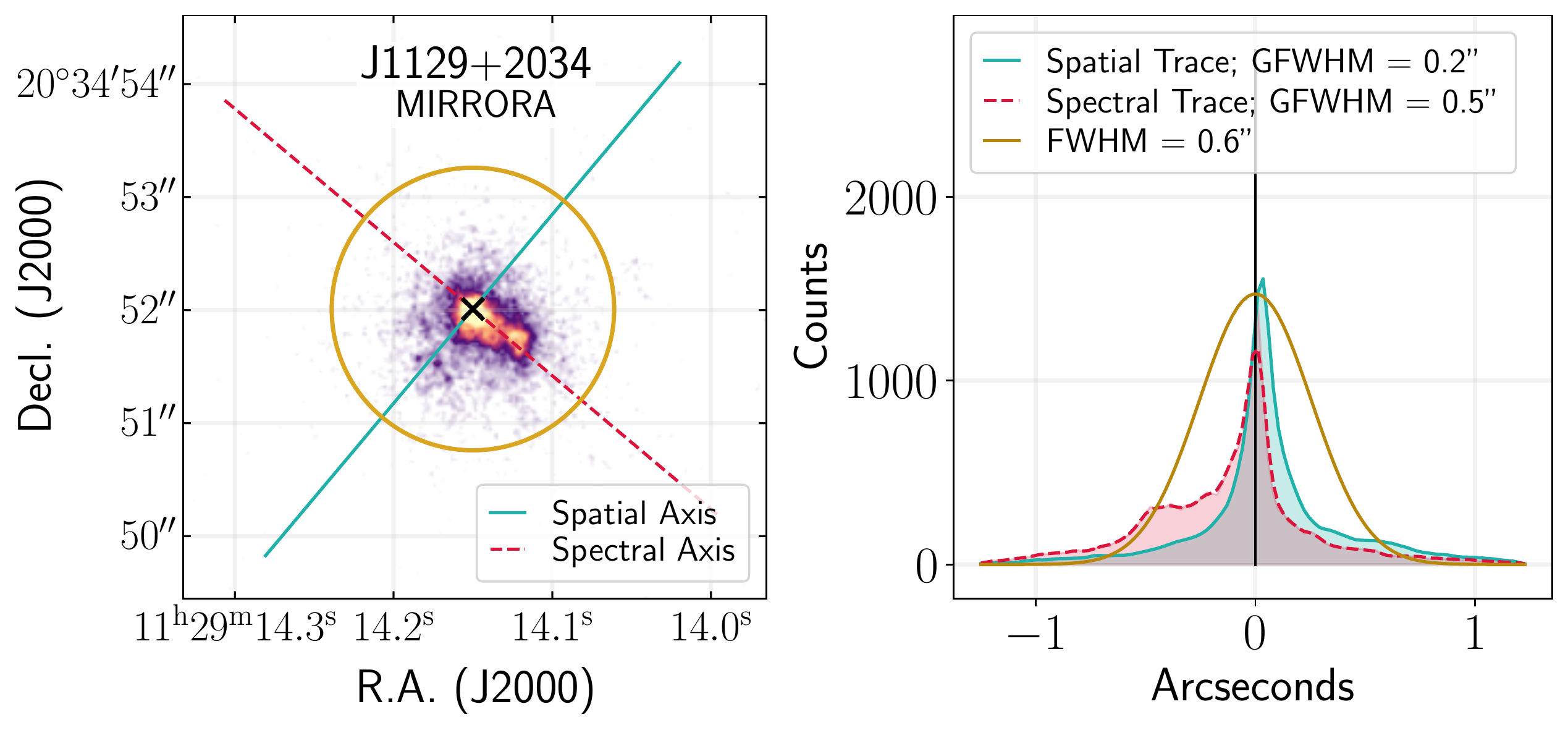}\\ \vspace{-1ex}
\includegraphics[width = 3.0in]{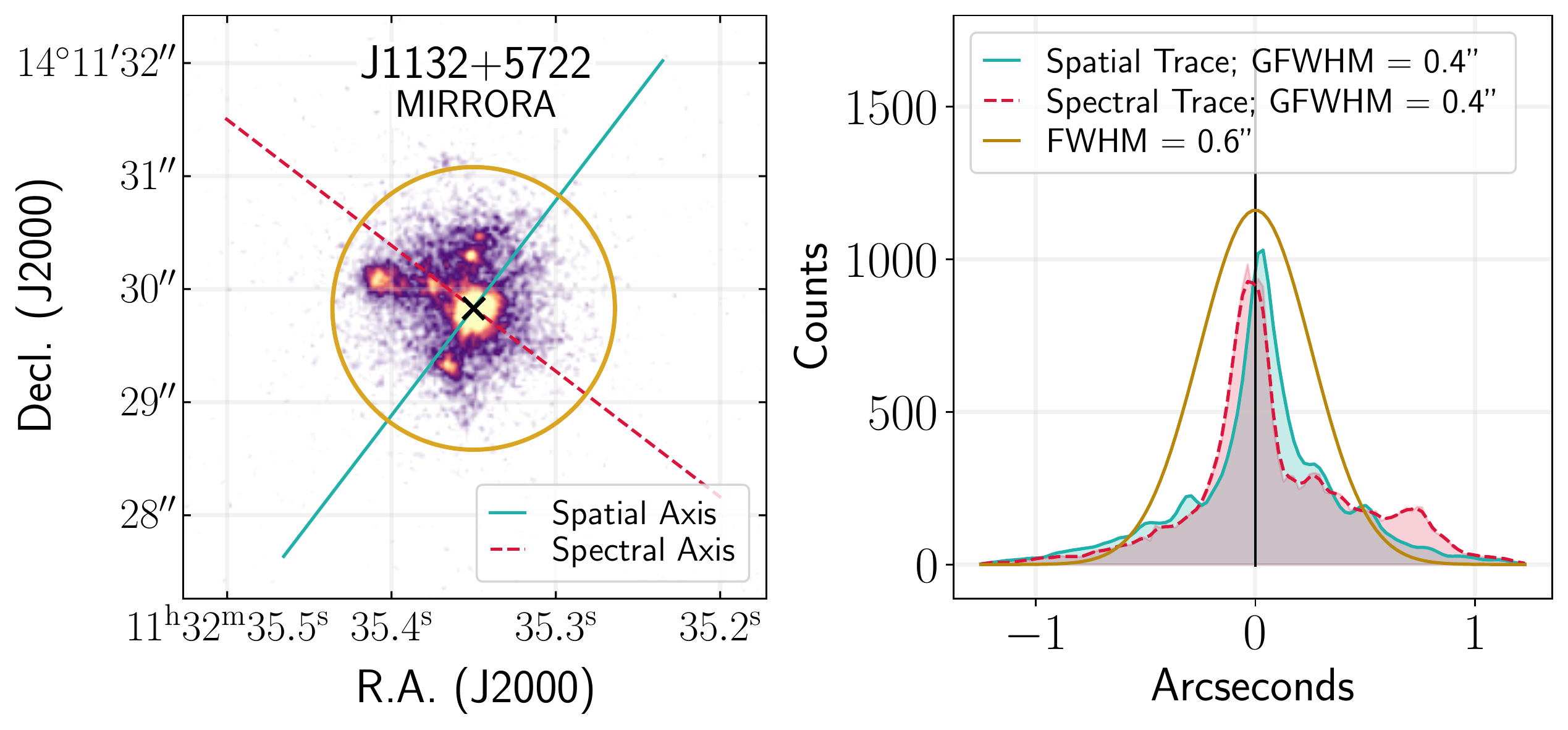} &
\includegraphics[width = 3.0in]{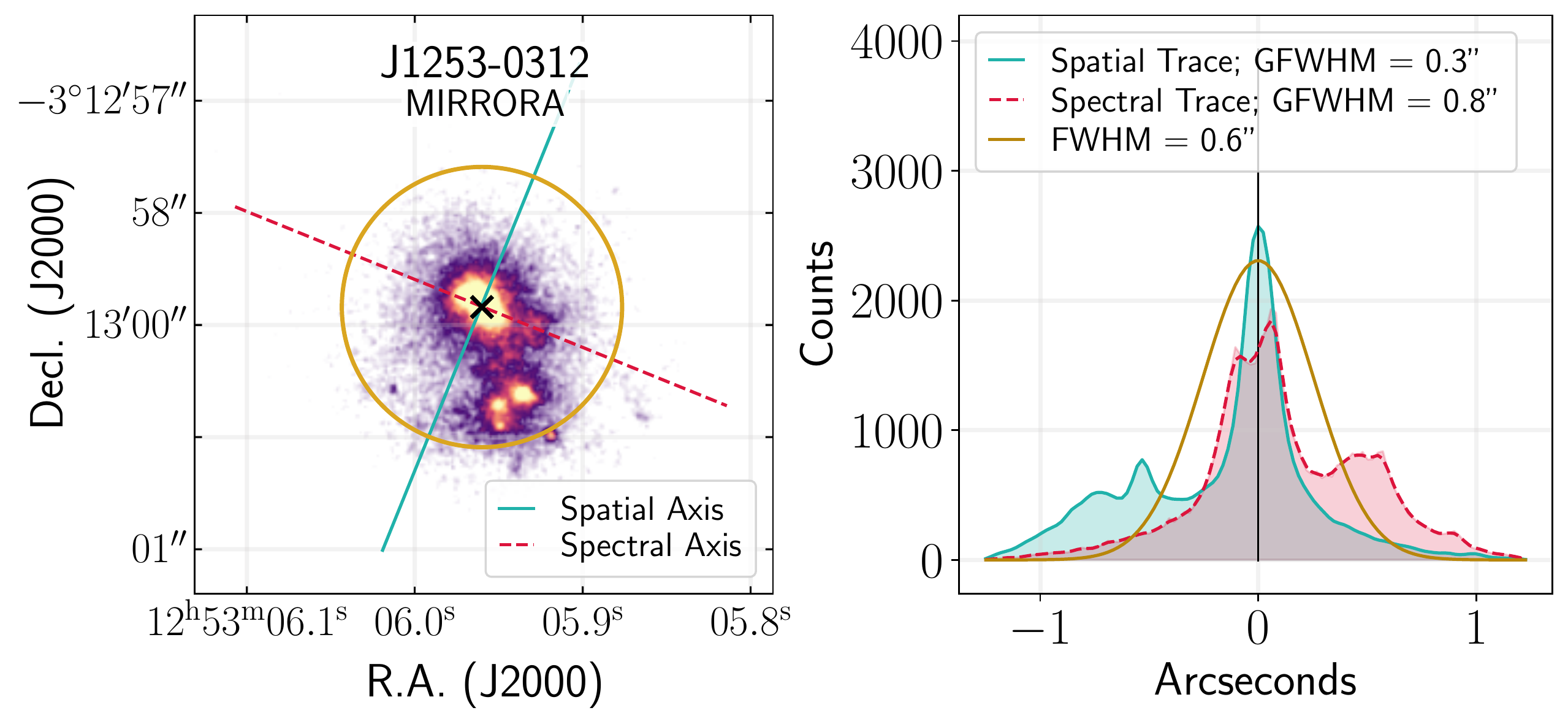}\\ \vspace{-1ex}
\includegraphics[width = 3.0in]{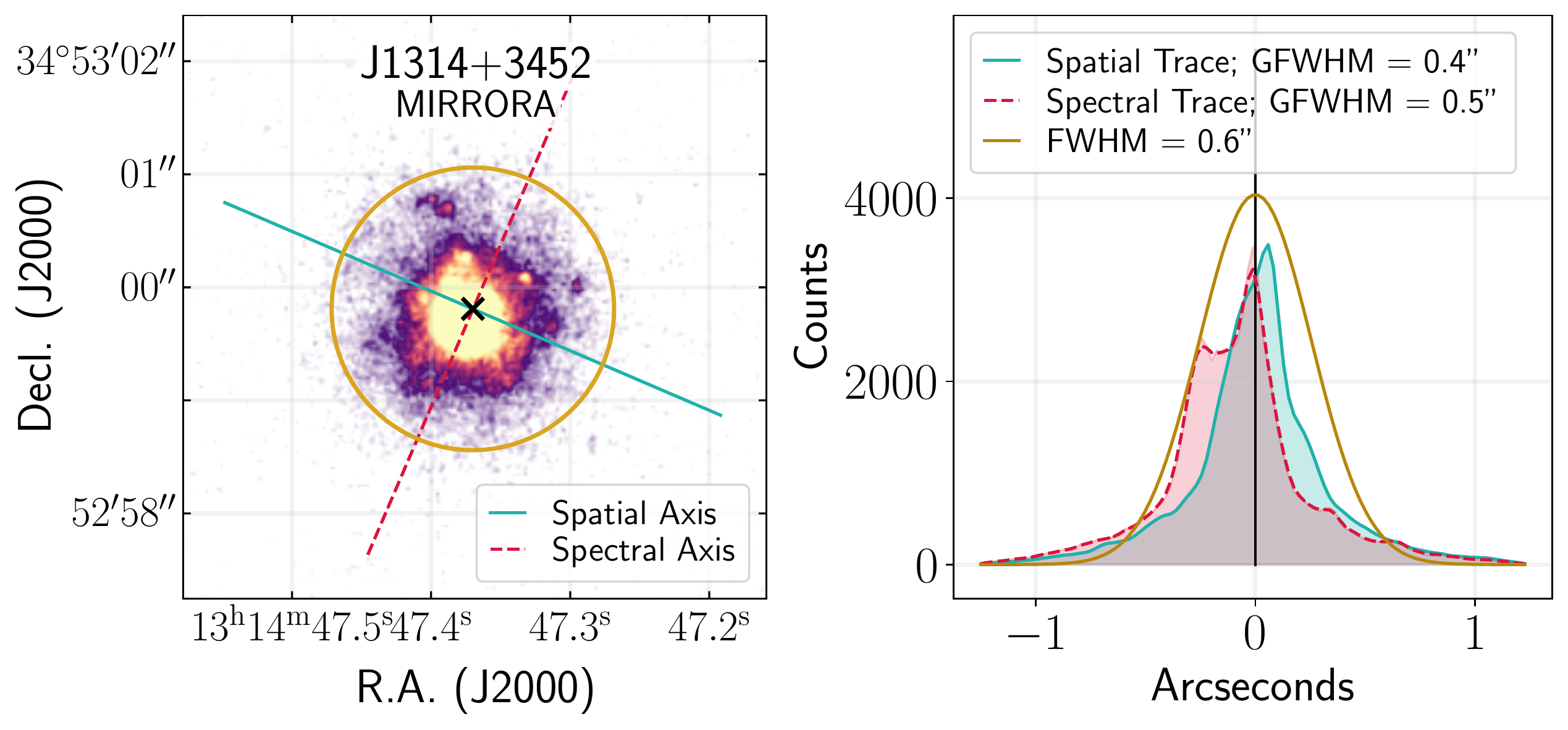} &
\includegraphics[width = 3.0in]{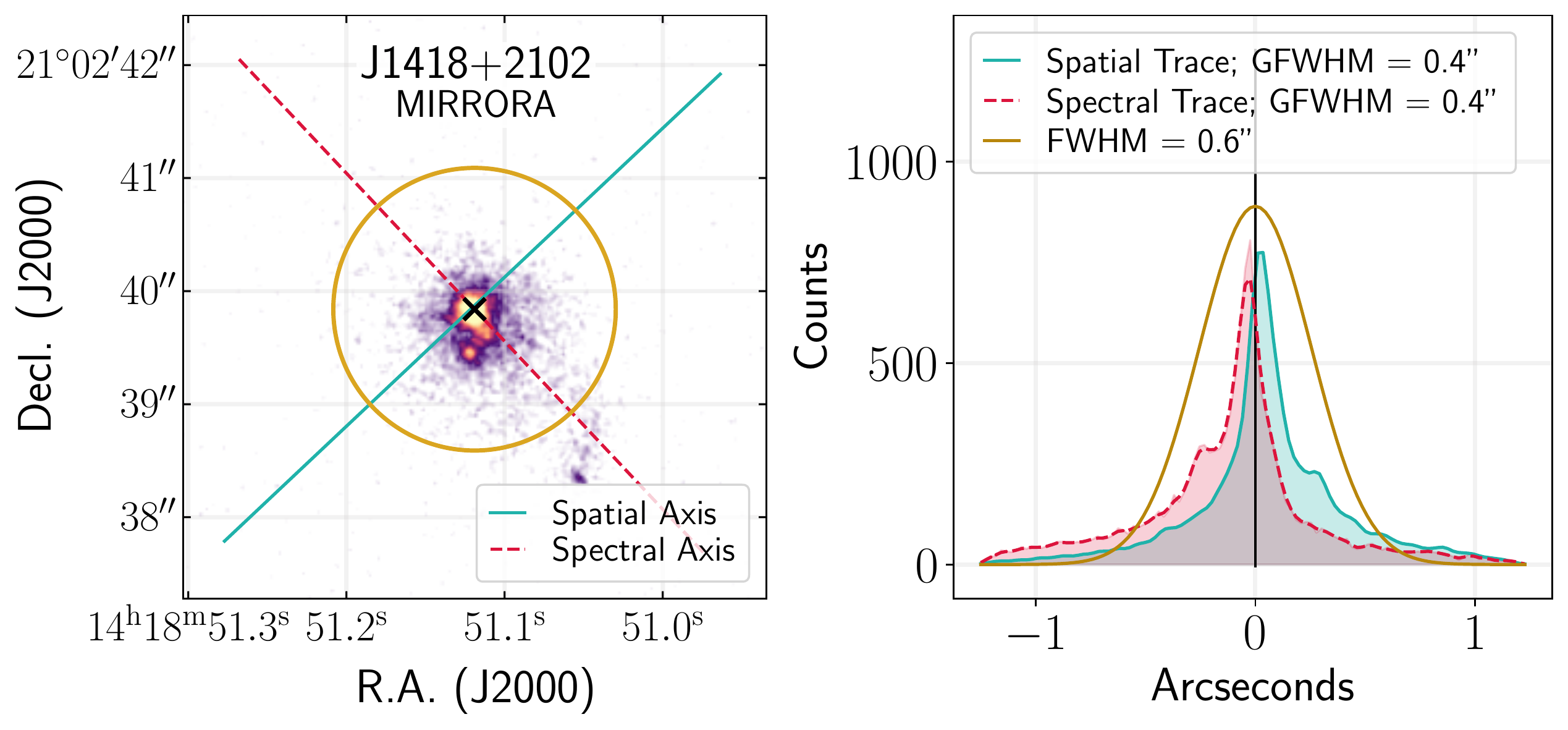}\\ \vspace{-1ex}
\includegraphics[width = 3.0in]{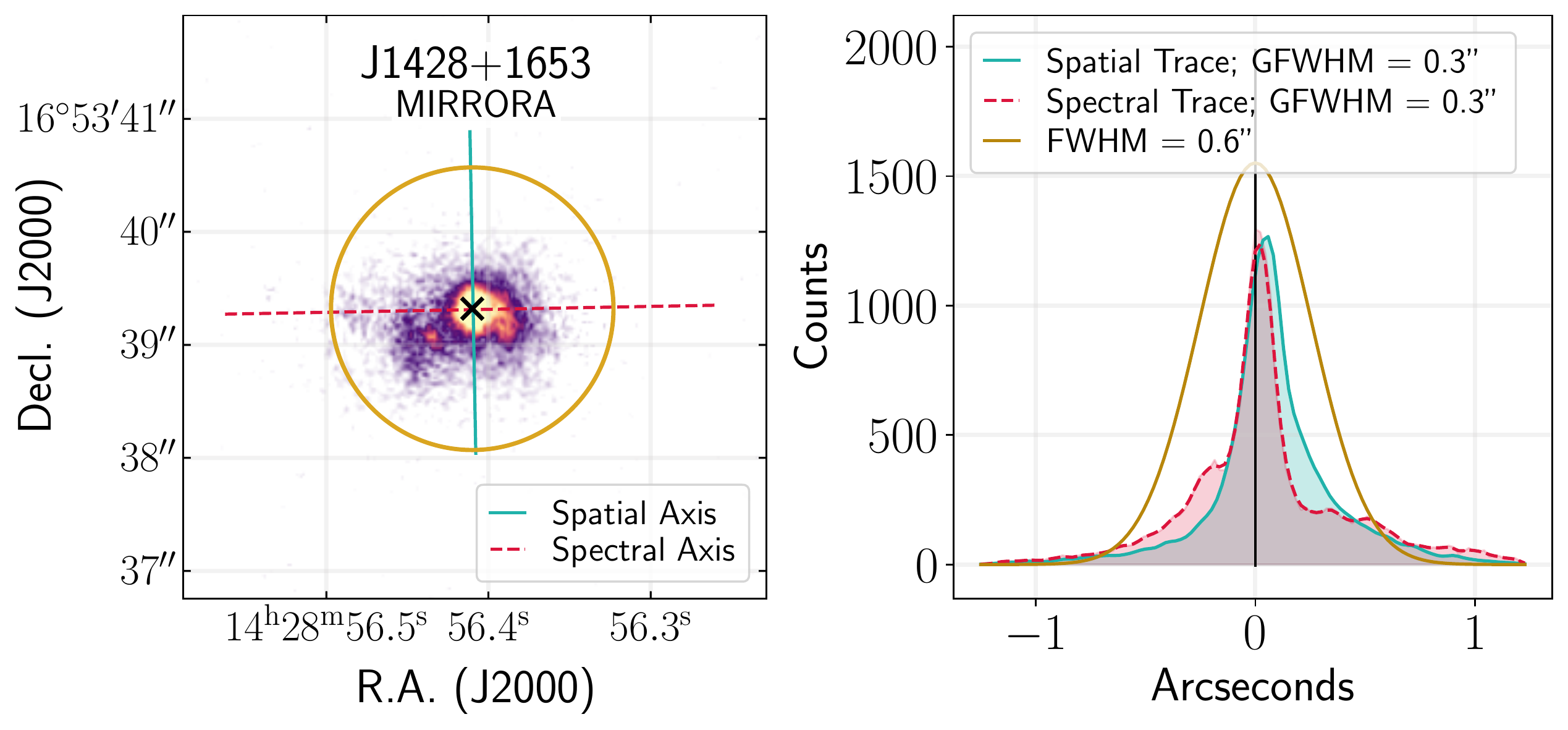} &
\includegraphics[width = 3.0in]{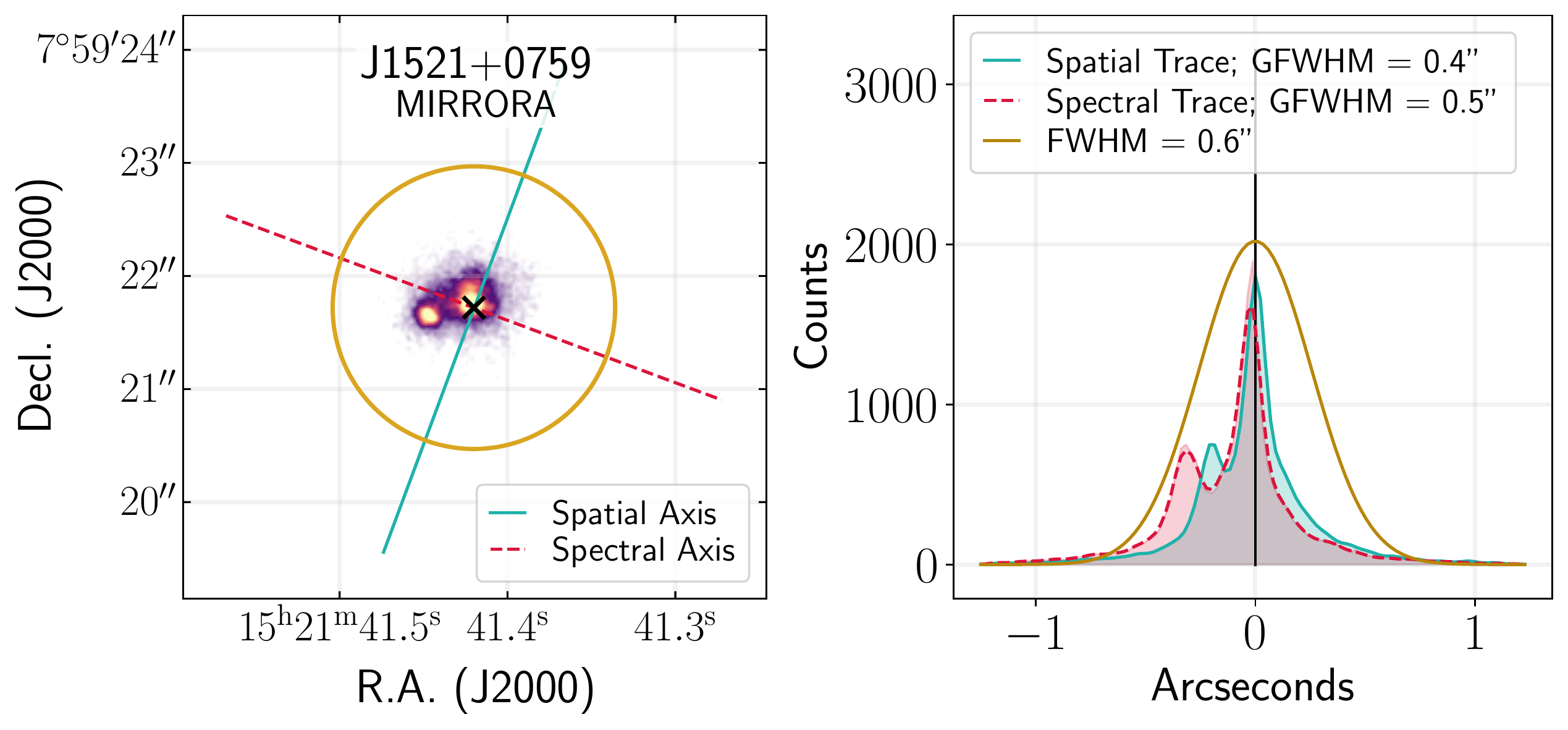}\\ \vspace{-1ex}
\includegraphics[width = 3.0in]{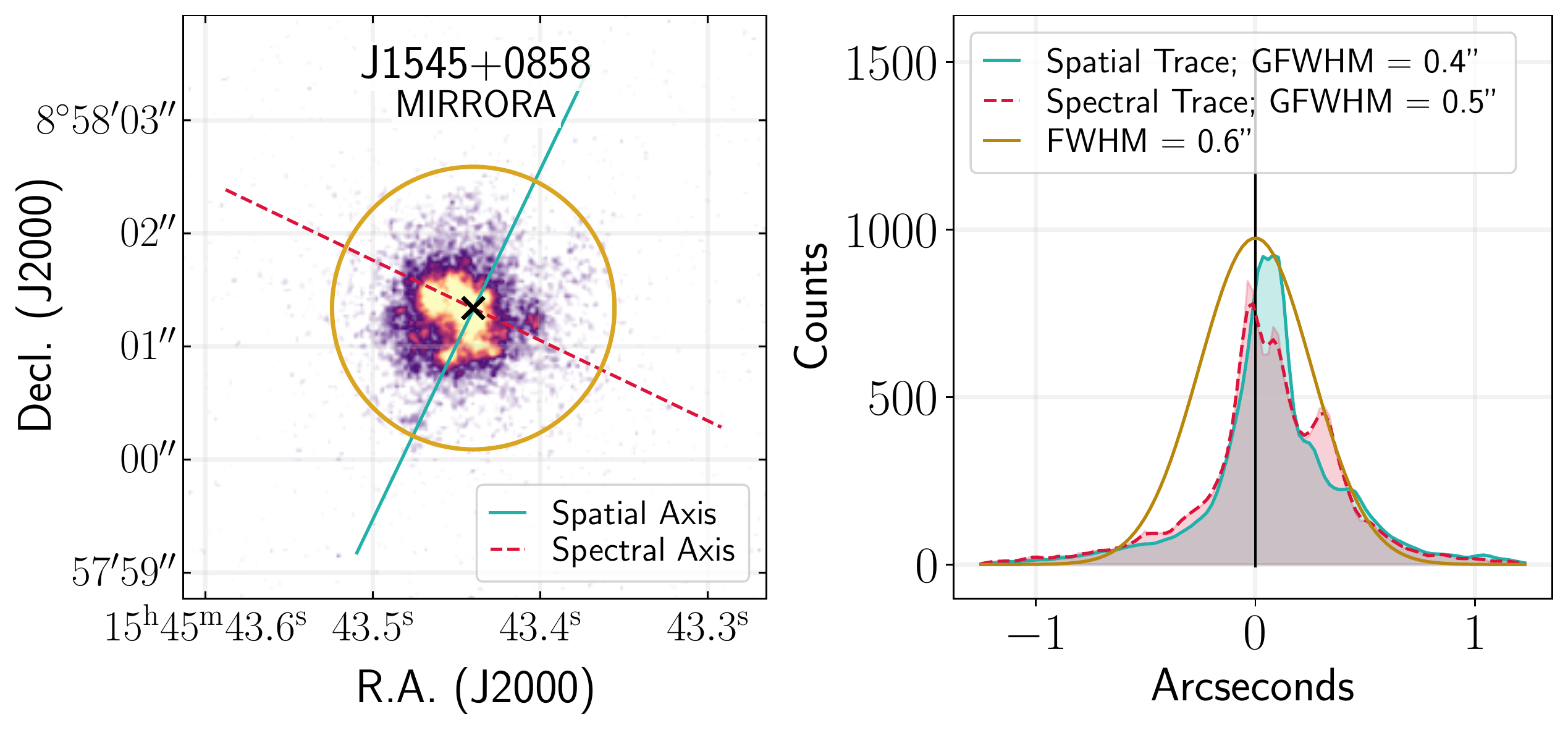} & \\
\end{tabular}
\caption{Same as Figure~\ref{fig:acq_PS}, here showing targets with multi-component profiles (defined as having additional NUV light components beyond 0\farcs4 from the center).}
\label{fig:acq_MC}
\end{figure*}

\section{Data Reduction}\label{sec:reduction}

Observation details of all the datasets included within CLASSY can be found in \citetalias{berg22}, including dataset IDs, gratings, \texttt{CENWAVE}s, position angles, exposure times and rest wavelength coverage. All targets covered by CLASSY are presented in Figures~\ref{fig:acq_PS}--\ref{fig:acq_MC}, which also show the placement of the COS aperture and the position angle used.

Upon retrieval from the archive, all raw data were reduced locally using the COS data-reduction package CalCOS v.3.3.11\footnote{\url{https://github.com/spacetelescope/calcos/releases}}. This includes both new data attributable to the CLASSY survey itself (PID: 15840, PI: Berg) and all archival data included in CLASSY, as detailed in \citetalias{berg22}. It is the combination of both data from PID:15840 and archival data that we define as CLASSY data hereon and it should be noted here that \textit{all} CLASSY datasets were processed in a self-consistent way, in that they were all reduced via the most recent version of the CalCOS pipeline. With regards to the homogeneity of the data reduction procedures employed, it may also be useful to note that CLASSY contains data observed at different Lifetime Positions on the COS FUV detector. As detailed in table~2 of \citetalias{berg22}, 15 galaxies were observed at LP 1/2 for G130M or G160M, 28 galaxies occur at LPs 3/4 for G130M and G160M and 2 galaxies contain coadded data with a mix of LPs 1/2 or 3/4. All new CLASSY data was executed before the commissioning of LP5.

The CalCOS pipeline consists of three main components that calibrate COS data by (1) correcting for instrumental effects such as thermal drifts, geometric distortion corrections, Doppler corrections, and pixel-to-pixel variations in sensitivity, (2) generating an exposure-specific wavelength-calibrated scale, and (3) extracting and producing a final (one-dimensional) flux-calibrated (summed) spectrum for the entire observation. While CalCOS typically produces `ready to go' final spectra, it was necessary to re-reduce a large portion of our datasets due to the extended nature of $\sim$60\%\ of the CLASSY targets.

HST/COS is a spectrograph optimized for UV spectroscopy of point sources. This is evident in several aspects of the COS pipeline, most notably in the \texttt{TWOZONE} extraction technique, which utilizes a point-source profile for the extraction of the data on the FUV detector (815--2050\AA). The NUV detector (1700--3200\AA) instead uses a fixed extraction box, which we describe in detail below.\footnote{The \texttt{TWOZONE} extraction is implemented on the FUV detector only because it was designed to reliably extract spectra close to regions suffering from gain sag (which primarily result from bright \Lya\ airglow emission). This extraction method is only implemented in the CalCOS pipeline for data at Lifetime Positions 3-5. The NUV detector does not suffer from gain sag and does not require optimized extraction.} The full details on the extraction methodologies employed by CalCOS can be found in the COS data handbook (DHB)\footnote{\url{https://hst-docs.stsci.edu/cosdhb}, \citet{COSDHB}}. Owing to the extended nature of $\sim$60\%\ of our targets, where non-negligible amounts of flux can extend beyond the extraction profiles used in CalCOS, it was necessary to test different extraction techniques with the aim of increasing the signal-to-noise (S/N) of the CLASSY spectra of extended sources. In order to determine whether a source was classified as compact, extended and/or had multiple component (MC) light profiles, we inspected each of the COS NUV target acquisition images by collapsing the light profiles along the cross-dispersion and dispersion direction, and subsequently fitting the spatial and spectral traces with single Gaussian profiles. The width of these profiles, which will be broadened in the cases of extended and/or MC light profiles, can help determine whether a source is considered to be a non-compact source. It is important to note that determining whether a target is extended or not also has implications on the spectral resolution of the spectra, in that extended sources will have a decreased spectral resolution compared to that of a point source (see Section~\ref{sec:3.2}). As such, for data quality purposes, we take both profiles into account when classifying the morphological nature of the CLASSY targets. 

\renewcommand{\thefigure}{4}
\begin{figure*} 
 \centering
\includegraphics[scale=0.25]{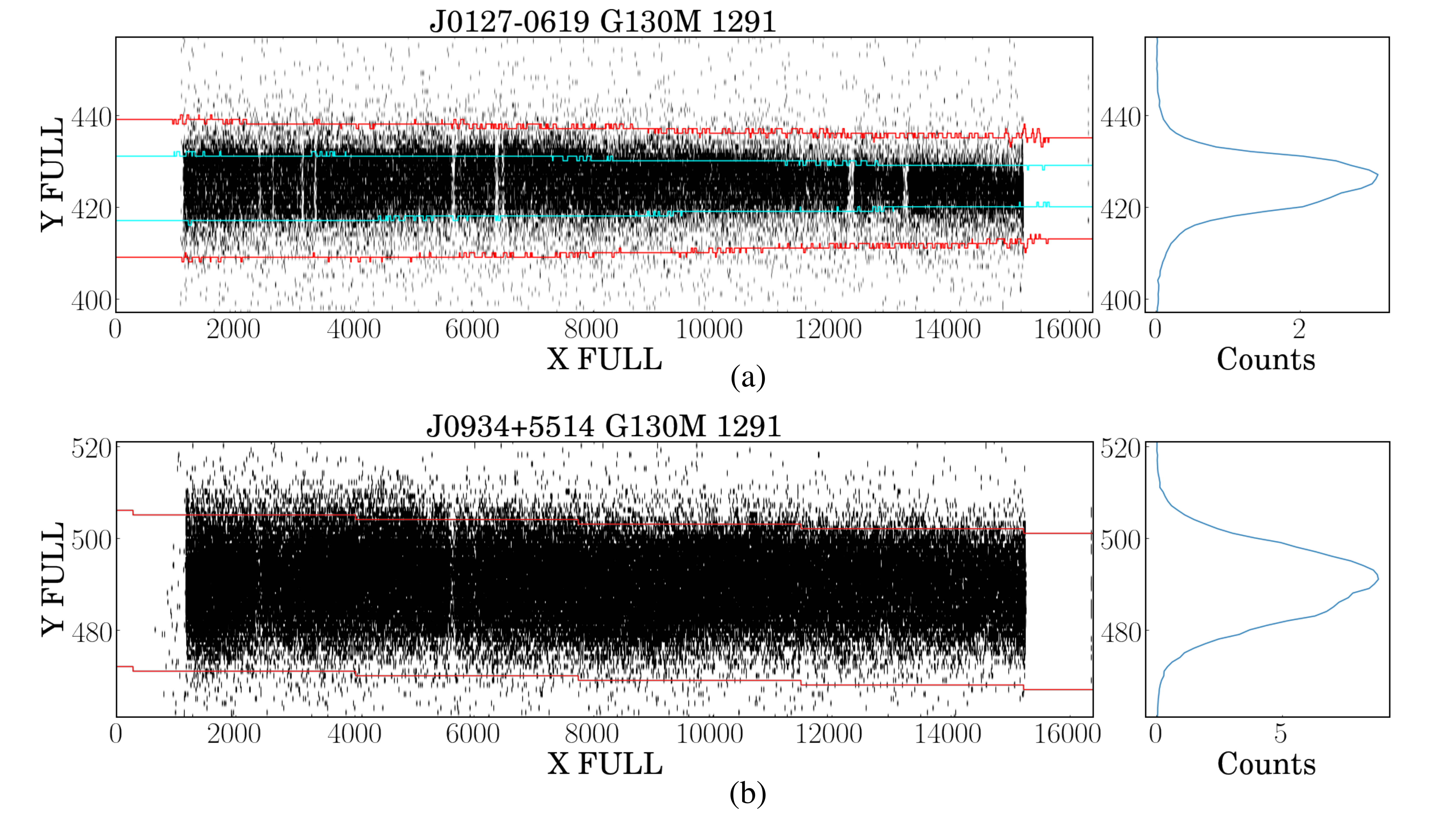}
\caption{Cross-dispersion light profiles for a compact (e.g., J0127-0619, top-panel) and an extended source (e.g., J0934+5514, bottom panel), as seen on Segment A of the COS FUV detector, here for G130M/1291 observational setup. The light profile collapsed along the x-direction is shown in the right panels. In each panel we show the extraction profiles used. In the top panel, which concerns a \texttt{TWOZONE} extraction, we show the inner and outer regions in red and blue, respectively (see Section~\ref{sec:fuv_ext} for details). On the bottom panel we show a \texttt{BOXCAR} extraction profile, where only an outer region is utilized. A \texttt{BOXCAR} extraction was the default setting for this specific observation because it was executed at Lifetime Position 1. However, due to the extensive width of the J0934+5514 light profile on the detector, a \texttt{BOXCAR} extraction may always be required to retrieve the full extent of the flux. }\label{fig:FUVprofiles}
\end{figure*}
Targets whose spatial or spectral profile had GFWHM $>$ 0\farcs6 (i.e., when the distribution of incoming light on the COS/FUV detector is significantly wider than that of a point source) were classified as `extended sources'. Targets whose spatial or spectral profile had GFWHM $<$ 0\farcs6 but showed more than one peak in their profile were classified as being MC sources. Targets whose spatial or spectral profile had GFWHM $<$ 0\farcs6, with single peaked profiles, are classified as the CLASSY `compact sources'. This classification scheme was further verified by inspecting the half-light radii (r$_{50}$) measured from the COS acquisition images (provided in Xu et al, 2022, in-prep). All galaxies that were classified as `compact' have r$_{50}<$0\farcs3, `extended' have r$_{50}>$0\farcs6 and `MC' sources typically have 0\farcs3$<$r$_{50}<$0\farcs6. However, it should be noted that these surface brightness profiles can be underestimated for extended sources in that the COS aperture (1) does not cover the entire galaxy and (2) is affected by flux vignetting beyond a radius of 0\farcs4. As such r$_{50}$ profiles measured from COS acquisitions images provide limited guidance for this purpose. In support of this, we additionally consulted the r$_{50}$ values presented in \citet{berg22}, which were measured from optical images. The optical r$_{50}$ values are indeed much larger than those measured from COS acquisition images because they encompass the entire galaxy, are affected by ground-based seeing, and trace the more extended optical emission rather than the highly-ionized UV emission which is typically highly concentrated around the central ionizing source. Thus being, without in-hand high spatial resolution, non-vignetted UV images of each source, we feel that the spatial and spectral light profiles from the COS acquisition images provide the best guidance on source classification currently available to our study.

 As such, we proceed with the classification of CLASSY galaxies into `compact', `extended', and `multi-component' based on the GFWHM cut described above. The target-acquisition images for all CLASSY targets are shown in Figures~\ref{fig:acq_PS}--~\ref{fig:acq_MC}, along with the spectral and spatial traces. The targets with compact light profiles are shown in Figure~\ref{fig:acq_PS}, those with extended single-component light profiles are in Figure~\ref{fig:acq_ext}, and those with MC light profiles in Figure~\ref{fig:acq_MC}. Within these figures, we additionally display the collapsed light profiles along the dispersion and cross-dispersion axes. However, it should be noted that due to the slitless design of COS, all
light within the COS aperture will contribute to the two-dimensional spectra, not just
the light indicated by the spatial trace. It is also important to note that all aberration and vingetting effects can be seen within the acquisition images simultaneously. For example, light can be seen beyond the 1\farcs25 radius aperture because light at the aperture plane has not been corrected for spherical aberration, which takes place at spectroscopic gratings. To this end, light beyond the 2\farcs5 diameter can reach the COS detector.

The majority of COS target acquisitions are taken with the MIRRORA configuration (this applies to 87 of our datasets). However, if a target is particularly bright, an attenuated light configuration is utilized, called MIRRORB, which serves to protect the detector. MIRRORB, was employed for ten galaxies in CLASSY (each one labelled accordingly in Figures~\ref{fig:acq_PS}--\ref{fig:acq_MC}, that are especially bright in the UV. Unfortunately, MIRRORB has the drawback of producing a double peaked image that impedes our interpretation of the spatial distribution of light. This can be seen most clearly in some of our compact targets acquired with MIRRORB (e.g., J0036-3333 and J0808+3948, Figure~\ref{fig:acq_PS}), where the fainter secondary image (marked with a gray box) can be seen in the direction of the spatial trace. For these specific cases extra care was taken when assessing the distribution of light and classifying the galaxy as extended, MC, or compact, and additional pre-existing optical imaging was inspected (when available).


\renewcommand{\thefigure}{5}
\begin{figure*} 
 \centering
\includegraphics[width=0.6\textwidth]{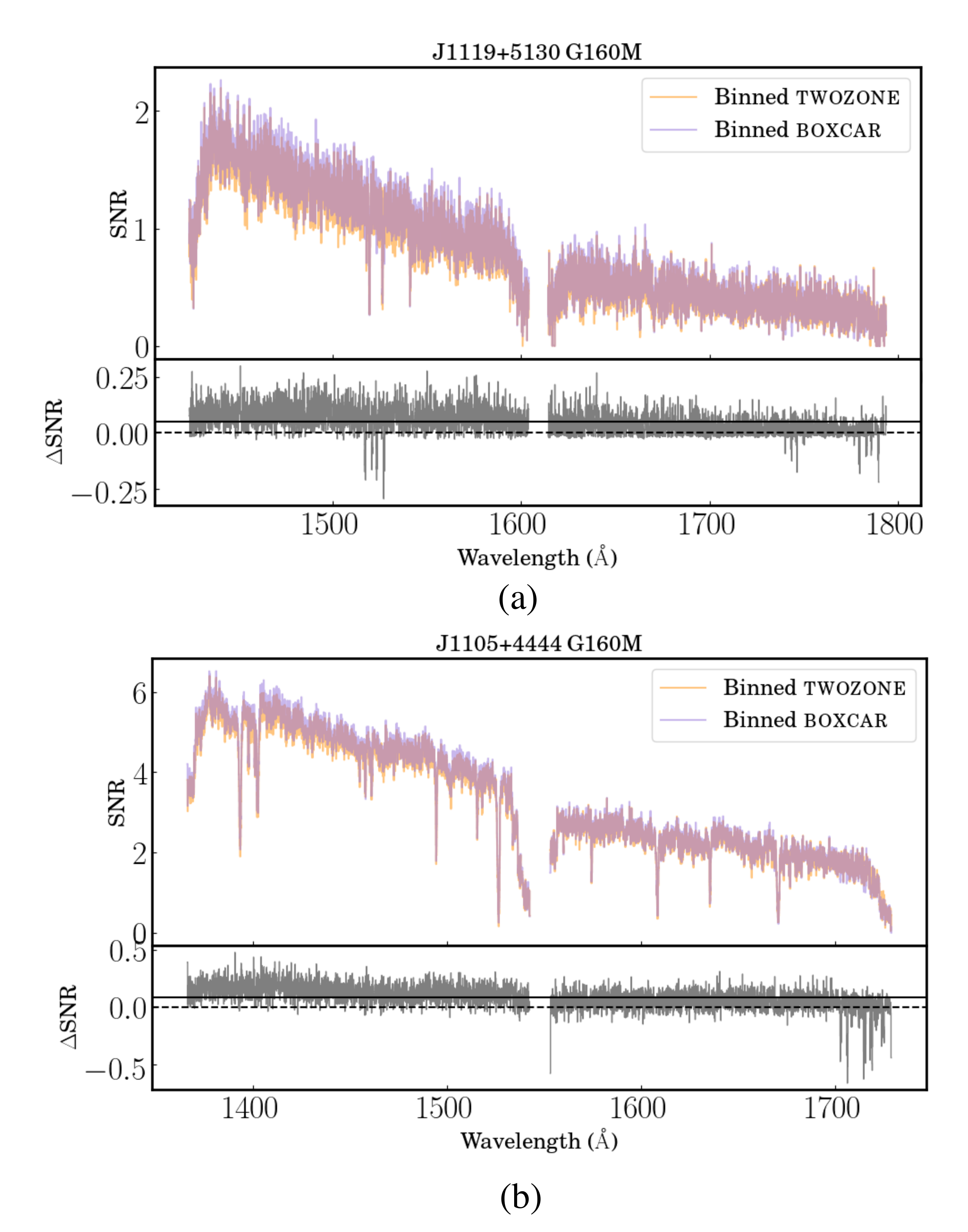} 

\caption{Signal-to-noise ratio (per 6~pixel resolution element) as a function of observed wavelength for the \texttt{TWOZONE} and \texttt{BOXCAR} extraction techniques (see text for details) for single visit G160M spectra of two example targets, J1119+5130 and J1105+4444.  Despite these sources being classed as `extended' in their target acquisition light profiles, only a very minor increase in S/N is seen from the \texttt{BOXCAR} extractions. This may be due to the fact that despite the extended profiles, the vast majority of light is still contained within the central 0\farcs8 diameter. These specific targets and gratings were chosen because they showed the largest change in S/N (shown in the bottom panel, $\Delta SNR=S/N_{BOXCAR}-S/N_{TWOZONE}$) of all the extended sources. Dashed lines display $\Delta$SNR=0, solid black lines display the mean $\Delta$SNR values of 0.05 and 0.08 for J1119+5130 and J1105+4444, respectively.}\label{fig:SNRplots}
\end{figure*}

\subsection{FUV Extraction}\label{sec:fuv_ext} 
As discussed above, the observations of extended sources by COS can require a customized extraction. To illustrate the difference between the cross-dispersion profile of a compact and an extended source, in Figure~\ref{fig:FUVprofiles} we show G130M/1291/FUVA profiles of J0127-0619 (GFWHM$\sim$0\farcs1) and J0934+5514 (GFWHM$\sim$1\farcs5). Due to the width of the cross-dispersion  profile of J0934+5514, compared to that of a compact (e.g., J0127-0619), an extended extraction box may be necessary to collect the full extent of its flux. As part of this customized extraction, as suggested by the COS DHB, the extended source CLASSY datasets were re-processed using a \texttt{BOXCAR} extraction technique. This involves a box of fixed height ($\sim27-45$ pixels depending on the \texttt{CENWAVE}), centered on the spectral profile defined within the extraction tab reference file (\texttt{XTRACTAB}), thus covering the entire cross-dispersion profile, including any low-level emission that may exist in the wings of the profile - simultaneously increasing the signal and potentially the noise. By comparison, the \texttt{TWOZONE} extraction method involves two extraction regions - an inner zone corresponding to the core of the profile, and an outer zone that covers the entire region used for the spectral extraction (i.e., outer+inner zone). The boundaries for these zones refer to the fraction of enclosed energy within the cross-dispersion profile of a point source (80\%\ and 99\%\ for the inner and outer zones, respectively) and are a function of wavelength. The separation of the profile into an outer and inner zone enables the COS pipeline to only reject wavelength bins if bad pixels occur within the \textit{core} region, rather than by those that lie in the outer wings of a profile. By comparison, in a \texttt{BOXCAR} extraction, \textit{any} bad pixels that lie within the box will result in rejected wavelength bins, which can lead to a loss of S/N or gaps in the data if the \texttt{BOXCAR} extraction region extends into regions of gain sag on the detector (where the local sensitivity of the detector is decreased).

To establish whether the \texttt{BOXCAR} extractions were an improvement over the default \texttt{TWOZONE} extractions adopted for spectra taken at LPs 3+4, we compared the signal-to-noise ratios (S/N) of the binned spectra for each FUV grating and \texttt{CENWAVE}. Only very minor increases ($\lesssim$~5\%) were found in the \texttt{BOXCAR} extractions, which we show examples of in Figure~\ref{fig:SNRplots} for single visit spectra (i.e., before coaddition occurs). In some cases, no improvement in S/N was found when employing the \texttt{BOXCAR} technique, which may be due to only low levels of light falling beyond the central 0\farcs4 radius beyond which the flux is vignetted (discussed in detail in Section~\ref{sec:vignet}), or due to the naturally large width of the light profile in the cross-dispersion direction for the specific \texttt{CENWAVE} (e.g., G130M/1222) which is fully encompassed by both the \texttt{TWOZONE} and \texttt{BOXCAR} extractions. On the other hand, the effects of an increased number of bad pixels lying within the \texttt{BOXCAR} extraction region compared to the \texttt{TWOZONE} were noticeable, which can be seen as negative spikes of $\Delta$S/N in the lower panels of Figure~\ref{fig:SNRplots}. 

Checks were also performed on the equivalent widths (EW) of the emission lines found within this regime, to assess whether the narrow \texttt{TWOZONE} extraction profiles were excluding any extended nebular emission that would have been extracted via a \texttt{BOXCAR} extraction, and thus consequently underestimating the EW measurements of those emission lines. Measurements were made on the \sfoiii~$\lambda\lambda$1660,66 emission line doublet with both extraction methodologies applied to the G160M datasets. In all cases, the EW(\sfoiii) remained unchanged within the uncertainties of the measurement. 
While this test could have also been performed with \heii~$\lambda$1640 and \Lya\, we note that EW measurement uncertainties on these lines will be impacted by the spectral continuum fit and subtraction to the extracted spectra, thus preventing a true comparison of the two extraction techniques. 

\renewcommand{\thefigure}{6}
\begin{figure*}
\centering
\includegraphics[width=1.0\textwidth]{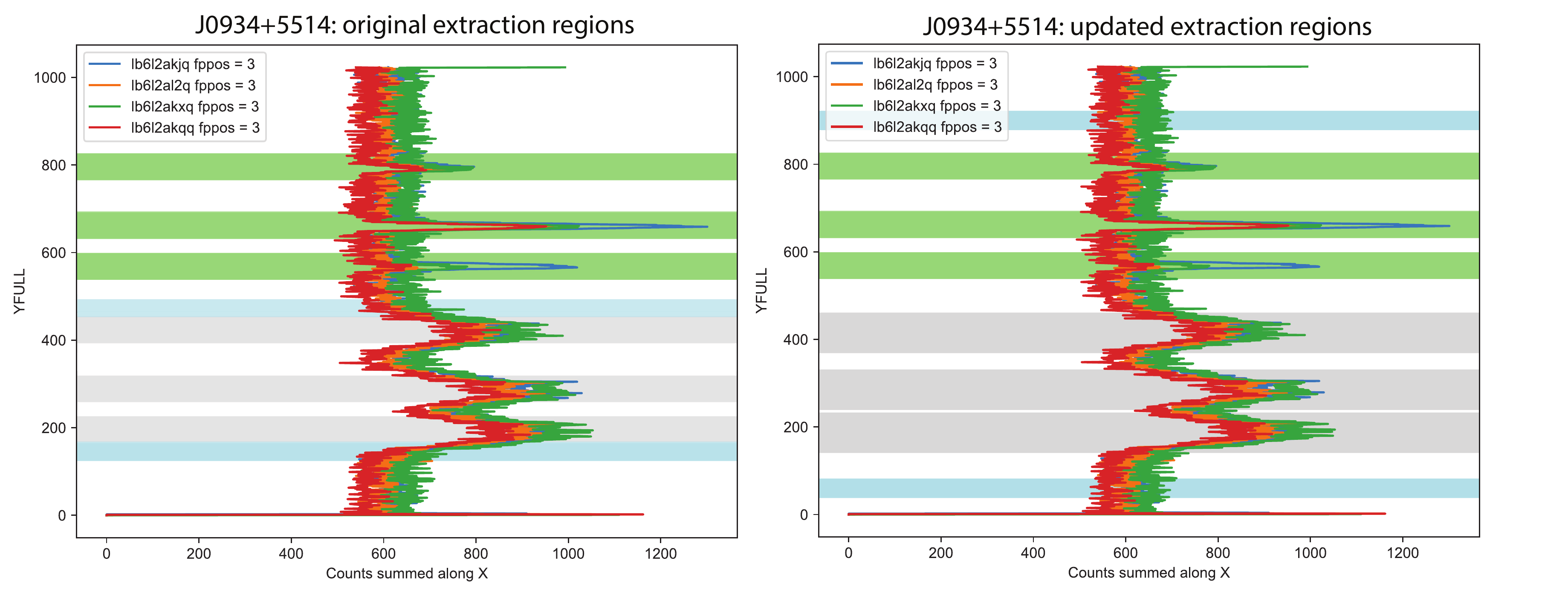}
\caption{Collapsed counts from the flat-fielded science images along the dispersion direction on the NUV detector for J0934+5514 (IZw18), in YFULL coordinates (Y pixels corrected for thermal and geometric distortion, walk, and offset in the cross-dispersion direction, based on the wavecal spectrum). Overlaid on each plot are the extraction boxes for the wavelength calibration spectrum (green), the background (blue) and the science spectrum (grey). In the left-hand panel we show the standard extraction boxes, as defined in the {\sc xtractab} reference file. In the right-hand panel we show the updated extraction boxes, where the background extraction boxes have been moved away from the source spectrum, and the science spectrum extraction regions have been widened to account for the extended light profile in the cross-dispersion direction. Credit: HST/COS Helpdesk Team }
\label{fig:NUV_ext}
\end{figure*}
Similarly, a too-narrow extraction aperture may miss the full extent of the continuum flux in extended sources, consequently leading to flux calibration issues. Indeed, several extended source datasets required minor scaling factors to align the continuum flux between gratings, which we describe fully in Section~\ref{sec:fluxcalib}. Here we note that while a \texttt{BOXCAR} extraction may have helped minimize minor flux calibration offsets between gratings, since the vast majority of FUV science utilizes continuum-normalized spectra, we felt that contiguous and optimized S/N data was of more benefit to the user. As such, we decided to remain with the \texttt{TWOZONE} extraction for all the FUV extended source datasets for which it is available (i.e., those that were taken at at Lifetime Positions 3 and 4). 

From our findings, we recommend that COS users targeting extended sources (GFWHM $>$ 0\farcs6)  
can remain with the default \texttt{TWOZONE} extraction method without impacting the S/N of their extracted spectra. However, this can have consequences on the flux calibration of the spectra, which we discuss further in Section~\ref{sec:fluxcalib}.

\subsection{NUV Extraction}\label{sec:nuv_ext}
For COS NUV data (i.e., data taken with the G185M and G225M gratings), CalCOS employs a \texttt{BOXCAR} extraction by default (57 pixels in height). However, again due to the extended nature of our sources, in several cases the spectral profiles were not sufficiently covered by the size of the extraction boxes. As such, it was necessary to re-process our NUV datasets with ad-hoc extraction profiles, centered on the source. To do this, we inspected the collapsed light profile on each of the NUV detector stripes (A, B, C) and adjusted the default height of the extraction boxes in the \texttt{XTRACTAB} reference file. Extraction boxes 95 pixels in height were chosen such that they sufficiently covered the width of the profile, without extending into the light profile of the adjacent stripe. We show an example plot used for this assessment in Figure~\ref{fig:NUV_ext}, where the collapsed counts in the dispersion direction (i.e., X pixels) are shown as a function of Y pixels on the NUV detector. We also denote the extraction regions for stripes A, B, and C (grey boxes, where A is the lowest on the detector and C the highest), the three respective wavelength calibration regions (green boxes) and the background extraction regions (blue boxes). Upon inspection of these plots, it was found that only targets classified as extended sources (Figure~\ref{fig:acq_ext}) required larger extraction boxes. Once the datasets had been re-reduced with updated extraction heights, the final summed flux were found to increase by factors of up to two times the original flux for the extended sources. As such, these datasets were now in better agreement with the FUV flux levels seen in the corresponding G130M and G160M datasets for the same targets (as shown in Fig.~\ref{fig:NUV_extract}). 

In order to ensure that the default extraction boxes were not excluding potentially extended nebular emission in our sources, we additionally measured the equivalent width of \sfciii~$\lambda\lambda$1907, 09 with the default and widened extraction boxes in \textit{all} of our sources. A widened extraction box was deemed suitable if the change in EW(\sfciii) was larger than the uncertainty on the measurements.  In addition to the extended sources identified previously identified, we additionally found one compact source (J1024+0524) whose EW(\sfciii) increased significantly with the widened extraction due to extended nebular emission. We discuss the scientific implications of this change in  EW(\sfciii) in Section~\ref{sec:em_fit}.

When extending the extraction boxes, care was also taken to avoid contamination from spectra extracted from a neighbouring stripe. The highest level of contamination would be between stripe A and B, which are closest in location on the detector. The level of contaminating flux was estimated to be $<1$\%\ for all cases, with the exception of J0934+5514 which reached a contamination level of $\lesssim3$\%. This is typically well within the level of uncertainties on the emission line fluxes measured within this wavelength regime and as such poses only a negligible effect on the scientific results.

\renewcommand{\thefigure}{7}
\begin{figure}
\centering
\includegraphics[scale=0.6]{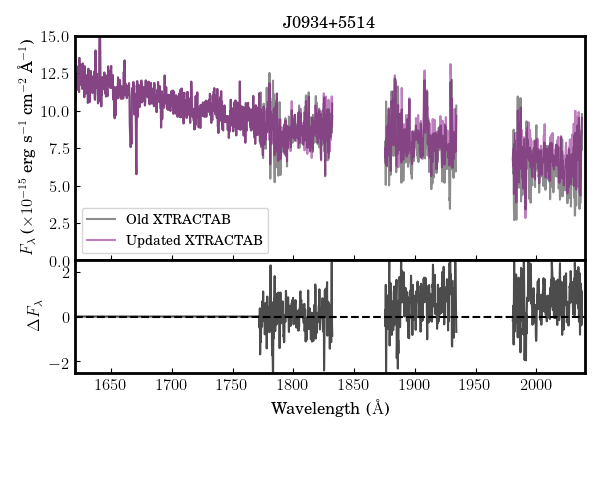}\vspace{-1cm}
\caption{Increasing the NUV extraction box height for extended sources (such as J0934+5514 shown here) enabled a better match in the flux calibration between the FUV and NUV datasets. Here we show the coadded data for J0934+5514 where the NUV data had been reduced with the default (grey spectrum) and extended (purple spectrum) extraction boxes. The difference in flux ($\Delta F_{\lambda}=F_{updated}-F_{old}$) is shown in the bottom panel, with a dashed line displaying $\Delta F_{\lambda}$=0.}
\label{fig:NUV_extract}
\end{figure} 
It should also be noted that it was necessary to re-locate the background extraction boxes for all of our NUV datasets. During our re-extraction process for the NUV datasets, the default back-to-back positioning of background extraction regions and the source extraction regions (as seen in Figure~\ref{fig:NUV_ext}a) were found to be problematic for the extended CLASSY sources in that the resultant background levels were too high due to the contribution of counts from the extended source profiles. To solve this issue, the COS team assisted the CLASSY collaboration in re-locating background extraction regions at the top and bottom of the detector, shown in Figure~\ref{fig:NUV_ext}b, which were sufficiently distanced from the source (a distance of $\sim$+400 pixels and $\sim -50$ away from the source for the upper and lower background extractions regions, respectively). While the original locations provided sufficient background subtraction for compact light profiles, this is not the case for extended sources and the COS Team hope to release a new NUV \texttt{XTRACTAB} in the near future with updated background regions for use with both point source and extended objects. In addition to this, the vignetting issues of NUV data, which affects the first $\sim$150 pixels of the NUV detector \citep[as detailed in the COS IHB][]{COSIHB}, were automatically removed during the CalCOS reduction process via data-quality masks located in the bad-pixel table (\texttt{BPIXTAB}).

From our findings we recommend that users of NUV observations of extended source profiles examine their NUV flat-fielded science images (\textsc{flt} images) and assess the width of the NUV extraction boxes with respect to the collapsed light profile, and adjust it to incorporate the entire cross-dispersion profile. Moreover, to ensure the most accurate NUV background subtraction is applied, we advise users to use updated background extraction regions, as detailed above. We discuss the effects of our optimized NUV extractions with respect to scientific application of the data in Section~\ref{sec:em_fit}.

\begin{table*}
\caption{Lines used for wavelength calibration.}
\begin{center}
\begin{tabular}{c|c|c|c|c}
\hline \hline
Line ID & Wavelength (\AA) & Type & Grating Coverage & $z^{\dagger}$ \\
\hline
{\heii}   & 1640.42   & Nebular       & G160M         & $<0.09$               \\
{\sfoiii} & 1660.81, 
            1666.15   & Nebular       & G160M         & $<0.08$               \\
{\sfciii} & 1906.68, 
            1908.73   & Nebular       & G185M         & $<0.12$               \vspace{2ex} \\
{\Ni}     & 1193.30   & ISM           & G130M, G160M  & $<0.23, 0.12-0.46$    \\
{\sIiii}  & 1206.50   & ISM           & G130M, G160M  & $<0.20$, $0.12-0.48$  \\
{\sIii}   & 1260.42   & ISM           & G130M, G160M  & $<0.15$, $0.06-0.42$  \\
{\oi}$^*$ & 1302.17   & ISM           & G130M, G160M  & $<0.11$, $0.03-0.37$  \\
{\cii}$^*$& 1334.53   & ISM           & G130M, G160M  & $<0.09$, $0.01-0.34$  \\
{\alii}   & 1670.79   & ISM           & G160M         & $<0.07$               \vspace{2ex} \\
{\ciii}   & 1175.53   & Photospheric  & G130M, G160M  & $<0.23$, $0.14-0.52$  \\
{\ciii}   & 1247.38   & Photospheric  & G130M, G160M  & $<0.16$, $0.08-0.44$  \\
{\sIiii}  & 1294.54   & Photospheric  & G130M, G160M  & $<0.12$, $0.04-0.38$  \\
{\oiv}    & 1341.64   & Photospheric  & G130M, G160M  & $<0.08$, $0.03-0.33$  \\
{\sv}     & 1501.76   & Photospheric  & G160M         & $<0.19$

\end{tabular}
\tablecomments{
Lines used for wavelength calibration check between gratings. 
Column 1 lists the spectroscopic ID of the line, 
Column 2 provides the vacuum line wavelength, and
Column 3 gives the primary type of the line.
Columns 4 and 5 list the gratings that these lines reside in 
and the corresponding redshift range for valid grating coverage.\\
$^*$ Isolated but typically saturated. \\
$^{\dagger}$ Calculated using the full wavelength range covered by all 
\texttt{CENWAVE}s within that grating.
}
\label{tab:lines}
\end{center}
\end{table*}

\section{Coaddition of Spectra}\label{sec:coadd}
The final calibrated product of spectroscopic reduction pipelines is traditionally a single spectrum. In most cases, this product combines the individual exposures within a set of observations, all of which have similar observational setup and conditions - i.e., the same filter, wavelength setting, sky-exposures etc. While coadding spectra of similar observational setups but different conditions can be relatively straightforward, the coaddition of spectra taken with different gratings, filters and, for the COS instrument, lifetime positions\footnote{The COS FUV detector is susceptible to gain sag, a reduction in the ability of the detector to convert incoming photons into electrons. A strategy used to mitigate the effects of this on COS data is to occasionally change the location along the cross-dispersion direction where spectra are recorded on the detector, which is defined as the lifetime position (LP).} on the FUV detector, can be far more problematic due to differences in resolution, wavelength solution and flux calibration accuracy. In this way, the HST/COS reduction pipeline, CalCOS, is no different in that it only produces a final \texttt{x1dsum} spectrum for each grating and \texttt{CENWAVE}, combined across the separate FP-POS positions\footnote{There are four FP-POS positions designed to provide an automatated spectral dithering technique which removes fixed pattern noise in COS observations, as detailed in the COS IHB, \citet{COSIHB}} for each observational visit. 

Several extensions exist within each \texttt{x1dsum} spectrum, including flux, wavelength, error, upper- and lower-bound error estimates, and information on counts, background variance, and data-quality flags. It should be noted that CalCOS delivers spectra with approximate 1-$\sigma$ statistical errors calculated using numerical approximation of the \citet{Gehrels:1986} analytic functions, where it combines the counts in the re-sampled, linearized pixels and take the error of that re-sampled sum. As shown in \citep{JohnsonISR:2021}, errors can become asymmetric in the low count regime of $<$20-30 counts/pixel and assuming symmetry in this regime can lead to severely over-estimated lower confidence limits.  While this count threshold is exceeded in the majority (65\%) of CLASSY datasets, there are instances of G130M and G160M spectra where the count level proceeds below 20 net counts per native COS pixel ($\sim$84\%\ and 44\%\ of G160M FUVA and FUVB raw datasets, respectively, and $\sim$25\% of both FUVA and FUVB G130M datasets). Interestingly, despite their somewhat lower S/N, only $\sim$10\%\ of the G185M/G225M datasets lie in this regime. Throughout the CLASSY coaddition process we extract and propagate the ERROR extension of the \texttt{x1dsum} data files (which pertains to the Gehrels upper confidence limit), and apply propagation techniques that assume a symmetric error distributions throughout. Within the N<20 regime, the statistical error is expected to be accurate to within 5-6\% \citep[][]{Gehrels:1986}, which we deem sufficiently accurate for the majority of the scientific application of the CLASSY data. The decision to assume symmetric errors for all datasets (which is appropriate for 65\%\ of all CLASSY raw datasets) is based on the need for a consistent treatment of errors throughout all datasets and the crucial requirement of propagating symmetric errors during re-sampling. Deriving a technique that correctly propagates asymmetric errors is beyond the scope of this study, and currently beyond the capabilities of the CalCOS pipeline due to its mathematical complexity. While we could adopt a method that approximates a symmetrical distribution from the separate upper and lower bound errors reported by CalCOS, the errors would themselves remain approximate in nature. As such, users whose scientific application of our data may be affected by our error propagation in these low count regimes are advised to revisit the original datasets and perform customized error propagation on those specific datasets. To aid users on this front, we provide a data flag in the final co-added spectrum (flag = 2) for wavelength regions where the original x1dsum datasets were below 20 counts per native pixel, where the statistical error is expected to be over-estimated and correct to within 5-6\%. Finally, it should also be noted that, along with the COS DHB \citep[][]{COSDHB}, we do not recommend using the raw, non-binned spectra in general. As such, the summed counts are higher per resolution element in the recommended binned spectra such that the error arrays are appropriate there.

In order to create the final CLASSY spectra spanning the G130M-G160M-G185M/G225M 
grating for each galaxy, a careful sequence of spectral coadditions was used.
Each coaddition (1) performed a common resampling of the wavelengths to the highest dispersion of a given combination of spectra and (2) used a combined normalized data quality weight (to filter out or de-weight photons correlated with anomalies) and exposure-time-weighted calibration curve (
see \S~\ref{sec:3.4}).
A detailed discussion of the main steps in the coaddition process is presented in 
\citetalias{berg22}, however, we provide a brief overview below.
Note that a main goal of the CLASSY templates is to provide useful spectra to the 
community for a wide variety of science objectives that have different data requirements.
Therefore, the steps of CLASSY coaddition process were ordered to allow for 
templates with different spectral resolutions to be produced. 
\begin{enumerate}
    \item {\it Joining grating segments and stripes:} 
    The first step of the coaddition process unifies the spectral segments/stripes 
    for each grating dataset while accounting for the wavelength-dependent dispersion.
    To do so, the \texttt{python} \texttt{pysynphot} package \citep{lim13} was used to interpolate the flux and error arrays of each segment onto a new, uniform wavelength grid defined by the largest step in the original wavelength array. Error arrays were resampled using the standard error propagation techniques (i.e., in quadrature).

    \item {\it Coadding single grating datasets:} 
    The spectral resolutions of the CLASSY COS gratings decrease with wavelength
    such that they are ordered from highest to lowest resolution as
    G130M, G160M, G185M, G225M, and G140L. 
    Therefore, to preserve the highest spectral resolution for a given template,
    multiple datasets of a given grating were coadded.

    \item {\it Coadding multiple grating datasets:} 
    Multiple templates with decreasing spectral resolution were created
    for each galaxy by progressively including lower resolution gratings
    in the coadd process.
    The resulting templates are 
    (1) the Very High Resolution (VHR) coadds: consisting of only G130M spectra, 
    (2) the High Resolution (HR) coadds: consisting of the CLASSY medium-resolution 
    FUV gratings, or G130M$+$G160M spectra, 
    (3) the Moderate Resolution (MR) coadds: consisting of the CLASSY FUV$+$NUV medium 
    resolution gratings, or G130M$+$G160M$+$G185M$+$G225M spectra, and
    (4) the Low Resolution (LR) coadds: consisting of the CLASSY medium and low 
    resolution gratings. Possible grating combinations include G130M$+$G160M$+$G140L 
    and G130M$+$G160M$+$G185M$+$G225M$+$G140L.

    \item {\it Binning the spectra:} 
    The COS medium-resolution gratings have a resolving power of $R\sim15,000$ 
    for a perfect point source, which corresponds to six FUV detector pixels 
    in the dispersion direction. 
    All coadds were binned by a factor of six to reflect this
    using the \texttt{SpectRes} \citep{carnall17} \texttt{python} function, 
    which efficiently resamples spectra and their associated uncertainties onto an arbitrary wavelength grid while preserving flux. This function employs Gaussian error propagation, which can be inappropriate in the very low S/N regime where errors become asymmetrical (i.e., <20 counts per native COS pixel). As discussed above, this regime occurs in most of our raw (pre-coaddition) G160M datasets and is flagged accordingly in our ERROR mask array.
    The final high-quality CLASSY coadded spectra are shown in the Appendix of \citetalias{berg22}. 

\end{enumerate}
The final coadded CLASSY spectral templates are HLSP multi-extension fits files
that include different extensions for observed-frame and rest-frame wavelength spectra, 
observed and Galactic reddening-corrected spectra, original and binned data. 
We direct the reader to \citetalias{berg22} for a full description of each of these extensions.  

The final CLASSY coaddition process outlined above was designed based on 
a series of intermediate analyses to test the impact of different methods
on the reduced spectra.
Below we describe the methodologies employed to test or account for the 
wavelength calibration (\S~\ref{sec:wavelength}),
differing spectral dispersions (\S~\ref{sec:3.2}), 
flux calibration (\S~\ref{sec:fluxcalib}),  
coaddition weighting (\S~\ref{sec:3.4}), and 
vignetting and optical aberrations (\S~\ref{sec:vignet}).

\subsection{Wavelength Calibration} \label{sec:wavelength}
While the wavelength solution of the medium-resolution grating COS spectra is expected to be within $\sim$15~\kms, and monitored regularly throughout the calendar year, differences \textit{between} the gratings are not monitored and could potentially diminish the wavelength accuracy of the final coadded data. This would be evident, for example, in velocity differences between the NUV and FUV nebular emission lines, or incorrect outflow velocities from FUV absorption lines calculated relative to NUV nebular emission lines. In order to check the wavelength solution alignment between gratings, we inspected a suite of lines within each spectrum by overlaying them in velocity space before performing any coaddition. Since ISM absorption lines are typically found to be blueshifted with respect to the nebular gas \citep[e.g., ][]{shapley03, steidel10, heckman02}, it is imperative that velocity comparisons are only made between the same \textit{type} of feature (i.e., nebular to nebular). The lines used for the grating-to-grating wavelength calibration check are listed in Table~\ref{tab:lines}.

Relatively-strong, non-resonant, nebular emission lines are in many of the CLASSY datasets. These lines enabled robust comparisons between the G160M and G185M/G225M datasets.
For 12 galaxies, no suitable emission lines were present for a comparison between the 
G160M and G185M/G225M gratings and, therefore, the wavelength calibration of the 
G185M/G225M gratings remains unchecked.
However, the only features of importance in the G185M/G225M gratings are nebular emission lines,
and so the lack of lines/wavelength calibration has no impact on the scientific analysis of these objects.
For comparison between the G130M and G160M datasets,
no pairs of strong, non-resonant emission lines are present.
Instead, isolated, unsaturated ISM absorption lines of similar ionization states were utilized, 
as listed in Table~\ref{tab:lines}. 
A comparison between the photospheric line profiles was also made between the G130M and G160M gratings
when possible.

Upon inspection, differences of $<15$~\kms, were found between the gratings (see e.g., Figure~\ref{fig:line_profiles} b--d). Since this is within the wavelength accuracy expected from each grating, it was not necessary to adjust the wavelength scale of the data before proceeding with the coaddition procedure. However, for one particular extended target, J1016+3753, an offset of $\sim0.5$~\AA\ ($\sim$107~\kms) was observed in the \sIiv\ line profiles between the G130M and G160M data (Figure~\ref{fig:line_profiles}a). In this case, the two exposures were taken with very different position angles ($\Delta$PA$\sim100^{\circ}$) and as such, this wavelength offset represents a real offset in the velocity of the gas observed at the different PAs and is not due to wavelength calibration issues. Since these are physical differences, we do not adjust the coadded data for wavelengths offsets such as this. As can be seen in the final coadded spectrum of J1016+3753, despite the velocity offset the combined features in this overlap regime are not significantly broadened due to the weighting effects of our coaddition process. However, velocity offsets may be seen when measuring lines of the same origin (i.e., ISM, nebular, photospheric) in the final CLASSY coadded data of J1016+3753 if they lie within different gratings.

As a general recommendation for COS data, the wavelength calibration offsets between gratings 
should be inspected using the methodology described above (if possible with respect to line detection). 
However, wavelength calibration offsets are likely negligible if the targets are compact sources 
or are extended targets where the individual datasets have been taken with the same position 
angle and pointing. 
\renewcommand{\thefigure}{8}
\begin{figure*} 
 \centering
\includegraphics[width=\textwidth]{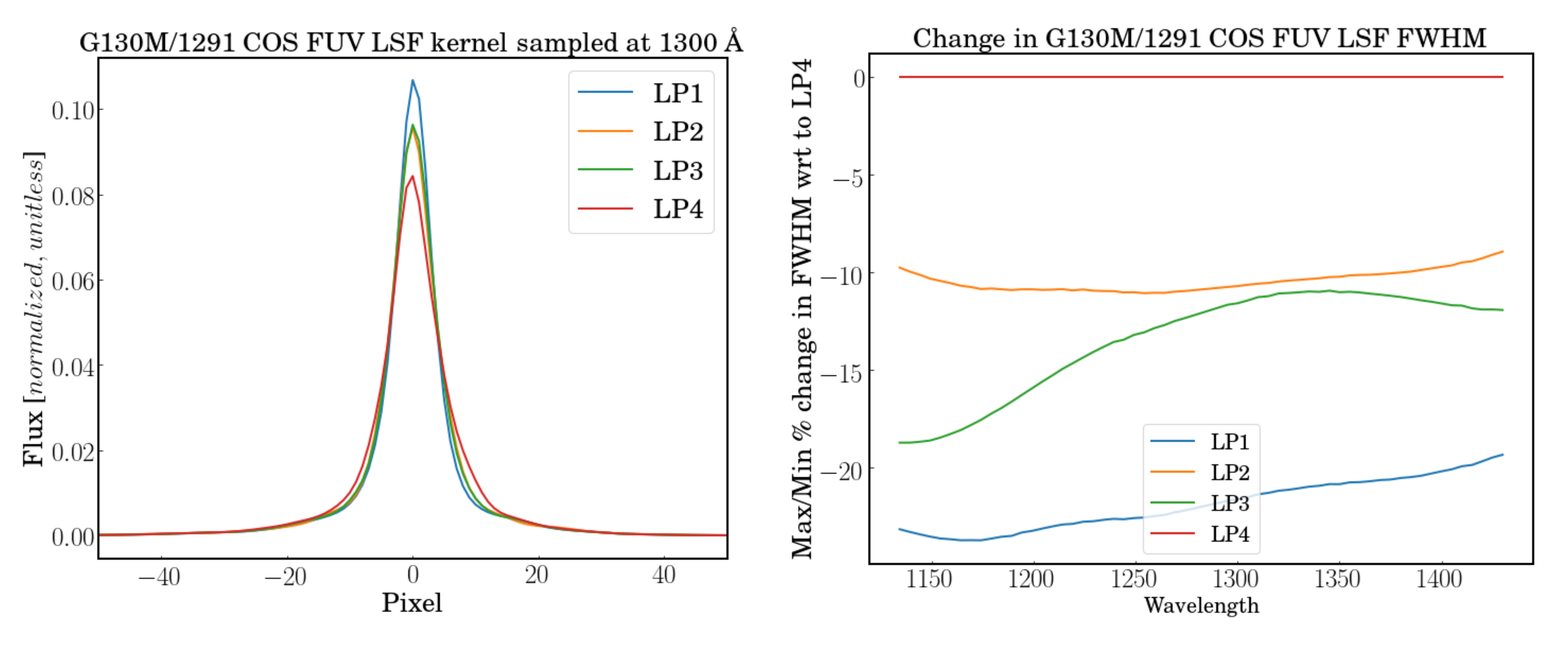} 
\caption{The COS line spread functions change with position on the COS FUV detector, which need to be taken into account when combining data that have been taken at different lifetime positions (LPs). As an example, we show the modeled COS LSFs for the G130M/1291 \texttt{CENWAVE} at 1300~\AA\ at each of the four LPs (\textit{left-hand panel}) and the  percentage decrease in the LSF FWHM relative to LP4 (i.e., corresponding to an increase in resolution relative to LP4) as a function of wavelength (\textit{right-hand panel}). }\label{fig:LSFs}
\end{figure*}

\begin{table*}
\caption{Properties of the four final CLASSY coadded spectra provided within the separate extensions of each CLASSY spectral data product: very high resolution (VHR), high resolution (HR), medium resolution (MR) and low resolution (LR).}
\begin{center}
\begin{tabular}{c|c|c|c|c}
\hline \hline
CLASSY coadd ID & Gratings Combined & Dispersion  & Dispersion  & Observed $\lambda$ coverage  \\
 & & (m\AA/pixel) & (\AA/resel$^*$) & (\AA) \\
\hline
VHR & G130M & 9.97 & 0.060  & $\sim$1200--1430 \\
HR & G130M+G160M & 12.23 & 0.073 & $\sim$1200--1750 \\
MR & G130M+G160M+G185M & 33.0 & 0.200 & $\sim$1200--2000 \\
LR & G130M+G160M+G140L+G185M/G225M & 80.3 & 0.482 &$\sim$1200--2200 \\ 
\hline
\end{tabular}
\end{center}
$^*1$ resel $=6$ pixels, as defined by the COS Data Handbook for the FUV detector \citep{COSDHB}.
\label{tab:spec}
\end{table*}
\subsection{Accounting for Resolution}\label{sec:3.2}
Accounting for the change in spectral resolution between individual gratings, 
and datasets, is a particularly difficult task, especially since resolution 
can depend on several different factors. 
First, the resolution is dependent on the extent of the UV light profile and, 
consequently, the position angle of the observation if the object is not azimuthally symmetric. 
The position angles of all observations included within CLASSY can be found in \citetalias{berg22}, while the position angles of all G130M observations are shown in Figures~\ref{fig:acq_PS}--\ref{fig:acq_MC}).  Secondly, resolution changes as both a function of wavelength across the detector and as a function of lifetime position (LP) on the COS FUV detector.  Within a single G130M \texttt{CENWAVE}, resolution can change by $\Delta R\sim5,000$ as a function of wavelength, and at a fixed wavelength can change by a further $\Delta R\sim5,000$ between different positions on the detector (i.e., at different LPs) - resulting in an overall (point source) resolution range of $R\sim11,000-20,000$ for G130M/1291 alone within the CLASSY datasets due to the use of both archival and new data taken at different LPs. G160M is somewhat similar to G130M, with a (point source) resolution range of $R\sim13,000-24,000$ across all \texttt{CENWAVE}s. The resolution of NUV datasets are somewhat different, $R\sim16,000-20,000$ for G185M and $R\sim20,000-24,000$ for G225M. CLASSY also utilizes the COS low-resolution gratings in some cases, G140L, which has a resolution range of $R\sim1,500-4,000$. However, readers are reminded that the spectral resolution of CLASSY spectra for targets that have been classified as having extended and/or MC light profiles will be significantly reduced compared to targets with compact or point-source light profiles.

Conserving the resolution to account for optical effects, such as changing line spread functions (LSFs), within coadded data would require a convolution kernel to match the wavelength-dependent resolution of one observation to that of another. In the left-hand panel of Figure~\ref{fig:LSFs} we plot the model LSF for G130M/1291 at 1300~\AA\ at the four different LPs on the COS detector, for a point source centered in the aperture\footnote{Available at \url{https://www.stsci.edu/hst/instrumentation/cos/performance/spectral-resolution}}, which are generated from two-dimensional, theoretical point spread functions (discussed in Section~\ref{sec:vignet}). The percentage change in LSF, relative to LP4 (where $\sim$60\% of the CLASSY FUV data is obtained) as a function of wavelength is plotted in the right-hand panel of Figure~\ref{fig:LSFs}. The resolution of LP4 spectra decreases by up to 25\%\ compared to LP1 spectra. Consisting of a combination of both new and archival data, CLASSY has several targets where data are coadded across multiple LPs. 

However, given the moderate S/N of the CLASSY datasets, and due to the extended nature of many of our sources (whose broad light profiles would have a dominant effect on the LSF resolution), changes in LSF resolution between LP were not accounted for when coadding datasets that were taken across separate LPs at the same grating and \texttt{CENWAVE}. We explore the scientific effects of using a non-optimized LSF on absorption line fitting in  Section~\ref{sec:abs_fit}. It should be noted, however, that this may not be the case for high S/N observations of point-source targets and in such cases, the change in resolution should be taken into account before the coaddition of data across different LPs.

\renewcommand{\thefigure}{9}
\begin{figure*}
\centering
\includegraphics[width = 6.5in]{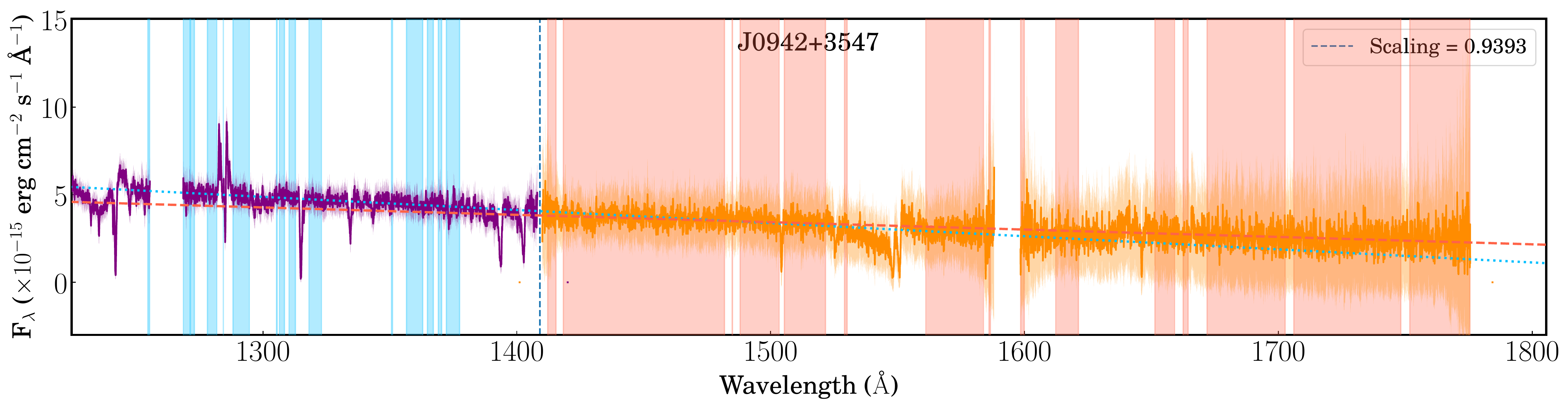}
\includegraphics[width = 6.5in]{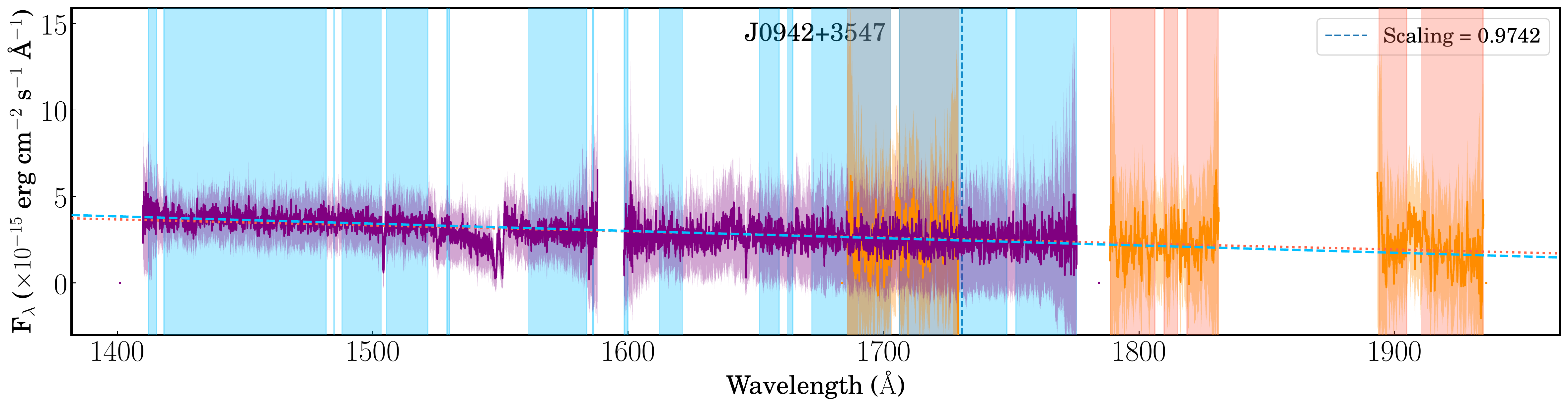}
\caption{
A demonstration of the flux calibration {\it between} the gratings of the CLASSY spectra
of J0942+3547.
These flux calibrations were performed to all of the datasets via small scaling-factors
to ensure highly accurate flux calibrations between gratings. 
{\it Upper panel}: Featureless continuum regions in the G130M/G160M spectrum (blue/orange shading) 
were fit with a 1D spline (blue-dotted/red-dashed line) and used to scale the G130M spectrum to the G160M spectrum at their mid-way intercept point (vertical blue dashed line).
{\it Lower panel}: Similarly, the featureless continuum regions in the G160M/G185M spectrum 
(blue/orange shading) were fit with a 1D spline (blue-dotted/red-dashed line) and used to 
scale the G185M spectrum to the G160M spectrum at their mid-way intercept point (vertical blue dashed line).
The final coadded dataset and scaling factor (used to multiply the G130M and G185M continuum flux) is shown in Figure~\ref{fig:fuv_scaling}. In each panel we show the flux$\pm1\sigma$ error spectrum, binned by 6 pixels. Scaling figures for all galaxies within the CLASSY sample are provided in Figure Set~\ref{fig:AppA}.}
\label{fig:scaling}
\end{figure*}

The spectral dispersion of the gratings differ considerably, with averages of 9.97 (0.060), 12.23 (0.073), 33.0 (0.200), and 80.3 (0.482) m\AA/pixel (\AA/resel) for G130M, G160M, G185M, and G140L, respectively, where we define one `resel' (or resolution element) as being six pixels\footnote{This is the fiducial FUV resolution element as defined by the COS Data Handbook. The fiducial NUV resolution element is instead three pixels, however we choose to adopt a 6-pixel resel here for our NUV datasets in order to provide a uniformly binned data product to the users. While the actual size of a resolution element changes as a function of wavelength and grating, we adopt the `resel' nomenclature to mean the fiducial case, in concordance with the COS DHB.}. To take this into account, at each stage of the coadding process, we resample the spectra to the largest average dispersion available within that combination of gratings. This results in four different types of coadded data (VHR, HR, MR, and LR), whose properties we show in Table~\ref{tab:spec}. Each are provided to the community within separate extensions of the final coadded CLASSY spectra, in addition to the original, unbinned data. This combination of data products enables users of CLASSY data to access both full-wavelength spectra of each object, albeit at lower resolution, in addition to the higher resolution spectra which may be more appropriate for their scientific goals.

\renewcommand{\thefigure}{10}
\begin{figure*}
\centering
\includegraphics[width=0.8\textwidth]{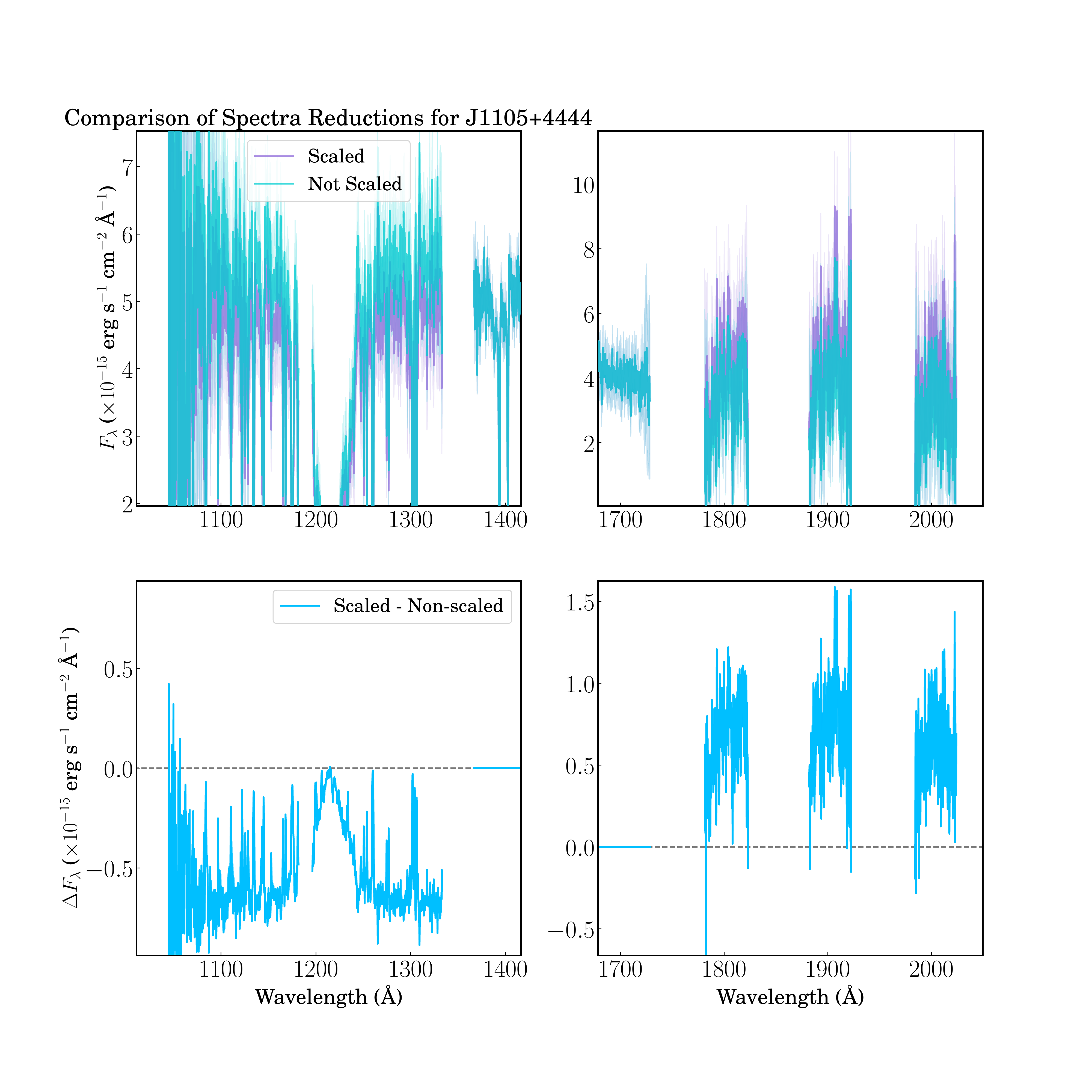} 
\caption{Differences in observational set-up between the datasets used for the CLASSY coadded data sometimes required small scale factors being applied to the data. Here we show three examples of coadded data before (blue) and after (purple) scaling (\textit{top two panels}) with the difference between the scaled versus non-scaled data (\textit{lower two panels}). The difference in flux offsets seen in J1105+4444 (a) and J1444+4237 (b) demonstrates that the observed offset depends on both $\Delta$PA and also the specific shape of the light distribution of each target. Despite J0942+3547 (c) being a compact source, an offset in flux of $-0.1$-- $-0.4\times10^{-15} erg s^{-1} cm^{-2}$\AA$^{-1}$ is seen, which is within the uncertainty of the continuum. The top panels show the flux$\pm1\sigma$ error spectrum, binned by 6 pixels. Here we show J1105+4444, an extended source, where the G130M and G185M data were offset from the G160M data by PA of $\sim$ 45$^{\circ}$. Scaling figures for all galaxies within the CLASSY sample are provided in Figure Set~\ref{fig:AppA}.}
\label{fig:fuv_scaling}
\end{figure*}
\renewcommand{\thefigure}{10}
\begin{figure*}
\centering
\includegraphics[width=0.8\textwidth]{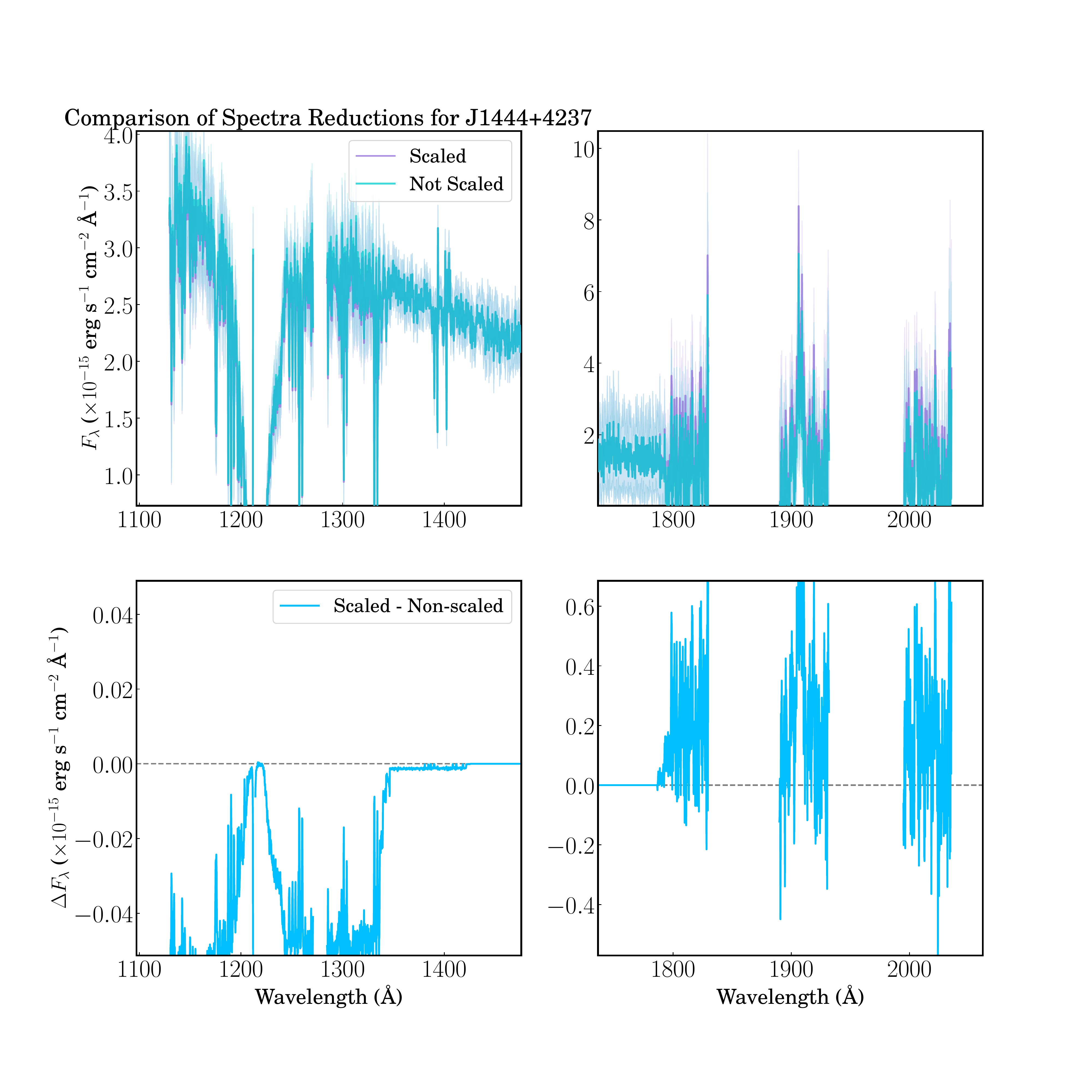} 
\caption{\textit{- continued}: J1444+4237, a multi-component source, where the G130M, G160M, and G185M datasets were offset by $\Delta$(PA)$\sim$118 overall.}
\end{figure*}

\begin{figure*}
\ContinuedFloat
\centering
\includegraphics[width=0.8\textwidth]{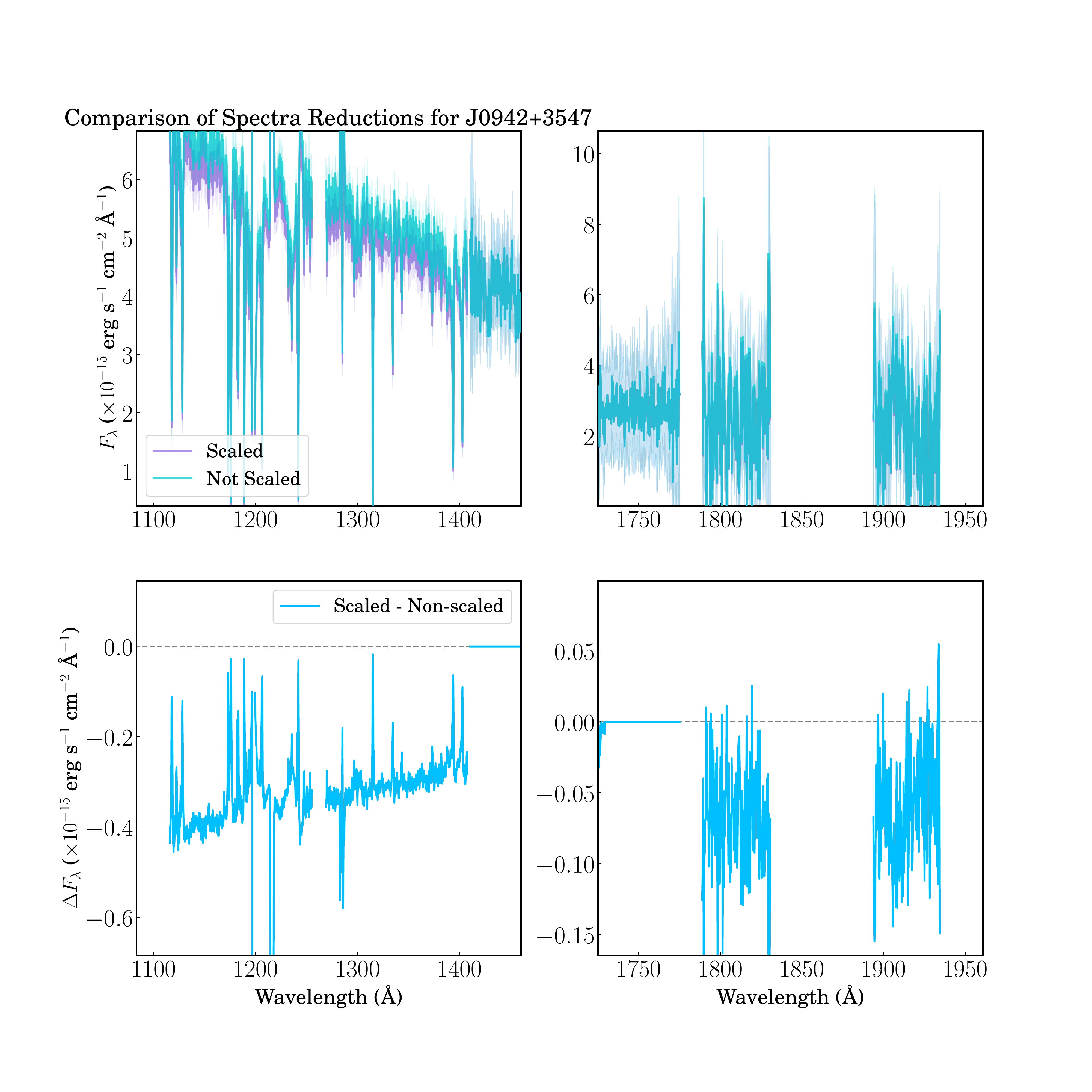}
\caption{\textit{- continued}: J0942+3547, a point-source target where the G130M data was offset in PA from the G160M+G185M data by 207$^{\circ}$.}
\end{figure*}

\subsection{Flux Calibration} \label{sec:fluxcalib}

Offsets in flux can sometimes exist when comparing observations of the same object taken within different programs, either between gratings, between \texttt{CENWAVE}s of the same grating, or observations of extended objects with different position angles. The majority of these offsets exist due to differences in the observational set-up of the object's target-acquisition (e.g., PA, pointing etc). For example, a difference in PA between observations of extended sources can result in altered light profiles in the cross-dispersion direction and the amount of flux extracted. While larger extraction regions have the potential to aid us here, we consider the optimization and conservation of S/N a higher priority than minor offsets in flux calibration, especially given the predominantly flux-normalized nature of UV-based science.

Moreover, due to the vignetting through the COS aperture (which is detailed in Section~\ref{sec:vignet}), slight offsets between COS pointings due to mis-matched centering on multi-component light profiles, can result in non-negligible differences (i.e., $>5\%$) in the amount of light reaching the detector. Along the same lines, if there are failures to re-acquire guide-stars during the observation's visit, the target can potentially drift within the aperture, causing a decrease in flux. Datasets suffering from failures such as these are flagged accordingly within the HST archive but not within the files themselves (such datasets were identified and removed from the CLASSY data sample). Changes in the LP of the observation (i.e., the position on the detector) do not account for flux offsets, since flux calibration is done self-consistently via the time dependent sensitivity calibrations performed as part of the standard COS calibration program\footnote{\url{https://www.stsci.edu/hst/instrumentation/cos/calibration}}. Flux offsets between gratings/\texttt{CENWAVE}s within the 2\% relative flux calibration accuracy achieved by COS are expected whereas the absolute flux calibration accuracy (i.e., between COS observations and models) is 5\% \citep[COS IHB, ][]{COSIHB}. However, given the vignetting through the COS aperture (detailed in Section~\ref{sec:vignet}), the accuracy of the absolute flux calibration ultimately depends on the particular source morphology.

Differences in flux were observed both within and between gratings for several of the CLASSY targets, for the specific reasons outlined above, including offsets within the relative flux calibration accuracy of COS. For scaling within the \textit{same} grating, a reference dataset was chosen (according to the most centered pointing and/or most common PA) and spectra were scaled towards the longest exposure time dataset within the reference set. Flux offsets within a grating were found to be minor ($\sim2$\% on average). For scaling between \textit{different} gratings, the G160M data was chosen to be the `nominal' dataset as it represents the middle wavelength range between the three gratings.\footnote{It should be noted that the choice of the nominal dataset is somewhat arbitrary for our needs here, mostly because FUV data is continuum normalized before measuring absorption lines and any line ratios between optical and UV data will be performed after scaling optical data to match the UV. However, for scientific use cases where the continuum level needs to be accurately evaluated, when coadding or comparing data taken at different PAs, the NUV light profiles should be taken into account and assessed before assigning a nominal dataset. }  Since the FUV data typically has more continuum and a higher S/N than the NUV data, G130M was first scaled to G160M, then G185M was scaled towards the G130M+G160M combination.  Scale factors were derived via a reduced $\chi^2$ 1D spline fit to regions of featureless high S/N continuum and comparing the flux offset in the mid-way interception point between the two fits (i.e., the midpoint of the overlap in wavelength ranges of the spectra), as demonstrated in Figure~\ref{fig:scaling}.  The error arrays were not included in the derivation of the scale factor. In some cases, due to small gaps in wavelength between G130M and G160M spectra or G160M spectra and the gaps between NUV stripes, the midway point may occur in a region that pertains to the extrapolated fits and thereby represents our best estimate of the flux in this wavelength region. Several fitting functions were tested for this process and a 1D spline was found to sufficiently represent the shape of the continuum for each target within the wavelength ranges covered. To accurately determine the slope of the continua, all spectral features were masked with velocity extents according to the specific type of feature: $-3500<v$(\kms)$<+2000$ for stellar wind lines, $-200<v$(\kms)$<+200$ for photospheric lines, $-1000<v$(\kms)$<+200$ for ISM lines, $-200<v$(\kms)$<+200$ for MW ISM lines,  $-100<v$(\kms)$<+100$ for fluorescent lines, and $-300<v$(\kms)$<+300$ for nebular lines, and geocoronal emission lines. The fluorescent lines are typically very narrow and do not need the wider masks used for strong nebular emission lines.

\renewcommand{\thefigure}{11}
\begin{figure*}
\centering
\includegraphics[width=0.8\textwidth]{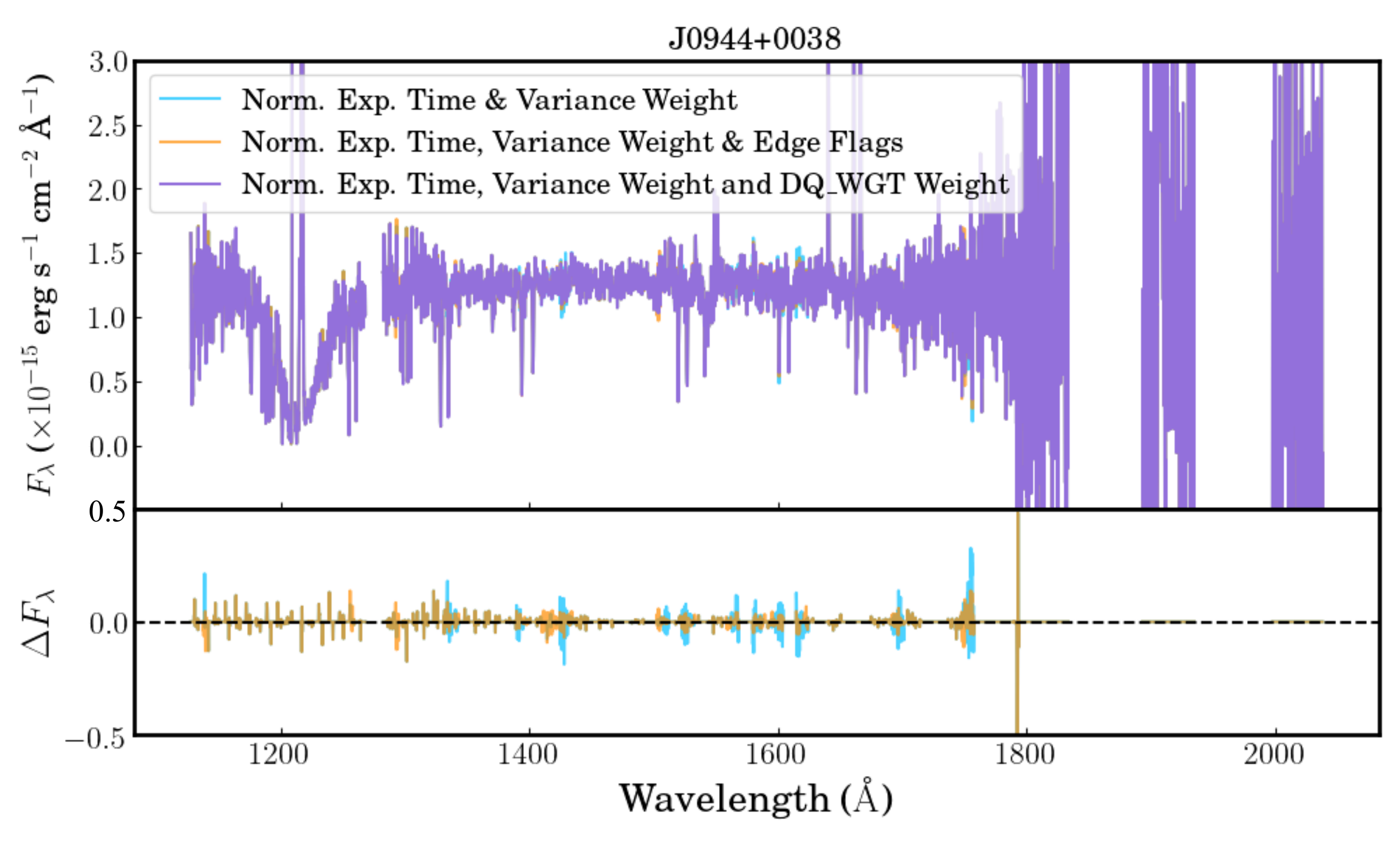}
\caption{A demonstration of different coadding techniques explored for the CLASSY final coadded spectra. Here we show the combined G130M+G160M+G185M data for J0944-0038 using three different coaddition techniques, consisting of normalized exposure time and variance weighting, edge flagging, and DQ-weighting (as described in the legend).
The difference in flux relative to the DQ-weighted coaddition is displayed in the bottom sub-panel, which illustrates the DQ-weighted coaddtion technique being most beneficial in removing noise in the overlap regions. Spectra have been corrected for Galactic reddening and redshifted to restframe wavelengths.}
\label{fig:coadd_tech}
\end{figure*}

For the majority of targets, a negligible offset in flux was found between the entire suite of G130M+G160M+G185M/G225M datasets (i.e., within the
relative $<2$\%\ flux accuracy of COS). However, flux offsets in overlapping wavelength regions between the gratings were found to be $\sim$3–15\%\ in 12 individual cases, which is outside the relative flux calibration of the COS instrument. These targets were
J0337-0502, 
J0942+3547, 
J1132+5722, 
J1157+3220, 
J0036-3333, 
J0021+0052, 
J0127-0619, 
J0944-0038, 
J1105+4444,
J1016+3754, and J1359+5726. The majority of these targets showed either extended or multi-component light distributions and changes in position angle between the observational setup (grating and/or \texttt{CENWAVE}). This resulted in differing cross-dispersion profiles on the detector, and thus slightly different levels of signal extracted during the \texttt{TWOZONE} extraction method. When inspecting the \texttt{BOXCAR} versus \texttt{TWOZONE} extractions for these targets, it could be seen that the \texttt{BOXCAR} flux levels were in agreement within the wavelength overlap regions. This is because a circular aperture, with a symmetric vignetting function, collects the same amount of light for the same pointing, regardless of changes in PA.
 However, as described in Section~\ref{sec:fuv_ext}, the increase in S/N was not sufficient enough to warrant a \texttt{BOXCAR} extraction for these sources, especially given the increased number of bad pixels that were included within the \texttt{BOXCAR} extraction box, which resulted in data gaps. 


Observations of extended/multi-component targets at the \textit{same} PA or \textit{compact sources} at different PAs do not result in large flux offsets. In Figure~\ref{fig:fuv_scaling} we show an example of the small flux offset observed between the G130M, G160M, and G185M gratings for J0942+3547. This target is a compact source where the G130M data were offset in PA from the G160M+G185M data by 207$^{\circ}$, where roughly the same level of signal should be expected. The scale factors were found to be $\sim$3--6\%\ of the G160M continuum flux) and may be attributable to absolute flux calibration offsets intrinsic to the COS instrument itself. 

In some faint targets (e.g., J0944-0038), it should be noted that small sections of G185M/G225M spectra lie within the unphysical negative flux regime. Upon inspection of the original, unscaled, spectra, these particular wavelength regions were already extending below zero despite the improved (lower) background estimate discussed in Section~\ref{sec:nuv_ext}. While the (sometimes) downward scaling of our G185M/G225M will potentially exaggerate the issue, it should be noted that the original cause is due to background subtraction issues. The uncertainties are particularly large in these wavelength regimes and may be asymmetric (and are flagged accordingly). The calculation of scale factors involve featureless regions of continuum and do not include uncertainties on the data and thus provide the most accurate flux calibration available between the gratings.

From our findings we conclude that inspecting the flux alignment between individual gratings is imperative, especially when working with spectroscopy of extended or multi-component sources. Moreover, the position angle of observations of extended or multi-component sources should be kept constant between observations in order to achieve an accurate flux calibration between datasets, while also optimizing the S/N of the dataset. We discuss the scientific impacts of correcting for flux offsets between gratings in Section~\ref{sec:cont_fit}.

\subsection{Coadd Weighting Techniques}\label{sec:3.4}

After accounting for any issues related to wavelength and flux calibration, 
and differences in spectral dispersion owing to different observational setups (i.e., grating and LP), 
the individual datasets are considered ready for coaddition. 
Coaddition was performed as a weighted mean of the flux and error arrays (propagated in quadrature),
where the weighting of the error arrays was modified if the weights included
the error or variance.
Several weighting techniques exist for coadding data, such as weighting by exposure time, 
data quality (DQ) arrays, error arrays, S/N, and various combinations. 
Other methodologies, such as weighting by the variance, squared S/N, inverse variance,  
and modified exposure time (where flagged pixels are devalued) have also been explored for COS data \citep[e.g.,][]{danforth10}. 

Each weighting technique was investigated as part of the CLASSY coadded process 
by assessing the overall S/N of the final spectra and checking for spurious features.  
In particular, regions of increased noise can appear in the overlap regions of gratings, 
due to large uncertainties at the edge of a gratings's bandpass where the throughput decreases. 
These edge effects are flagged within the COS data with both DQ flags 
(integer values depending on the quality condition) and \texttt{DQ\_WGT} flags 
(0 to 1 in the individual \texttt{x1d} files to indicate whether an event should be 
included in the final \texttt{x1dsum} data and values of 0 to N in the \texttt{x1dsum} data, 
where N is the number of spectra used to create the \texttt{x1dsum}). 
\texttt{DQ\_WGT} arrays are derived from serious data quality flags (\texttt{SDQFLAGS}), 
which will set the \texttt{DQ\_WGT} values to 0 for photon events that lie near the edge of the detector, dead spots, hot spots etc.

A comparison of three different coaddition weighting methods is presented in Figure~\ref{fig:coadd_tech}, consisting of a normalized exposure time $+$ variance weight method, which is then modified to include an edge flagging technique and a DQ-weighting technique. In the bottom panel of Figure~\ref{fig:coadd_tech} we show the difference between each technique relative to the DQ-weighting method. Spikes in noise can be seen throughout, which are most apparent at the spectral overlap regions between the individual datasets. The CLASSY coadds were, therefore, constructed using the best resulting method that weights by the
(1) normalized exposure time, 
(2) normalized \texttt{DQ\_WGT} arrays, and
(3) variance arrays, 
resulting in spectra with maximal S/N and smooth transitions between gratings. 
Since the coaddition takes place on \texttt{x1dsum} data, 
the \texttt{DQ\_WGT} arrays were normalized by the maximum value within each array 
before being utilized as a weight.  

\renewcommand{\thefigure}{12}
\begin{figure*}
\centering
\begin{tabular}{cc}
\includegraphics[width = 3.5in]{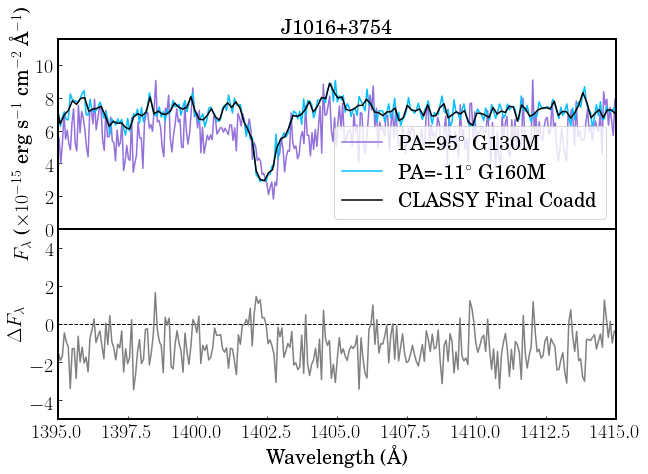} &
\includegraphics[width = 3.5in]{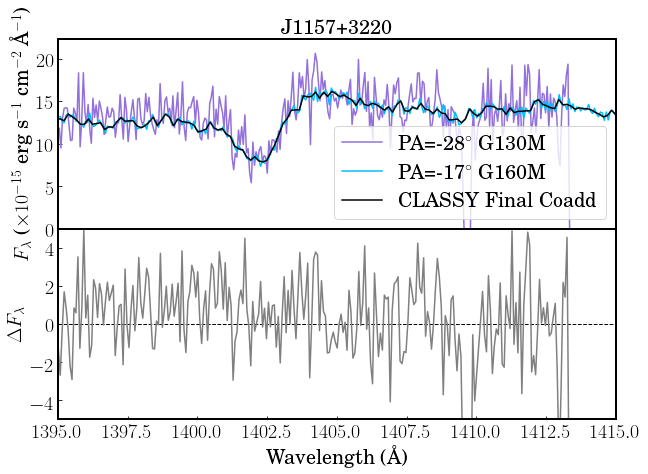}\\
\includegraphics[width = 3.5in]{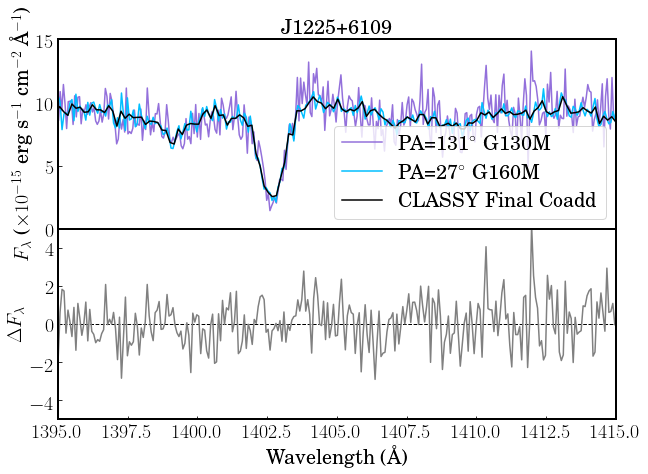} &
\includegraphics[width = 3.5in]{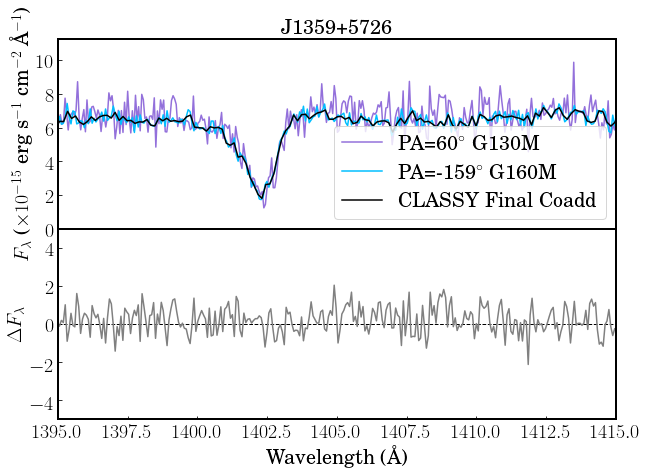}\\
\end{tabular}
\caption{Spectral overlaps for two extended sources (J1016+3754 and J1157+3220; top panels) and 
two sources with multi-component light profiles (J1225+6109 and J1359+5726; bottom panels) where the G130M and G160M data were observed at different position angles, and the effects of optical aberrations and/or different cross-dispersion profiles would be most apparent. Only a minor difference in the continuum flux is seen for each object (shown as $\Delta F=F_{\rm G130M}-F_{\rm G160M}$ in the bottom of each panel) and a $\sim$0.5~\AA\ ($\sim107$~\kms) wavelength (velocity) offset is seen in J1016+3754 due to a change in the dispersion profile on the detector. In the upper panel of each sub-plot we show the spectra from the original datasets (purple/cyan) overlaid with the final CLASSY coadded spectrum (black).}
\label{fig:line_profiles}
\end{figure*}

\subsection{Vignetting \& Aberrations} \label{sec:vignet}
As discussed in Section~\ref{sec:intro}, the COS spectrograph was designed to perform optimally for point-source targets at the nominal (off-axis) location of COS in the optical telescope assembly (OTA) focal plane, which corresponds to the first lifetime position of COS (LP1, May 2009 until July 2012). The COS FUV grating corrects for the OTA spherical aberration while dispersing the light and corrects for some of the astigmatism inherent in the Rowland Circle design with a single optical element when a target is at this position. The NUV channel does a similar correction using multiple optical elements.

During a COS FUV observation, the selected aperture (PSA or BOA) is moved to a unique offset position for each LP. Because the COS aperture is encountered before the correction for the OTA’s spherical aberration is made, the aperture does not function as a traditional field stop with sharp edges, and thus the effects of the aperture on the light entering the spectrograph are more complicated than in a typical optical system. Light coming from more than ~0.4” from the center of the aperture will be partially vignetted, and some light beyond 1.25” from the center (nominally outside of the aperture), reaches the detector (as demonstrated in the acquisition images shown in Figures~\ref{fig:acq_PS}--\ref{fig:acq_MC}).

Because of this complex vignetting function, the distribution of counts on the detector for a non-symmetric extended source in an NUV image (taken at LP1) does not directly reflect the distribution of light on the sky. In addition, since vignetting effects vary with LP, there is no direct way to map its effects from one LP to another. This means that the 2D distribution of light passing through the aperture (and ultimately reaching the detector) will be slightly different at each LP. A further difference between NUV and FUV is that the mid-frequency wavelength errors (MFWEs) on the primary and secondary mirrors of HST cause a broadening of the PSF which is a function of wavelength \citep[as detailed in the COS IHB,][]{COSIHB}. Although the MFWE effects were not accounted for in the design of the spectrograph, they are included in the theoretical PSFs provided by the COS team and discussed in Section \ref{sec:3.2}. In summary, NUV images of extended objects can not be used to correct for the effects seen in spectra.


The instrumental focus is set to maximize the performance for a point-source target in the center of the aperture for each LP. Light entering the aperture but not at the center, such as that from an extended object, will necessarily be out of focus, and these effects will depend on the position angle. In addition, the vignetting effects will selectively remove light in a complicated way, further distorting the two-dimensional PSF at the detector.

Since individual spectra cannot be corrected for these effects, caution should be used when combining datasets taken at different pointings, especially for extended sources whose light profiles extend beyond the central $\sim$0\farcs4 region of the aperture. For example, the same extended source observed with one pointing may have different cross-dispersion heights and/or different dispersion profiles on the detector, leading to changes in spectral resolution. Additionally, changes in cross-dispersion profile heights can lead to flux offsets when adopting the default extraction technique. For several of the extended CLASSY targets where datasets were observed at different PAs, flux offsets between the spectra were observed due to these effects and had to be scaled accordingly (as discussed in Section~\ref{sec:fluxcalib}). 

We assessed these effects in the regions of overlapping wavelength coverage between extended source datasets observed at different PAs. In Figure~\ref{fig:line_profiles} we show the \sIiv~$\lambda\lambda$1393, 1402 line profiles of four example targets where G130M and G160M data were taken at different PAs: two extended sources J1016+3754 and J1157+3220,  and two multi-component sources, J1225+6109 and J1359+5726.

The offsets in PA resulted in changes in flux by factors of $\sim$1--20\%\ of the continuum flux for extended sources (due to the extracted flux along the different cross-dispersion profiles). In J1016+3754 a shift of $\sim$0.5\AA\ ($\sim107$~\kms) is observed between the two datasets, most likely due to the fact that this is an extended object, and the change in aperture position angle modifies its dispersion profile on the detector (see Section~\ref{sec:wavelength}). As such, we advise that care should be taken when averaging data for extended sources when different position angles were utilized for the observations. Inspecting the data for offsets in wavelength and/or flux is recommended before coadding datasets of this nature. 

When it comes to scientific application of the data, it should be noted that aberration effects are accounted for with point sources via the LP-dependent line-spread functions (available on the COS website\footnote{\url{https://www.stsci.edu/hst/instrumentation/cos/performance/spectral-resolution}}) which should be convolved with the model spectra or model line profile before comparing them with COS spectra (i.e., during line profile fitting). For extended sources, however, the effects of vignetting and aberrations in the dispersion direction can affect the shape of the LSF. While LSF profiles are available for point sources, we did not attempt to create ad-hoc LSFs for each of the extended sources as a function of wavelength. While the vignetting affects both the acquisition images and spectrographic light profiles in the same way (both are taken through the same aperture), since it varies with LP, one could \textit{only} use the vignetted image to model the extended-source LSF, and apply it to the spectral profile on the FUV detector, if the spectrum was taken at the same LP as the acquisition images (LP1). This accounts for only 8/177 datasets included within CLASSY.

Additionally, it should be noted that photometric measurements of extended sources will also be affected by vignetting, since the light near the center is weighted more heavily than the light at larger radii. Along the same lines, studies of nebular lines may also be affected, in that emission from ions near the center of a light profile will be more heavily weighted than ions further out. As such, care should be taken when comparing photometric and/or nebular emission line measurements of extended sources obtained from COS images/spectra against data obtained with non-vignetted apertures. However, this should only be applicable when a significant amount of the source's flux extends beyond the central $\sim$0\farcs4 radius of the aperture.

\section{Results \&\ Discussion}\label{sec:science}
The addition of archival and new observations of 45 nearby star-forming galaxies has resulted in the first high-resolution and high-S/N spectroscopic atlas covering the entire COS UV wavelength range. In order to provide the best possible spectroscopic dataset to the community, the utmost care was taken to extract, reduce, and combine each target's suite of observations, resulting in a catalogue of high-S/N medium-to-high resolution spectra for each target. The final coadded CLASSY spectroscopic datasets are described in detail in \citetalias{berg22} and are available for download at \url{https://mast.stsci.edu/CLASSY}. A thorough documentation of the reduction and coadding procedures used to create this dataset are detailed in the above section, along with guidelines and best practices for performing these procedures on COS spectroscopic datasets in general. During these procedures, enabling the maximum scientific accuracy of our data was of prime importance. However, as with all spectroscopic observations, there are several aspects of the data reduction and coaddition procedures that have the potential to affect the scientific results obtained from the data. Here we use a range of typical spectroscopic analysis techniques such as continuum fitting, emission line fitting, and absorption line fitting to understand the extent to which our data reduction techniques affect the scientific output of the CLASSY data.

\renewcommand{\thefigure}{13}
\begin{figure}
\centering
\includegraphics[width=0.4\textwidth]{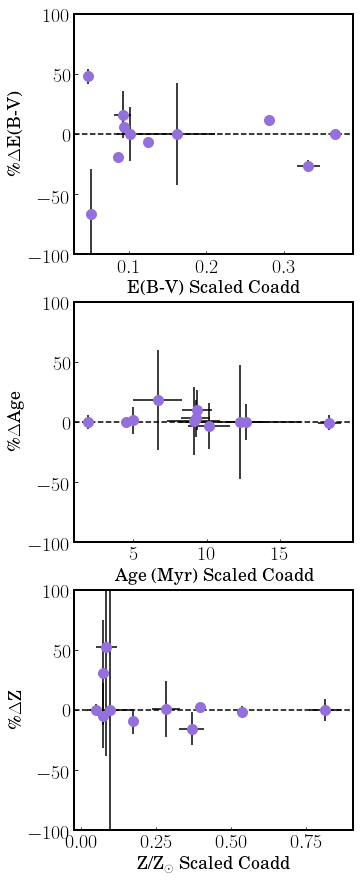}
\caption{
Parameters derived from stellar continuum fitting of CLASSY coadded spectra 
with and without scaling between the individual grating datasets, expressed as a percentage change relative to the parameter derived before scaling.
The top panel demonstrates that the reddening value due to dust, $E(B-V)$, 
derived from the stellar continuum fits is quite sensitive to relative
flux calibration.
On the other hand, the middle and lower panels show little change in the 
luminosity-weighted stellar population metallicity relative to solar, 
age, and $Z/Z_{\odot}$, respectively. }
\label{fig:test_cont}
\end{figure}

\subsection{Continuum fitting} \label{sec:cont_fit}
The majority of UV spectral analyses require robust stellar continuum fits, 
either for continuum normalization or for determining the properties of the 
stellar population. 
For the latter application, the properties derived from continuum fitting 
with simple population synthesis (SPS) models are directly dependent on the shape 
of the continuum.
Reversed, the observed UV continuum shape depends both on the intrinsic 
stellar continuum slope and the reddening due to dust.
Given that the dust is intrinsic to the system observed, we can investigate
the effect of the relative flux calibration on the derived stellar population 
properties. 

\renewcommand{\thefigure}{14}
\begin{figure*} 
\centering
\includegraphics[scale=0.3]{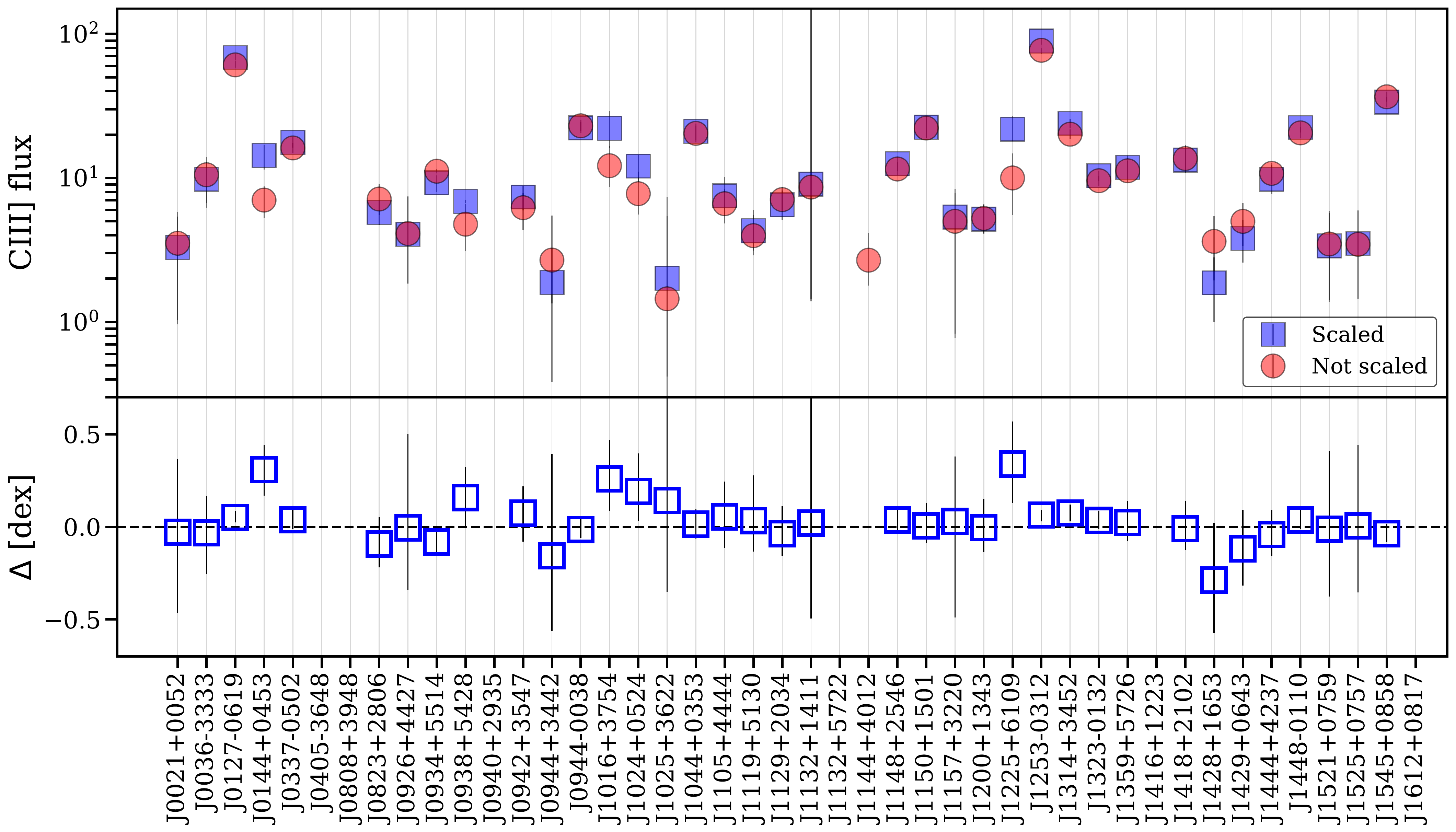} 
\caption{The effects of relative flux scaling on the NUV nebular emission lines.
The flux measurements for the \sfciii\W1907,1909 emission-line doublet
are plotted for the CLASSY coadded spectra before and after flux scaling was applied 
between the FUV and NUV datasets. The bottom panel shows the difference in \sfciii\ flux (scaled-unscaled) in dex for each target.} 
\label{fig:CIII_flux}
\end{figure*}
\renewcommand{\thefigure}{15}
\begin{figure*} 
 \centering
\includegraphics[trim={17cm 0 0 0},clip,width=0.5\textwidth]{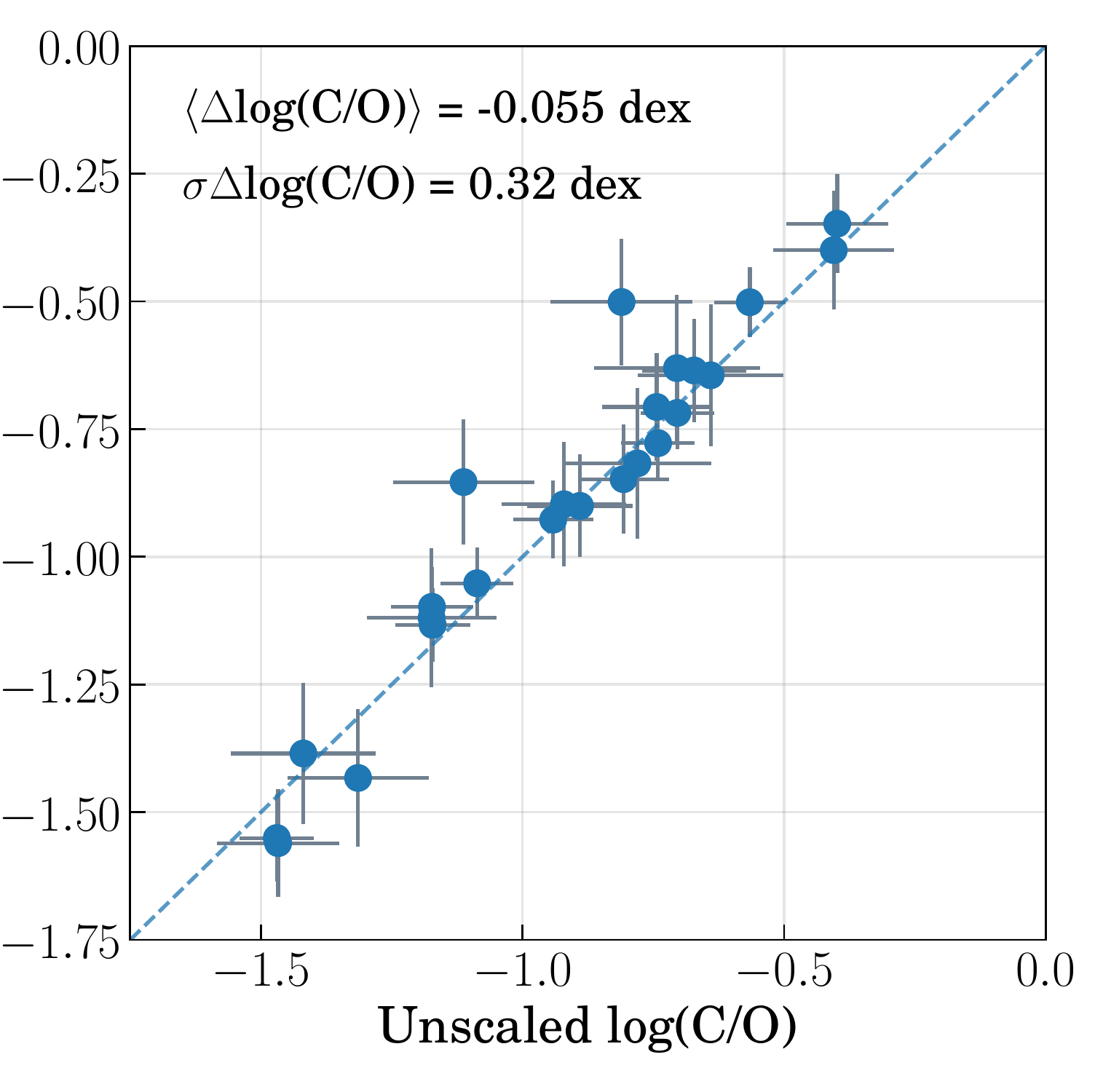} 
\caption{The effects of relative flux scaling on the NUV nebular emission lines.
The effects of relative flux scaling on the direct-method C/O abundances, derived using the \sfciii\W1907,1909 to
\sfoiii\W1666 line ratio, from the scaled and unscaled coadds.
In general, scaling the NUV gratings to the FUV gratings results in slightly decreased 
\sfciii\W1907,1909 C/O abundances. The dashed-blue line represents the 1:1 ratio.} 
\label{fig:test_fluxes}
\end{figure*}
As discussed in Section~\ref{sec:fluxcalib}, the continuum level of all gratings of an 
individual galaxy were scaled to its corresponding G160M grating continuum.
While the scaling factors between the G130M, G160M, and G185M datasets were usually 
relatively small (at most a factor of 1.7), if the CLASSY data had not been aligned 
correctly there may have been adverse affects on the accuracy of properties derived 
from fitting the stellar continuum. 
To test the effects of this scaling method, 
we used a subset of 11 CLASSY galaxies with larger than average scaling parameters
and created two sets of coadded spectra, one with the optimal scalings and one 
with no scaling.
We then performed stellar continuum fits using the methodology detailed in 
\citet{chisholm19}, 
which fits a linear combination of single-age and single-metallicity 
\textsc{Starburst99} SPS models \citep{leitherer99, leitherer10} to 
determine which combination of models best fits the observations, 
while simultaneously accounting for the dust attenuation ($E(B-V)$).
The attenuation law from \citet{reddy16} was used for all fits. 
From this best-fit, we derive $E(B-V)$ values and the light-weighted ages 
and metallicities of the ionizing stellar population. Uncertainties on these parameters were derived via a Monte Carlo technique, modulating the observed flux with a Gaussian kernel centered on zero with a width equal to the error on the flux. The stellar continuum was then re-fitted and the $E(B-V)$, age, and metallicity values tabulated. This process was repeated 100 times to build a distribution of parameters from which we calculated the standard deviation of the distribution as the error on the estimated parameters.

In Figure~\ref{fig:test_cont} we compare the derived properties from
the stellar continuum fits of our scaled and non-scaled samples. 
The most pronounced difference is in the best-fit reddening values
(see top panel of Figure~\ref{fig:test_cont}), 
with 0.0$<\Delta$E(B-V)$~\lesssim$0.12, and a (weighted) average percentage reduction of $\sim$14\%.  The difference in the luminosity-weighted stellar population age (middle panel of Figure~\ref{fig:test_cont}) was less pronounced, however, with a (weighted) mean difference of only $\sim$5\%\ (corresponding to $\langle \Delta$age$\rangle \sim$0.17~Myr). The scaling also had a negligible effect on the stellar metallicity (bottom panel of Figure~\ref{fig:test_cont}), with a (weighted) mean percentage difference of only $\sim$0.3\% overall (corresponding to $\sim$0.004\Zsol ).
These difference in effects on each parameter are due to the fact that the stellar wind and photospheric features in the FUV that constrain the metallicity and age are not very sensitive to changes in the continuum slope. 
These results suggest that an accurate alignment of the continuum flux across the full FUV to NUV wavelength range is imperative for deriving an accurate reddening of the stellar light. 
However, flux calibration of the continuum is less important for deriving
accurate metallicity and age estimates of the stellar population. 

\subsection{Emission line measurements}\label{sec:em_fit}

Nebular UV emission lines have a plethora of scientific applications, including the determination of physical and chemical conditions, the level of ionization of a source, and diagnostics of stellar age and metallicity. CLASSY data have a suite of nebular emission line detections, including \heii~$\lambda$1640\footnote{Both a narrow nebular and a broad stellar component are often seen in \heii\ emission. As such, all emission line fitting of this line takes place on stellar continuum subtracted data, as detailed in Mingozzi et al. 2022, in-prep.}, \sfoiii~$\lambda\lambda$1660,66, \fciii,\sfciii~$\lambda\lambda$1907, 1909 (referred to as \sfciii\ hereafter), and nebular components of \civ~$\lambda\lambda$1548,1551. Emission line fit parameters from CLASSY data will potentially be affected by two components of the procedures utilized to make the CLASSY coadded data: (i) the scaling between the NUV and FUV gratings; and (ii) the optimized extraction of the NUV datasets for extended sources. We explore each of these below.

To assess the effects of flux scaling between the FUV and NUV datasets 
on the nebular emission-line science, we fit the \sfciii\ emission-line 
doublets in the CLASSY coadded spectra, with and without relative
scalings applied. 
Flux measurements were made for galaxies where \sfciii\ and \sfoiii\ were detected, producing a sample of 37 galaxies with \sfciii\ and 29 galaxies with \sfoiii. In Figure~\ref{fig:CIII_flux} we show the \sfciii\ flux before and after scaling for the individual sources (whose scale factors can be found in Figure~\ref{fig:AppA}). Overall we found a median increase of 5\%\ in the \sfciii\ flux. Since the \sfoiii\ emission lies within the G160M grating, which was 
chosen as the nominal dataset towards which the G130M and G185M data were 
scaled, their fluxes do not change before and after scaling.

Next, to test the scientific effect of the \sfciii\ flux scaling, 
we computed the C/O abundance ratios for both the scaled and unscaled 
flux values. 
We determine direct-method C/O abundances using the method outlined in 
\citet{berg19a} with values from \citetalias{berg22} for the electron 
temperatures and densities, log([\ion{O}{3}] \W5007/[\ion{O}{2}] \W3727), 
and 12+log(O/H) values.
Briefly, this method uses the \ion{C}{3}]\W\W1907,1909/\ion{O}{3}]\W1666
line ratio and the high-ionization zone temperature and density with the
\texttt{pyneb} \texttt{python} package to calculate the 
C$^{+2}$/O$^{+2}$ ion abundance.
C$^{+2}$/O$^{+2}$ is then converted to a total C/O abundance using an
ionization correction factor inferred from the 
log([\ion{O}{3}] \W5007/[\ion{O}{2}] \W3727), and 12+log(O/H) values.

The resulting C/O abundances are plotted in Figure~\ref{fig:test_fluxes}, where we see that the effect of the relative grating scalings is to slightly decrease the C/O abundances.
On average, the C/O abundances decreased by $\langle \Delta$log(C/O)$\rangle$ = -0.055 dex,
with a dispersion of $\sigma\Delta$log(C/O) = 0.32 dex.
Given the $\sim0.1$ dex magnitude of this effect, its relevance
will depend on the S/N of emission lines of a given spectrum and 
the subsequent uncertainty on the C/O value.

\renewcommand{\thefigure}{16}
\begin{figure*}
\centering
\includegraphics[scale=0.3]{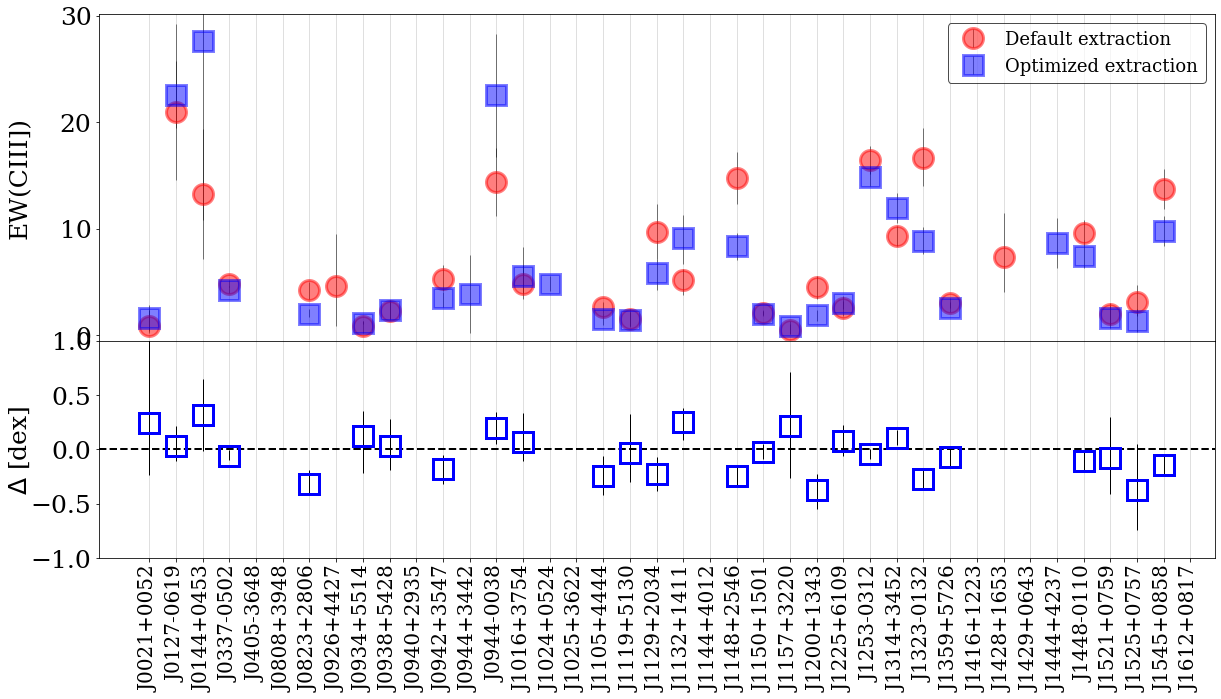} 
\caption{The effect of COS NUV spectral extraction method on the \ion{C}{3} \W\W1907,1909 equivalent widths. In the top panel we show the equivalent widths (in units of \AA) of the \sfciii~\W\W1907,1909 emission-line doublet from the default CalCOS and optimised extraction method utilized by CLASSY. The bottom panel shows the corresponding difference in EW for each system. All galaxies classified as extended sources were re-processed with the optimized extraction for the CLASSY coadd data, along with any systems that showed a change in EW(\sfciii) outside the uncertainties in the measurement.}
\label{fig:EW_test}
\end{figure*}

The customized extraction of the NUV data for extended sources that CLASSY utilized will affect both the continuum and emission line flux within the NUV wavelength range. In order to assess the effects of our methodology, we measured the equivalent width of the \sfciii~$\lambda\lambda$1907, 1909 emission line doublet. Being a ratio between the emission line flux and the continuum level flux, the equivalent width provides an unbiased view of the NUV extraction method in that it is unaffected by the additional flux scaling that was applied to several of the NUV datasets. In Figure~\ref{fig:EW_test} we compare the EW(\sfciii) of all the sources, measured on both the original NUV datasets and with widened NUV extraction boxes. It can be seen that the optimized extraction of the NUV datasets has a range of effects on the EW(\sfciii), with an overall mean decrease of $\sim-3$\%\ but a mean increase of $\sim17$\%\ if we consider just the extended sources. For each case, if the target was extended \text{or} change in EW(\sfciii) was larger than the uncertainties on the measurement, we adopted the optimized extraction. For the majority of cases we see a decrease in the EW measurement, owing to the fact that non-optimized NUV extraction (1) did not fully encompass the cross-dispersion profile for extended and MC sources and (2) over-estimated the background levels (as shown in Figures~\ref{fig:NUV_ext} and \ref{fig:NUV_extract}). As such, one would expect an overall increase in the continuum level and subsequent decrease in the EW values. However, for several extended source targets, we instead see an \textit{increase} in the EW(\sfciii) - this suggests that in these cases, the nebular emission is more spatially extended than the continuum emission, resulting in larger net increase in emission line flux compared to the continuum. The larger uncertainties in the optimized extraction are due to an increased level of noise resulting from a larger extraction box. 


With regards to scientific impact, the \sfciii\ EW has shown to have a dependence on stellar age and nebular metallicity, in that the EW decreases as a function of age and metallicity \citep[e.g.,][]{erb10, stark14, rigby15, jaskot16, senchyna17, nakajima18}, although it remains unclear as to whether this dependence holds for extremely metal poor systems such as SBS~0335-052 \citep[][]{wofford21}. Thus being, if only the \sfciii\ lines were used to constrain these parameters (i.e., in high-$z$ systems where only bright UV nebular emission lines may be observable in the optical, or low-$z$ systems with only UV wavelength coverage), an over/underestimate of the \sfciii\ EW would lead to under/overestimated metallicities and stellar ages for a particular target. We therefore highly recommend that users of NUV data of extended sources perform a customized extraction of the NUV source profiles (e.g., using the methodology described in Section~\ref{sec:nuv_ext} in the case of low-$z$ systems with COS observations).


\subsection{Absorption line fitting}\label{sec:abs_fit}

As discussed in Section~\ref{sec:vignet}, and shown in Figure~\ref{fig:LSFs}, the point source LSF of the COS data is dependent on the lifetime position of the observation, and changes in resolution of up to 25\%\ can occur between data obtained at LP1 and LP4. As such, when fitting absorption line profiles, users need to incorporate the correct LSF for the particular dataset. However, since several of our coadded spectra consist of data taken across multiple lifetime positions on the detector, incorporating several LSFs into a line fitting procedure would be complex and inefficient. Also, a change of $<$25\%\ in the LSF may not be noticeable given that (1) $\sim60\%$ of our sources are extended such that their spectral resolution will be lower than that achieved by a point source profile and (2) the S/N of the CLASSY data is $<20$. 

To assess the effects of changes in LSF within our datasets we fit a range of isolated ISM and photospheric absorption lines (\sIiv~$\lambda$1402, \sIii~$\lambda$1260, \sIii~$\lambda$1304, and \cii~$\lambda$1334) using both the correct LSF (i.e., the LSF for the LP of that particular dataset) and an LP4 LSF. An LP4 LSF was chosen as a comparison case because the majority of the CLASSY data are taken at LP4 (October 2, 2017 until October 2021) and as such, applying an LP4 LSF to all ISM measurements of CLASSY data would be straightforward for users of CLASSY data if no differences are found. In both cases we convolve the chosen LSF with a Gaussian profile with a width equal to that measured from the COS acquisition images, thus creating a non-point-source LSF. Lines were fit using a Gaussian line profile (as detailed in Xu et al., in-prep) on stellar continuum-subtracted data (as detailed in Chisholm et al. and Senchyna et al., both in-prep), binned by 15 pixels, using non-point-source LSFs binned by the same amount. The velocity of the outflow (i.e., the line centroid) was measured using the correct LSF and an LP4 LSF, for four lines: \sIii~$\lambda$1260, \sIii~$\lambda$1304, \cii~$\lambda$1334 and \sIiv~$\lambda$1402. The change in velocity measurement was found to be negligible, with mean differences ranging between 0.4--2.8~\kms\, which is within the uncertainties of the measurements. A similar test was performed by calculating the FWHM of the same line profiles measured using both the correct LSF and the LP4 LSF. Again, the difference was found to be minimal, with mean differences ranging between 3--7~\kms\ which is within the COS calibration uncertainties.  

This test shows that for this particular scientific application of the CLASSY data, it is not necessary for the user to adopt the LSF particular to that dataset and users can convolve their CLASSY coadded spectra with any of the COS LSFs (LP1 to LP4) without affecting the line properties themselves. Of course, this advice is applicable to the CLASSY data specifically. For COS observations of point sources, for very high S/N data (S/N$>$20), and/or data at the native resolution of COS (i.e., binned to only 6 pixels) the effect of using the correct LSF may be far more apparent, and users are advised to use the correct LSF according to the LP of their data.

\section{Conclusions}

The COS Legacy Archive Spectroscopic SurveY will provide the astronomical community with the first high-resolution UV spectroscopic atlas of nearby star-forming galaxies. With chemical and physical properties akin to systems at high-$z$, the galaxies were chosen to provide a UV spectroscopic training set for observations of galactic systems in the early Universe. The design and full description of CLASSY and its primary HLSPs can be found in \citetalias{berg22}. Here we provide technical details of the data reduction and coaddition strategies utilized by CLASSY, with particular attention paid to the extended source profiles for which COS is not optimized. As such, this paper hopes to serve as a technical guide for UV spectroscopic observations in general, with a specific focus on observations of extended sources with HST/COS.  We additionally explore the scientific implications of our strategies. In summary:

\textit{Data reduction:}
 For sources with extended and/or multi-component light profiles, no significant change in the S/N was found in FUV spectra using a \texttt{BOXCAR} extraction method over the default optimized extraction method. This is due to the fact that the majority of the source light profile was contained within the aperture's central 0\farcs4 radius. Moreover, spikes of decreased S/N were seen due to an increased number of bad pixels being included within the \texttt{BOXCAR} extraction. As such, the default \texttt{TWOZONE} extraction was adopted for all sources for which it was available (i.e., datasets taken at LPs 3 and 4).

Improvements in the S/N and flux calibration of NUV data of extended/multi-component sources were obtained by performing ad-hoc spectral extractions according to the width of the cross-dispersion profile of each source. The effects of the optimized NUV extraction can be seen in the equivalent width measurements of \sfciii\, which had a mean change of $\sim-3$\%\ for all sources, and an increase of 17\%\ if we consider just the extended/MC sources. Without this extraction method, EW(\sfciii) measurements, which can be used to constrain nebular metallicity and stellar age, may have resulted in poorly estimated metallicities and stellar ages for these particular targets according to an empirical relation found at high redshift. 
    
\textit{Data coaddition:}
    The primary CLASSY HLSP for each galaxy, the coadded data product covering the full UV wavelength range, involves the combination of spectra taken with several different gratings, many of which were taken at different position angles and LPs on the COS detector. As such, it was necessary to perform several checks on the alignment of data before the final coaddition could take place, such as wavelength alignment, spectral dispersion, flux calibration, PA or pointing offsets. 
    
    Firstly, the agreement in wavelength solution between the gratings was investigated by comparing the alignment of a suite of isolated emission and ISM absorption lines that were coincident within the datasets.  An excellent agreement was found in the wavelength scale between the COS gratings and no adjustment was necessary. 
    
    Secondly, changes in spectral dispersion between gratings were accounted for by re-binning the data to the largest dispersion value of the dataset combination. While changes in spectral resolution of up to $\sim$25\%\ between datasets observed with different LSFs within the coadds (due to the change in illumination angle at each LP on the detector) could not be accounted for, such differences were found to have a negligible affect on the velocity and outflow measurements of isolated absorption lines. This is mostly due to the S/N of the CLASSY data (S/N$<$20) and binning of the data beyond the native resolution of COS. This may not be the case for high S/N datasets (S/N$>$20) and should be accounted for accordingly before coadding data across different LPs on the COS detector.
    
     Thirdly, the flux calibration within and between the FUV and NUV gratings was inspected. Offsets of up to 20\%\ in flux were found for extended/MC sources where datasets were observed with different PAs. Such offsets were accounted for by scaling towards a nominal dataset (in this case the G160M grating, which represents the middle wavelength regime). For the majority of UV-spectroscopic science cases, this approach is suitable in that the FUV spectra will be continuum normalized and the continuum level itself is arbitrary. However, properties of the massive star populations can be affected by the shape of the continuum, in that reddening estimates would have changed by up to $\Delta E(B-V)\sim0.12$ if the FUV data had not been scaled accordingly before coaddition took place. Conversely, no effects were seen on the metallicity estimate of the stellar population between scaled and unscaled data. The effects of flux calibration between the FUV and NUV data were also seen in the \sfciii\ emission line fluxes, which resulted in a mean offset of $\sim-$0.1~dex in the C/O abundance ratios.

One important aspect of the coaddition process concerned the propagation of uncertainties on the flux. During this process we assumed a symmetrical flux distribution, which pertains to the Gehrels upper confidence limit reported by CalCOS. This is appropriate for 65\%\ of all CLASSY raw datasets. For those remaining cases in the low count regime, i.e., where errors were in fact asymmetric, we caution that the errors are accurate within 5-6\%, and flag them accordingly in the HLSP. As an issue that consistently plagues observations of this type, we hope that the re-sampling of asymmetric errors becomes a focus of future efforts in the community.

When combining datasets for extended/MC sources taken with different PAs, changes in the dispersion and cross-dispersion profile may have an effect on absorption line profiles and/or continuum flux level.  The latter effect can be mitigated by utilizing a \texttt{BOXCAR} extraction, albeit at a cost to S/N. We advise users to use a constant orientation when observing extended/MC sources with COS if they intend to combine datasets at the highest possible S/N. 

Finally, after accounting for wavelength, flux, and spectral dispersion offsets between individual datasets, it was found that a coaddition technique that incorporates weighting by both the exposure time (squared) and the data-quality flag of each pixel provided coadded data products with maximum S/N, with smooth transitions between spectra from different gratings.

\vspace{5mm}
In summary, the data reduction and coaddition processes described here are aimed to guide users of UV spectroscopic data towards creating accurately calibrated and combined spectroscopic datasets, ready for scientific application. In particular, we highlight the care and attention that needs to be employed when dealing with extended/MC source observations made with an instrument optimized for point source targets and the scientific impacts of those specialized reductions.

As a result of the technical decisions and techniques that were carefully assessed and adopted throughout this process, CLASSY will not only deliver the first high-resolution UV spectral atlas of star-forming galaxies, but also provide the astronomical community with a suite of high level science products, including properties of the massive star population, Lyman continuum escape fractions, outflows, metallicities, ionization strengths, UV-based nebular diagnostics and much, much more.  With a sample designed to mimic the extreme specific SFRs and low-metallicities of high-$z$ systems, the CLASSY spectral atlas and its plethora of scientific products will provide an essential tool-kit for interpreting the high-$z$ systems observable in the upcoming JWST/ELT era. 

\begin{acknowledgments}\nolinenumbers
BLJ is thankful for support from the European Space Agency (ESA).
DAB is grateful for the support for this program, HST-GO-15840, that was provided by 
NASA through a grant from the Space Telescope Science Institute, 
which is operated by the Associations of Universities for Research in Astronomy, 
Incorporated, under NASA contract NAS5-26555. 

The CLASSY collaboration extends special gratitude to the Lorentz Center for useful discussions 
during the "Characterizing Galaxies with Spectroscopy with a view for JWST" 2017 workshop that led 
to the formation of the CLASSY collaboration and survey.
The CLASSY collaboration thanks the COS team for all their assistance and advice in the 
reduction of the COS data, with particular thanks to Elaine Frazer for NUV data extraction and Christian Johnson in assisting with error analysis.

\end{acknowledgments}

\vspace{5mm}
\facilities{HST (COS)}
\software{
astropy (The Astropy Collaboration 2013, 2018)
CalCOS (STScI),
dustmaps (Green 2018),
jupyter (Kluyver 2016),
Photutils (Bradley 2021),
PYNEB (Luridiana 2012; 2015),
python,
pysynphot (STScI Development Team)}

\bibliography{CLASSY}

\begin{thebibliography}{}
\expandafter\ifx\csname natexlab\endcsname\relax\def\natexlab#1{#1}\fi
\providecommand{\url}[1]{\href{#1}{#1}}
\providecommand{\dodoi}[1]{doi:~\href{http://doi.org/#1}{\nolinkurl{#1}}}
\providecommand{\doeprint}[1]{\href{http://ascl.net/#1}{\nolinkurl{http://ascl.net/#1}}}
\providecommand{\doarXiv}[1]{\href{https://arxiv.org/abs/#1}{\nolinkurl{https://arxiv.org/abs/#1}}}

\bibitem[{{Berg} {et~al.}(2019){Berg}, {Erb}, {Henry}, {Skillman}, \&
  {McQuinn}}]{berg19a}
{Berg}, D.~A., {Erb}, D.~K., {Henry}, R.~B.~C., {Skillman}, E.~D., \&
  {McQuinn}, K.~B.~W. 2019, \apj, 874, 93, \dodoi{10.3847/1538-4357/ab020a}

\bibitem[{{Berg} {et~al.}(2022){Berg}, {James}, {Chisholm}, {Heckman}, \&
  {Martin}}]{berg22}
{Berg}, D.~A., {James}, B.~L., {Chisholm}, J., {Heckman}, T., \& {Martin},
  C.~o. 2022, \apj

\bibitem[{{Carnall}(2017)}]{carnall17}
{Carnall}, A.~C. 2017, arXiv e-prints, arXiv:1705.05165.
\newblock \doarXiv{1705.05165}

\bibitem[{{Chisholm} {et~al.}(2019){Chisholm}, {Rigby}, {Bayliss}, {Berg},
  {Dahle}, {Gladders}, \& {Sharon}}]{chisholm19}
{Chisholm}, J., {Rigby}, J.~R., {Bayliss}, M., {et~al.} 2019, \apj, 882, 182,
  \dodoi{10.3847/1538-4357/ab3104}

\bibitem[{{Danforth} {et~al.}(2010){Danforth}, {Keeney}, {Stocke}, {Shull}, \&
  {Yao}}]{danforth10}
{Danforth}, C.~W., {Keeney}, B.~A., {Stocke}, J.~T., {Shull}, J.~M., \& {Yao},
  Y. 2010, \apj, 720, 976, \dodoi{10.1088/0004-637X/720/1/976}

\bibitem[{{Erb} {et~al.}(2010){Erb}, {Pettini}, {Shapley}, {et~al.}}]{erb10}
{Erb}, D.~K., {Pettini}, M., {Shapley}, A.~E., {et~al.} 2010, \apj, 719, 1168

\bibitem[{{Gehrels}(1986)}]{Gehrels:1986}
{Gehrels}, N. 1986, \apj, 303, 336, \dodoi{10.1086/164079}

\bibitem[{{Green} {et~al.}(2012){Green}, {Froning}, {Osterman},
  {et~al.}}]{green12}
{Green}, J.~C., {Froning}, C.~S., {Osterman}, S., {et~al.} 2012, \apj, 744, 60,
  \dodoi{10.1088/0004-637X/744/1/60}

\bibitem[{{Heckman}(2002)}]{heckman02}
{Heckman}, T.~M. 2002, in Astronomical Society of the Pacific Conference
  Series, Vol. 254, Extragalactic Gas at Low Redshift, ed. J.~S. {Mulchaey} \&
  J.~T. {Stocke}, 292

\bibitem[{{Hirschauer}(2021)}]{COSIHB}
{Hirschauer}, A.~S. 2021, {COS Instrument Handbook v. 13.0}, 13

\bibitem[{{James} {et~al.}(2021){James}, {Berg}, {King}, {Chisholm}, {Heckman},
  {et~al.}}]{james21}
{James}, B.~L., {Berg}, D.~A., {King}, T., {et~al.} 2021, \apj

\bibitem[{{Jaskot} \& {Ravindranath}(2016)}]{jaskot16}
{Jaskot}, A.~E., \& {Ravindranath}, S. 2016, \apj, 833, 136

\bibitem[{{Johnson} {et~al.}(2021){Johnson}, {Plesha}, {Jedrzejewski},
  {Frazer}, \& {Dashtamirova}}]{JohnsonISR:2021}
{Johnson}, C.~I., {Plesha}, R., {Jedrzejewski}, R., {Frazer}, E., \&
  {Dashtamirova}, D. 2021, {Updated Flux Error Calculations for CalCOS},
  Instrument Science Report COS 2021-03

\bibitem[{{Le F{\`e}vre} {et~al.}(2015){Le F{\`e}vre}, {Tasca}, {Cassata},
  {Garilli}, {Le Brun}, {et~al.}}]{lefevre15}
{Le F{\`e}vre}, O., {Tasca}, L.~A.~M., {Cassata}, P., {et~al.} 2015, \aap, 576,
  A79, \dodoi{10.1051/0004-6361/201423829}

\bibitem[{{Leitherer} {et~al.}(2010){Leitherer}, {Ortiz Ot{\'a}lvaro}, \&
  {Bresolin}}]{leitherer10}
{Leitherer}, C., {Ortiz Ot{\'a}lvaro}, P.~A., \& {Bresolin}, F.~o. 2010, \apjs,
  189, 309, \dodoi{10.1088/0067-0049/189/2/309}

\bibitem[{{Leitherer} {et~al.}(1999){Leitherer}, {Schaerer}, {Goldader},
  {et~al.}}]{leitherer99}
{Leitherer}, C., {Schaerer}, D., {Goldader}, J.~D., {et~al.} 1999, \apjs, 123,
  3

\bibitem[{{McLure} {et~al.}(2018){McLure}, {Pentericci}, {Cimatti}, {Dunlop},
  {Elbaz}, {et~al.}}]{mclure18}
{McLure}, R.~J., {Pentericci}, L., {Cimatti}, A., {et~al.} 2018, \mnras, 479,
  25, \dodoi{10.1093/mnras/sty1213}

\bibitem[{{Nakajima} {et~al.}(2018){Nakajima}, {Schaerer}, {Le F{\`e}vre},
  {Amor{\'\i}n}, {Talia}, {et~al.}}]{nakajima18}
{Nakajima}, K., {Schaerer}, D., {Le F{\`e}vre}, O., {et~al.} 2018, \aap, 612,
  A94, \dodoi{10.1051/0004-6361/201731935}

\bibitem[{{Reddy} {et~al.}(2016){Reddy}, {Steidel}, {Pettini},
  {Bogosavljevi{\'c}}, \& {Shapley}}]{reddy16}
{Reddy}, N.~A., {Steidel}, C.~C., {Pettini}, M., {Bogosavljevi{\'c}}, M., \&
  {Shapley}, A.~E. 2016, \apj, 828, 108, \dodoi{10.3847/0004-637X/828/2/108}

\bibitem[{{Rigby} {et~al.}(2015){Rigby}, {Bayliss}, {Gladders}, \&
  {others}}]{rigby15}
{Rigby}, J.~R., {Bayliss}, M.~B., {Gladders}, M.~D., \& {others}. 2015, \apjl,
  814, L6

\bibitem[{{Senchyna} {et~al.}(2017){Senchyna}, {Stark}, {Vidal-Garc{\'{\i}}a},
  {et~al.}}]{senchyna17}
{Senchyna}, P., {Stark}, D.~P., {Vidal-Garc{\'{\i}}a}, A., {et~al.} 2017,
  \mnras, 472, 2608

\bibitem[{{Shapley} {et~al.}(2015){Shapley}, {Reddy}, {Kriek},
  {et~al.}}]{shapley15}
{Shapley}, A.~E., {Reddy}, N.~A., {Kriek}, M., {et~al.} 2015, \apj, 801, 88

\bibitem[{{Shapley} {et~al.}(2003){Shapley}, {Steidel}, {Pettini}, \&
  {Adelberger}}]{shapley03}
{Shapley}, A.~E., {Steidel}, C.~C., {Pettini}, M., \& {Adelberger}, K.~L. 2003,
  \apj, 588, 65

\bibitem[{{Soderblom}(2021)}]{COSDHB}
{Soderblom}, D.~R. 2021, {COS Data Handbook v. 5.0}, 5

\bibitem[{{Stark} {et~al.}(2014){Stark}, {Richard}, {Siana},
  {et~al.}}]{stark14}
{Stark}, D.~P., {Richard}, J., {Siana}, B., {et~al.} 2014, \mnras, 445, 3200

\bibitem[{{Steidel} {et~al.}(2010){Steidel}, {Erb}, {Shapley}, {Pettini},
  {et~al.}}]{steidel10}
{Steidel}, C.~C., {Erb}, D.~K., {Shapley}, A.~E., {Pettini}, M., {et~al.} 2010,
  \apj, 717, 289

\bibitem[{{Steidel} {et~al.}(2014){Steidel}, {Rudie}, {Strom},
  {et~al.}}]{steidel14}
{Steidel}, C.~C., {Rudie}, G.~C., {Strom}, A.~L., {et~al.} 2014, \apj, 795, 165

\bibitem[{{STScI Development Team: Lim, P. L., Diaz, R. I., \& Laidler,
  V.}(2013)}]{lim13}
{STScI Development Team: Lim, P. L., Diaz, R. I., \& Laidler, V.} 2013,
  {pysynphot: Synthetic photometry software package}.
\newblock \doeprint{1303.023}

\bibitem[{{Wofford} {et~al.}(2021){Wofford}, {Vidal-Garc{\'\i}a}, {Feltre},
  {Chevallard}, {et~al.}}]{wofford21}
{Wofford}, A., {Vidal-Garc{\'\i}a}, A., {Feltre}, A., {Chevallard}, J.,
  {et~al.} 2021, \mnras, 500, 2908, \dodoi{10.1093/mnras/staa3365}

\end{thebibliography}

\newpage
\appendix
\section{CLASSY Scaling Figures}\label{sec:AppA}
In the following figures we show the flux calibration between the gratings for each of the 45 CLASSY targets. A full description of this flux scaling process can be found in Section~\ref{sec:fluxcalib}, along with Figures~\ref{fig:scaling} and \ref{fig:fuv_scaling}. For each galaxy we show the final coadded dataset and scaling factor applied to the G130M and G185M/G225M gratings, along with the coadded data before and after scaling. Flux calibration procedures for all 45 galaxies in the CLASSY sample are provided in Figure Set 17, which is available in its complete form in the online journal.

\renewcommand{\thefigure}{17}
\begin{figure}[hbp]
\centering
\includegraphics[scale=0.5,trim=170mm 0mm 80mm 0mm,clip]{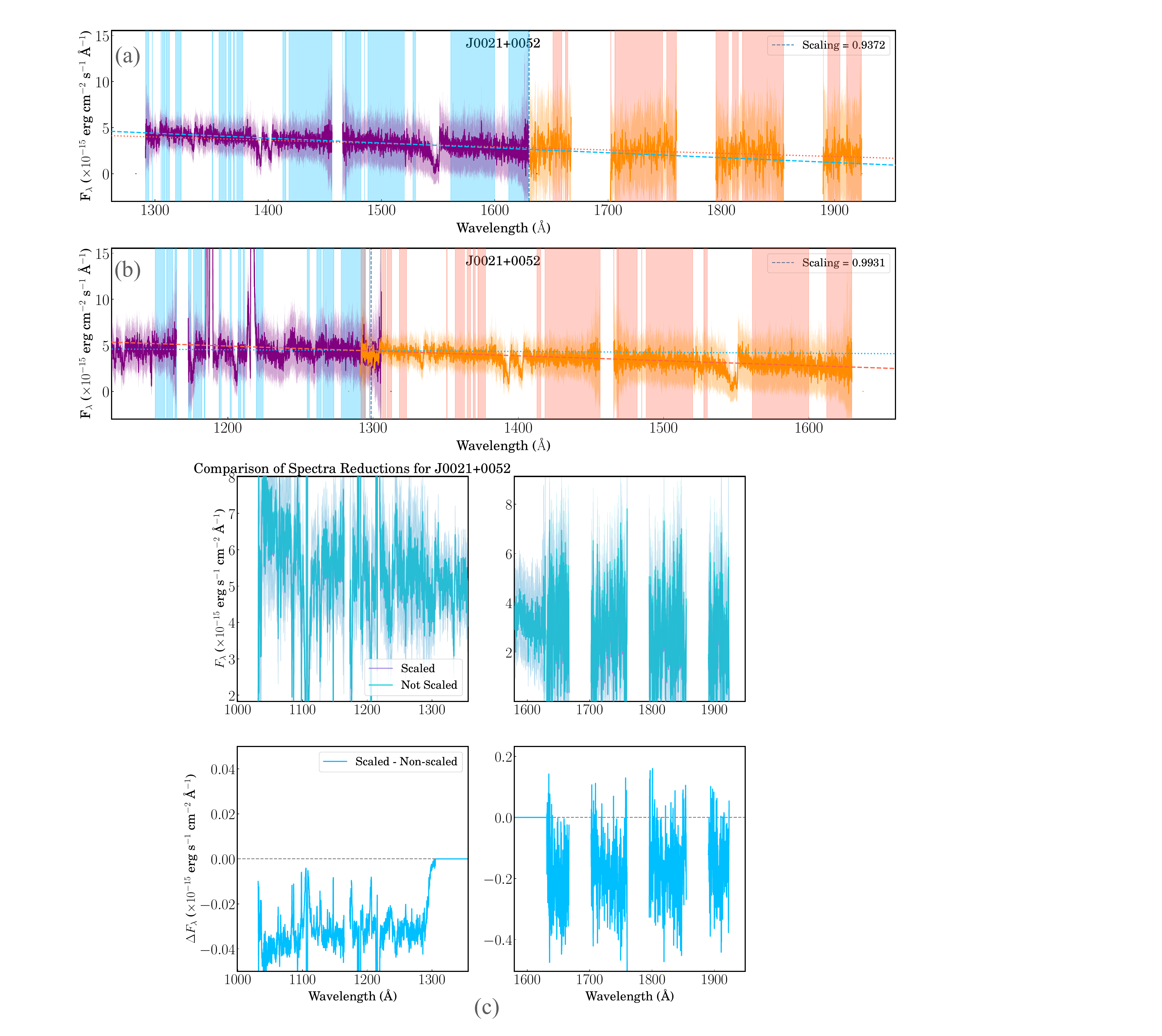}
\caption{Flux calibration procedure for J0021+0052: \textit{(a)} Featureless continuum regions in the G130M/G160M spectrum (blue/orange shading) were fit with a 1D spline (blue-dotted/red-dashed line) and used to scale the G130M spectrum to the G160M spectrum at their mid-way intercept
point (vertical blue dashed line) by the scale factor shown in the legend.; \textit{(b)} Same as in \textit{(a)}, here showing the 1D spline and derived scale factor used to scale the G185M spectrum to the G160M spectrum; \textit{(c)} Final coadded data before (blue) and after (purple) scaling. The top two panels show zoom-ins to the G130M and G185M/G225M datasets, with the corresponding difference between the scaled versus non-scaled data in the lower two panels. In each flux ($F_\lambda$) panel we show the flux$\pm1\sigma$ error spectrum, binned by 6 pixels. }

\label{fig:AppA}
\end{figure}

\figsetstart
\figsetnum{17}
\figsettitle{CLASSY Co-add Scaling: available in the online journal}

\figsetgrpstart
\figsetgrpnum{figurenumber.13}
\figsetgrptitle{J0942+3547}
\figsetplot{FIGURES/appendix/J0942+3547_G130M+G160M_scalings.pdf}
\figsetplot{FIGURES/appendix/J0942+3547_G185M+G160M_scalings.pdf}
\figsetplot{FIGURES/appendix/J0942+3547_scaling_comp.pdf}
\figsetgrpnote{}
\figsetgrpend

\figsetgrpstart
\figsetgrpnum{figurenumber.20}
\figsetgrptitle{J1105+4444}
\figsetplot{FIGURES/appendix/J1105+4444_G130M+G160M_scalings.pdf}
\figsetplot{FIGURES/appendix/J1105+4444_G185M+G160M_scalings.pdf}
\figsetplot{FIGURES/appendix/J1105+4444_scaling_comp.pdf}
\figsetgrpnote{}
\figsetgrpend

\figsetgrpstart
\figsetgrpnum{figurenumber.40}
\figsetgrptitle{J1444+4237}
\figsetplot{FIGURES/appendix/J1444+4237_G130M+G160M_scalings.pdf}
\figsetplot{FIGURES/appendix/J1444+4237_G185M+G160M_scalings.pdf}
\figsetplot{FIGURES/appendix/J1444+4237_scaling_comp.pdf}
\figsetgrpnote{}
\figsetgrpend

\figsetgrpstart
\figsetgrpnum{figurenumber.41}
\figsetgrptitle{J1448-0110}
\figsetplot{FIGURES/appendix/J1448-0110_G130M+G160M_scalings.pdf}
\figsetplot{FIGURES/appendix/J1448-0110_G185M+G160M_scalings.pdf}
\figsetplot{FIGURES/appendix/J1448-0110_scaling_comp.pdf}
\figsetgrpnote{}
\figsetgrpend

\figsetgrpstart
\figsetgrpnum{figurenumber.42}
\figsetgrptitle{J1521+0759}
\figsetplot{FIGURES/appendix/J1521+0759_G130M+G160M_scalings.pdf}
\figsetplot{FIGURES/appendix/J1521+0759_G185M+G160M_scalings.pdf}
\figsetplot{FIGURES/appendix/J1521+0759_scaling_comp.pdf}
\figsetgrpnote{}
\figsetgrpend

\figsetgrpstart
\figsetgrpnum{figurenumber.43}
\figsetgrptitle{J1525+0757}
\figsetplot{FIGURES/appendix/J1525+0757_G130M+G160M_scalings.pdf}
\figsetplot{FIGURES/appendix/J1525+0757_G185M+G160M_scalings.pdf}
\figsetplot{FIGURES/appendix/J1525+0757_scaling_comp.pdf}
\figsetgrpnote{}
\figsetgrpend

\figsetgrpstart
\figsetgrpnum{figurenumber.44}
\figsetgrptitle{J1545+0858}
\figsetplot{FIGURES/appendix/J1545+0858_G130M+G160M_scalings.pdf}
\figsetplot{FIGURES/appendix/J1545+0858_G185M+G160M_scalings.pdf}
\figsetplot{FIGURES/appendix/J1545+0858_scaling_comp.pdf}
\figsetgrpnote{}
\figsetgrpend

\figsetgrpstart
\figsetgrpnum{figurenumber.45}
\figsetgrptitle{J1612+0817}
\figsetplot{FIGURES/appendix/J1612+0817_G130M+G160M_scalings.pdf}
\figsetplot{FIGURES/appendix/J1612+0817_G185M+G160M_scalings.pdf}
\figsetplot{FIGURES/appendix/J1612+0817_G225M+G185M_scalings.pdf}
\figsetplot{FIGURES/appendix/J1612+0817_scaling_comp.pdf}
\figsetgrpnote{}
\figsetgrpend

\figsetend

\end{document}